%% file: main-arXiv.tex
\documentclass[12pt]{article}

\RequirePackage{graphicx}
\RequirePackage{amsthm}
\RequirePackage[cmex10]{amsmath}
\RequirePackage[colorlinks,citecolor=black,urlcolor=black]{hyperref}

\newtheorem{lemma}{Lemma}

\newtheorem*{claim*}{Claim}
\newtheorem{theorem}{Theorem}

\newtheorem{corollary}{Corollary}

\newtheorem{proposition}{Proposition}
\newtheorem{definition}{Definition}

\usepackage[top=1.25in, bottom=1.25in, left=1.25in, right=1.25in]{geometry}
\newtheorem*{example*}{Example}

\usepackage{txfonts}
\usepackage{amsfonts}
\usepackage{epsfig,setspace}
\usepackage{amsthm,array,geometry, wasysym,enumerate}
\usepackage{sgame}
\usepackage{subcaption}
\usepackage{comment}
\usepackage[normalem]{ulem}
\usepackage{amsmath,amssymb,amsthm}
\usepackage{setspace,graphicx}
\usepackage{lmodern}\usepackage{slantsc}\usepackage[cmtip,all]{xy}
\usepackage{accents}
\usepackage{color}
\usepackage[cmyk]{xcolor}
\usepackage{tikz}
\usepackage{pgfplots}
\usepackage{enumerate}
\usepackage{ifthen}
\usepackage{xfrac}
\usepackage{epstopdf}
\usepackage{times} 
\usepackage{titlesec}
\usepackage{verbatim}
\usepackage{mathtools}
\usepackage{natbib}
\DeclareMathOperator*{\argmax}{\mathrm{arg}\!\max\limits}

\newcommand{\supp}{\mathrm{supp}}
\newcommand{\idu}{V}

\newcommand{\qcost}{\kappa}\newcommand{\icost}{c}\newcommand{\ddd}{\,\mathrm{d}}\newcommand{\dd}{\mathrm{d}}\newcommand{\Signals}{\mathcal{I}}\newcommand{\Signal}{\mathcal{I}}\newcommand{\DMPrice}{P}\newcommand{\DMprice}{p}\newcommand{\DMpriceset}{\mathcal{P}}
\newcommand{\minidu}{\underline{u}}\newcommand{\thmax}{\bar{\theta}}\newcommand{\thmin}{\underline{\theta}}\newcommand{\Alloc}{\mathcal{Q}}

\newcommand{\ICC}{\mathcal{I}}

\newcommand{\costcoef}{\gamma}
\newcommand{\avgdist}{w}

\newcommand{\citeapos}[1]{\citeauthor{#1}'s (\citeyear{#1})}

\newcommand{\acomment}[1]{}
\newtheorem{condition}{Condition}

\usetikzlibrary{calc,arrows,automata,shapes.misc,shapes.arrows,chains,matrix,positioning,scopes,decorations.pathmorphing,shadows}

\makeatletter
  \renewcommand\@seccntformat[1]{\csname the#1\endcsname.{\hskip.7em\relax}} \makeatother

\title{{\bf Monopoly, Quality, and Costly Information
}\footnote{This paper supersedes two distinct papers: ``Monopoly, Product Quality, and Flexible Learning" by Mensch and Ravid, and ``Screening Costly Information" by Thereze. Mensch acknowledges support from the Israel Science Foundation (grant 798/18). We would like to thank Sylvain Chassang, Alex Gershkov, Elliot Lipnowski, Alessandro Lizzeri, Pietro Ortoleva, Wolfgang Pesendorfer, and Leeat Yariv for their helpful comments and feedback. We would also like to thank conference participants
at EEA-ESEM 2021, Stony Brook 2022, EEA-ESEM 2022, CMID 2022, EC 2024,  and seminar participants at the HUJI-TAU joint theory seminar and the Technion. Karen Wu and Guy Yanay provided superb research assistance.}}

\author{
\begin{minipage}{0.33\textwidth}\centering  
Jeffrey Mensch\footnote{\scriptsize{\texttt{jeffrey.mensch@mail.huji.ac.il,  https://sites.google.com/site/jeffreyimensch/.}}} \\ \centering \it \small Hebrew University of Jerusalem
\end{minipage}                  
\begin{minipage}{0.33\textwidth}\centering 
Doron Ravid\footnote{\scriptsize{\texttt{doronr@umich.edu, http://doronravid.com.}}}  \\ \centering \it \small University of Michigan
\end{minipage}
\begin{minipage}{0.33\textwidth}\centering
    João Thereze\footnote{\scriptsize{\texttt{joao.thereze@duke.edu, https://joaothereze.com.}}} \\ \centering \it \small Duke University 
\end{minipage}
}

\date{\vspace{0.8cm} \today}

\begin{document}
\maketitle

\begin{abstract}

\noindent 

A seller offers a buyer a schedule of transfers and associated product qualities. After observing this schedule, the buyer chooses a flexible costly signal about his type. We show it is without loss to focus on a class of allocations that compensate the buyer for his marginal learning costs. In the optimal menu, all types typically receive lower-than-efficient quality (including “at the top”). Moreover, profits are non-monotonic in the level of information costs, and the consumer may be better off when such costs are low than when information is free. When learning costs are steep, the optimal menu contains at most two purchasing options.

\vspace{.2cm}

\end{abstract}

\section{Introduction}

Consumers are often unsure about which product is right for them. Frequently, additional information that could aid the buyer's decision is available, but acquiring and processing it requires effort. In health insurance markets, for instance, potential buyers struggle to select the most suitable plan, incurring large losses for choosing the wrong product \citep{AG11,HKT15, BJ24}. Consumers may consider their past experience and investigate their family health history to forecast coverage needs, or they may compare how different plans cover various medical conditions. Whether they are assessing their personal needs or evaluating how different products align with their preferences, acquiring information entails costs. We study how this costly information acquisition in demand affects market outcomes when a firm screens consumers.

Understanding endogenous responses to information frictions is important for several reasons. First, extensive evidence shows that such frictions significantly affect consumer behavior \citep{AG11, T18, BJ24}. Second, supply responses shed light on producers' incentives to aid or hinder consumer learning. Indeed, insurance company websites offer tools that help consumers compare plans, suggesting that sellers act to influence information acquisition. Finally, information frictions justify public interventions---in health care, governments may provide instruments for price transparency \citep{B19}, or standardize and simplify information on insurance plans \citep{ES16}. The welfare consequences of these interventions cannot be ascertained without considering endogenous effects. In fact, we show that helping buyers learn may reduce consumer welfare.

\paragraph{The Model} We study a seller who decides what menu of vertically differentiated products to offer to a potential buyer. Unlike the classical models of \citet{mussa1978monopoly}
and \citet{maskin1984monopoly}, we do not assume the buyer possesses
private information when he first sees the seller's menu. Instead,
the buyer sees this menu, and then \emph{chooses} what to learn about his type. The buyer's information choice is flexible and costly; that is, the consumer is rationally inattentive \citep{sims2003implications}. Because learning takes place after the seller makes her offer, the menu of products affects information choices: the seller can influence which information the buyer acquires by choosing which products to provide at which prices. Our main interest
is in the structure of this menu and the efficiency of the resulting
allocation with respect to the buyer's chosen information.

Our key modeling assumption is that information costs are \emph{posterior-mean separable}. That is, we assume that for the buyer to learn a certain signal about his type, he faces a cost that is linear in the distribution of posterior means generated by that signal. With the standard payoff specification in \citet{mussa1978monopoly}, this assumption implies that the buyer's optimal learning program is a special case of mean-measurable information design \citep{gentzkow2016rothschild,dworczak2019simple,arieli2020optimal,kleiner2021extreme}. Unlike their more general counterpart, mean-measurable information design problems remain tractable even when the underlying state space is very large. Consequently, we can accommodate discrete and continuous prior distributions. When the prior distribution is binary, the class of cost functions we entertain includes the standard posterior-separable costs \citep{caplin2021rationally}, commonly used in the rational inattention literature.

Information acquisition brings a new dimension to the seller's problem: the product line must induce the buyer to learn what the seller wants him to. Optimal learning for the buyer then imposes joint constraints on the amount of product differentiation and the resulting information asymmetry. We show one can summarize these constraints by focusing on the class of \emph{cost-compensating} allocations. These allocations decompose the buyer's gross surplus into two parts: one part that compensates the buyer's cost of learning, and a residual that comes from the shadow value of information. Together, these two parts encode the restrictions imposed by the buyer's optimal learning constraint on the seller's product line: the buyer's chosen quality must increase at a rate no slower than the change in the buyer's learning costs, and can increase at faster rate only when the shadow value of information is positive.

The reduction to cost-compensating allocations enables us to reason about the seller-optimal outcome using two familiar-looking programs. The first program involves maximizing the seller's profits across all cost compensating allocations that induce the buyer to learn a certain signal. This program resembles the standard mechanism design problem. The second program searches for the most profitable signal for the seller among those signals that the buyer can be induced to learn given a fixed allocation. When this allocation is cost-compensating, the second program turns out to be a ``constrained" mean-measurable information design problem.

\paragraph{Results} We prove three main results about the seller's optimal outcome. First, we identify a sense in which information acquisition exacerbates the inefficiencies generated by information asymmetry alone. As in standard screening, the allocation of quality to consumers is always weakly below the efficient one. However, when information costs are steep, quality distortions are more widespread than in standard screening: agents with all valuations receive below-efficient quality levels---in contrast to the standard ``no distortion at the top'' result, where the highest-valuation buyer's quality is efficient.  

Our model highlights two novel forces leading to downward quality distortions. The first force is the need to deter the buyer from changing his information. In particular, the seller must consider the rents she provides to unrealized types. For intuition, consider the simple case in which information is symmetric at the optimum---i.e., the buyer chooses to remain uninformed.\footnote{We prove that information is indeed symmetric at the optimum when information costs are sufficiently high.} In standard screening, if information is symmetric, the seller offers a single product, with efficient quality tailored to the buyer's valuation. Moreover, the product is priced so  that the consumer is indifferent between buying or not. Such indifference cannot occur in our model. If the consumer is ex-ante indifferent between buying and not buying, he can strictly benefit from acquiring information: if he gets a signal indicating that his type is low, he can not purchase the product. Hence, the buyer avoids learning only if the seller offers a price discount, which creates a wedge between the seller's marginal profit from quality and quality's marginal social benefit. The resulting contract is therefore inefficient even though information is symmetric. This example highlights a key distinction between costly information and standard screening: the mere \emph{threat} of obtaining information is distortionary, even without the presence of private information.

A second novel distortionary force comes from the seller steering the buyer's information acquisition. For an illustration, suppose there are only two types, and that the seller finds it optimal to induce the buyer to acquire a signal with two realizations, neither of which fully reveals any type. By reducing the quality differential between the two realizations, the seller can induce the buyer to acquire a less informative signal that pools more of the high type into the signal realization that values quality less. This deviation increases the seller's profits from the low signal realization, which receives a higher quality at a  higher per-quality price. The impact on the profit from the high realization is less clear: while that realization values quality less and receives a lower quality product, it also receives lower information rents. It turns out that the social value of changing that realization's quality dominates: profits decrease at the margin only if the marginal unit of quality has positive social value. And, indeed, the seller's profit from the high-realization must go down; otherwise, this deviation would increase the seller's profits. Therefore, increasing the high-realization's quality must increase social surplus---that is, quality must be distorted downwards, even at the top.

The second main result is that profits and consumer surplus can be non-monotonic in the cost of learning. Profits increase and consumer surplus decreases when the scale of information costs is high, and they move in the opposite direction when such scale is low, under suitable conditions. Heuristically, this is because the value of the agent's preferred deviations varies as costs change. When information costs are high, the seller induces the buyer to acquire very little information. In that case, the buyer extracts rents by threatening to learn more than the seller wants. But that threat becomes less pressing when information costs grow, benefiting the seller at the buyer's expense. The symmetric logic delivers a similar result when information costs are low: in that region the buyer extracts rents by threatening to learn less, a threat that becomes more potent as information costs increase. Hence, profits fall with information costs when learning is cheap, but consumers receive further rents. In that case, the increased consumer rents are offset by the growth of information costs, and we characterize when the net consumer surplus grows or falls as a function of the shape of learning costs.

Finally, we show that simple menus are optimal when information costs are steep. In other words, the possibility of information acquisition by the buyer can make it optimal for the seller to offer just a few product options, even when the set of possible buyer types is very large. 
In particular, we prove that with steep information costs, the seller can always maximize profits by inducing a binary signal, and so a two-product menu must be optimal due to a revelation-style argument. Intuitively, the seller benefits from inducing a signal with many realizations only if that signal is sufficiently informative, which is too expensive to do when learning costs are steep. 
We further identify conditions in which the buyer's signal is binary, but the seller can attain the optimum offering exactly one purchasing option. This explanation for menu simplicity stands in contrast to the full-information case, where small menus typically arise only in restrictive cases, such as when the set of possible buyer types is small, or when the marginal production costs are constant in quality.

\bigskip

\paragraph{Related Literature} Our paper belongs to the literature on flexible information acquisition in trade \citep{condorelli2020information,ravid2022learning}. Closest to our work, \citet{mensch2022screening} studies the optimal sale of an indivisible good to a buyer who can flexibly acquire information about her value after observing the menu. In contrast to our setting, his model features posterior-separable---rather than \emph{posterior-mean} separable---information costs, and costless ``quality,'' interpreted as the probability of receiving the good. These differences generate distinct implications: whereas \citet{mensch2022screening} obtains efficiency at the highest signal, we often find inefficiently low quality provision at every signal, including the highest. We also derive comparative statics and conditions for the optimality of simple menus in environments with multiple states, which have no direct counterparts in his analysis. \citet{condorelli2020information}
and \citet{ravid2022learning} are also closely related to our work. Like \citet{mensch2022screening}, they study bilateral
trade with a single indivisible good and costly flexible learning. However, their models differ from ours in their timing assumptions. 

We are also related to the literature analyzing the intersection of information design and mechanism design \citep[e.g.,]{bergemann2015limits,Roesler2017,haghpanah2023pareto}. The most closely related paper is \citet{bergemann2026screening}, who study a different extension of the \citet{mussa1978monopoly} model in which the seller also chooses the buyer's information. They show that it is always optimal to provide the buyer with a partitional signal. Partitional signals do not generally arise in our setting because information acquisition is chosen by the buyer rather than designed by the seller.

More broadly, our paper contributes to the literature on rational inattention \citep{sims1998stickiness,sims2003implications} and flexible information acquisition \citep{caplin2013behavioral,caplin2015revealed,matvejka2015rational,caplin2021rationally}. We are particularly related to applications of costly information acquisition in strategic environments, including global games \citep{yang2015coordination,morris2021coordination,denti2022unrestricted}, bargaining \citep{ravid2020ultimatum}, attention management \citep{lipnowski2020attention}, and security design \citep{yang2019optimality}.

\section{Model \label{section:Model}}

A monopolistic seller (she) interacts with a buyer (he). The game begins with
the seller offering the buyer a menu, which is a compact set
of pairs, $M\subseteq\left[0,\bar{q}\right]\times\mathbb{R}$.\footnote{We require the menu to be compact to ensure existence of an optimal choice for the buyer.} Each
menu item $\left(q,t\right)\in M$ corresponds to a transfer of $t$
to be paid to the seller by the buyer, and the quality $q$ of
the product the buyer gets in exchange. The buyer's utility from $\left(q,t\right)$
depends on his quality valuation, $\boldsymbol{\theta}$, a random variable distributed
over $\Theta=[\underline{\theta},\bar{\theta}]\subseteq\mathbb{R}_{+}$
according to a CDF $F_{0}$, with mean $\theta_0$. Without loss of generality, $F_{0}$ includes $\thmin$ and $\thmax$ in its support. Given $\theta$, the buyer's utility from $(q,t)$ is 
\[
U\left(\theta,q,t\right)=\theta q-t.
\]
The seller's profit from the buyer's chosen menu item $\left(q,t\right)$
is 
\[
\Pi\left(q,t\right)=t-\qcost\left(q\right),
\]
where $\qcost:\mathbb{R}_{+}\rightarrow\mathbb{R}_{+}$ is an
increasing, twice continuously differentiable, and strongly convex function
satisfying $\qcost\left(0\right)=0$.\footnote{A twice continuously differentiable function $f$ is strongly convex if $\min_x f''(x) > 0$.} We also assume it is efficient to serve all types, that is $\qcost'(0) < \thmin$ (though we can relax this assumption---see Section~\ref{sec: Discussion}), and that $\qcost'(\bar{q})\geq \thmax$, meaning the set of available qualities is sufficiently wide.\footnote{This assumption means that $\bar{q}$ only serves the technical role of ensuring compactness of the set of possible qualities.}
The seller's menu must give the buyer the option of not buying anything; the menu $M$ must include $\left(0,0\right)$.
Both the seller and the buyer are risk-neutral expected
utility maximizers.

Neither the seller nor the buyer know $\boldsymbol{\theta}$,
but the buyer can choose to learn about it after observing the seller's
menu. The buyer's information acquisition is flexible: he
can use any signal $\mathbf{s}$ to learn about $\boldsymbol{\theta}$. 
The flexibility assumption expresses two ideas. First, the buyer has access to detailed data that helps him determine his exact value for the item. Second, the buyer can collect only the information he finds useful, avoiding any unnecessary effort to gather or process data he wishes to disregard.

The functional form of the buyer's utility means that his expected payoff
from any menu item depends on his posterior mean, $\mathbb{E}\left[\boldsymbol{\theta}|\mathbf{s}\right]$.
Therefore, the marginal distribution of $\mathbb{E}\left[\boldsymbol{\theta}|\mathbf{s}\right]$
pins down the buyer's expected trade surplus from any menu. This distribution also determines the probability the buyer purchases any menu item, which, in turn,
is sufficient for calculating the seller's profits and optimal
menu. In other words, trade outcomes depend only on the marginal distribution
of the buyer's posterior mean, and so we identify each signal with
the CDF of this marginal.\footnote{This method of modeling flexible information is common in the information-design
literature; see, for example, \citet{gentzkow2016rothschild},
\citet{Roesler2017}, \citet{kolotilin2018optimal}, and \citet{dworczak2019simple}.} More precisely, letting $\mathcal{F}$ be the set of all CDFs over
$\Theta$, we let the buyer choose any element of $\mathcal{F}$ that
can arise as the marginal CDF of $\mathbb{E}\left[\boldsymbol{\theta}|\mathbf{s}\right]$
for some $\mathbf{s}$. We denote this set by $\Signals$ and describe
it formally below.

As observed by \citet{gentzkow2016rothschild}, $F$ is the CDF of
the marginal distribution of the buyer's posterior mean for some signal
if and only if it is a mean-preserving contraction of the prior, $F_{0}$.
Recall that $F\in\mathcal{F}$ is a \textbf{mean-preserving spread} 
of $G\in\mathcal{F}$ (denoted by $F\succeq G$) if and only if
\[
\int_{\tilde{\theta}\leq\theta}\left(F-G\right)(\tilde{\theta})\dd\tilde{\theta}\geq0 \ \text{for all} \ \theta\in\Theta, \ \text{with equality at }\theta=\bar{\theta}.
\]
Therefore, one can describe the set of feasible posterior mean distributions via
\[
\Signals=\left\{ F\in\mathcal{F}: F_0 \succeq F \right\}.
\]
We refer to CDFs in $\Signals$ as \textbf{signals}. Given $F \in \Signals$, we denote the lowest and highest realizations $F$ can generate by $\thmin_F:=\min[\supp \ F]$ and $\thmax_F:=\max[\supp \ F]$, respectively. The integral $\int_{\tilde\theta\in[\underline{\theta},\theta]}\left(F_{0}-F\right)(\tilde{\theta})\mathrm{d}\tilde{\theta}$ measures the degree to which the signal that generates $F$ pools states above and below $\theta$. Indeed, this integral is zero if and only if every signal structure that generates $F$ does not pool types strictly below $\theta$ with types strictly above $\theta$ with positive probability; that is, the signal must separate types above and below $\theta$. We follow \cite{ravid2022learning} and refer to any such $\theta$ as $F$-\textbf{separating}, and refer to any $\theta$ that is not $F$-separating as $F$-\textbf{pooling}.

Information acquisition comes at a cost. In general,
different information structures generating the same distribution $F$
of posterior expectations might come at different costs. However,
because the buyer's expected payoff from trade depends only on this $F$, she would always
use the least expensive signal structure that leads to $F$. In fact,
the buyer may even get $F$ by randomizing. Thus, we can evaluate the
cost of $F$ by the expected cost of the cheapest randomization that
generates it, resulting in an indirect cost function, $C:\Signals\rightarrow\mathbb{R}_{+}$. 
We follow \citet{ravid2020learning} and state our assumptions directly
in terms of this $C$. We assume $C$ is continuous, affine,\footnote{Some readers may be interested in deriving these properties from a more primitive object, such as cost function defined over the distribution of the agent's posterior belief. To do so, one can assume the cost of a distribution over posteriors is posterior separable \citep{caplin2021rationally}, with a divergence function that depends only on posterior's mean. \cite{mensch2023posterior} provides a revealed preference characterization of such cost functions.} and strictly
increasing in informativeness; that is, $C\left(F\right)>C\left(G\right)$
whenever $F$ is a strict mean-preserving spread of $G$. 
In the online appendix, we prove these properties imply the existence of some continuous, 
strictly convex function $c:\Theta\rightarrow\mathbb{R}_{+}$ such that 
\[
C\left(F\right)=\int\icost(\theta)F\left(\dd \theta\right).
\]
Moreover, we show it is without loss for $\icost$ to attain its minimum at $\theta_0$. 

In addition to the above, we assume $c$ is thrice differentiable with a strictly positive and bounded second derivative. In Section~\ref{sec: Discussion}, we discuss which results extend to the case in which the derivatives of $\icost$ are unbounded.

To summarize, the game begins with the seller choosing a menu. 
Next, the buyer observes the menu, and chooses what signal $F\in\Signals$ to acquire. 
The buyer then sees his signal realization $\theta\in\Theta$, and chooses an item from the seller's menu. We are interested in the menu and signal that maximize the seller's expected profits, subject to the buyer behaving optimally, which exists by the following lemma.

\begin{lemma}
\label{Existence}A seller-optimal menu exists.
\end{lemma}

Our timing assumptions imply the buyer's interim expected payoff from each menu item is
fully determined by his posterior expectations, and so this expectations completely determine the buyer's decision from the seller's menu. The revelation
principle therefore enables us to focus on direct revelation
mechanisms in which the buyer reports his posterior mean. Such menus can be described with two maps,
\[
Q:\Theta\rightarrow\left[0,\bar{q}\right],\,T:\Theta\rightarrow\mathbb{R}_{+},
\]
where $Q(\theta)$ and $T(\theta)$ correspond
to the quality and transfer pair chosen by a buyer with posterior mean
$\theta$. These mappings must satisfy the standard incentive compatibility and individual
rationality constraints,\footnote{For every menu, every type $\theta \in \Theta$ admits some optimal selection from that menu. Letting $(Q(\theta),T(\theta))_{\theta \in \Theta}$ be this selection then delivers a $Q$ and a $T$ that satisfy \eqref{eq:IC} and \eqref{eq:IR}. Note these constraints are satisfied even for posterior means that do not arise under the buyer's chosen CDF.} 
\begin{align}
\theta Q(\theta)-T(\theta) & \geq\theta Q\left(\theta'\right)-T\left(\theta'\right)\quad\forall\theta,\theta'\in\Theta,\label{eq:IC}\tag{IC}\\
\theta Q(\theta)-T(\theta) & \geq0\quad\forall\theta\in\Theta.\label{eq:IR}\tag{IR}
\end{align}
Usual envelope-style reasoning \citep[][]{myerson1981optimal}
delivers that a $Q$ and $T$ satisfy the above two conditions if
and only if $Q$ is increasing and 
\begin{equation}
T(\theta)=\theta Q(\theta)-\int_{\underline{\theta}}^{\theta}Q(\tilde{\theta})\mathrm{d}\tilde{\theta}-\underline{u},\label{eq: EnvelopeWage}
\end{equation}
where $\underline{u}\geq 0$ is the utility granted to the lowest possible
type,
\[
\minidu=\underline{\theta}Q\left(\underline{\theta}\right)-T\left(\underline{\theta}\right).
\]
It follows that $\underline{u}$ and $Q$ are sufficient for pinning
down every feasible IC and IR menu. Let $\mathcal{Q}$
be the set of all increasing functions from $\Theta$ to $\left[0,\bar{q}\right]$.
We refer to a $Q\in\Alloc$ as an \textbf{allocation}. For essentially the usual reasons, setting $\underline{u}=0$ is always optimal for the seller. Therefore, the allocation alone pins down the payoffs of both players, and we henceforth omit $\underline{u}$. Given an allocation $Q$, we let $T_{Q}$ denote the transfer
implied by (\ref{eq: EnvelopeWage}) with $\underline{u}=0$. This description implies 
that a type-$\theta$ buyer's utility from truthful reporting
under $Q$ is
\[
\idu_{Q}(\theta):=\theta Q(\theta)-T_{Q}(\theta)=\int_{\underline{\theta}}^{\theta}Q(\tilde{\theta})\mathrm{d}\tilde{\theta}.
\]
We refer to the function $\idu_Q$ as the buyer's \textbf{gross value}. 
Given an allocation $Q$, let $\thmin_Q$ be the highest type that $Q$ excludes, and $\thmax_Q$ to be the lowest type to which $Q$ gives the highest quality. If no type is excluded (gets the highest quality), set $\thmin_Q=\thmin$ ($\thmax_Q=\thmax$).\footnote{Formally, set $\thmax_{Q} := \sup\left[\{\thmin\} \cup Q^{-1}[0,\bar{q})\right]$ and $\thmin_{Q} = \inf\left[\{\thmax\} \cup Q^{-1}(0,\bar{q}]\right]$.
}

We now state the seller's problem of choosing a
profit-maximizing mechanism. Given an allocation $Q$, the buyer's utility from using signal $F\in\Signals$ is given by his expected utility net of information costs. We refer to an allocation-signal tuple $\left(Q,F\right)$ as
\textbf{an outcome}, and say the outcome is \textbf{incentive compatible
}(IC) if $F$ maximizes the buyer's utility given $Q,$
\begin{equation}\label{eq: buyer's problem}
F\in\argmax_{\tilde{F}\in\Signals} \int \left(\idu_{Q} (\theta) - \icost(\theta) \right)\tilde{F}\left(\mathrm{d}\theta\right).
\end{equation}
Denote the seller's payoff when the buyer reports a signal realization of $\theta$ by
\[
\pi_{Q}(\theta):=T_{Q}(\theta)-\kappa\left(Q(\theta)\right)=\theta Q(\theta)-\idu_{Q}(\theta)-\qcost\left(Q(\theta)\right).
\]
Then the seller's expected profit from
offering $Q$ when the buyer uses $F$ is $
\int\pi_{Q}(\theta)F\left(\mathrm{d}\theta\right)$, 
and so the seller's program is given by
\begin{align*}
\max_{\left(Q,F\right)} & \int\pi_{Q}(\theta)F\left(\mathrm{d}\theta\right)\,\text{s.t. }\left(Q,F\right)\text{ is IC}.
\end{align*}
The goal of this paper is to study the above program. 

Before proceeding, we introduce two helpful definitions. Fix an increasing function $\varphi:\Theta \rightarrow [x,y]$. We say $\varphi$ is \textbf{constant around} $\theta$ whenever it is constant in some open neighborhood of $\theta$. If $\varphi$ is not constant around $\theta$, we say that $\varphi$ is \textbf{strictly increasing} at $\theta$. 

\paragraph{Binary types}
Throughout the paper we utilize the case in which $F_0$ is binary---$|\supp F_0|=2$---to help illustrate and provide intuition about the general problem. Formally, this case assumes $F_0$ is the CDF with mean $\theta_0$ and $\supp \ F_0 = \{\thmin,\thmax \}$. In this case, any CDF $F \in \mathcal{F}$ with mean $\theta_0$ is a mean-preserving contraction of $F_0$. That is:

$$\Signals = \left\{F \in \mathcal{F}: \int \tilde{\theta} F(d\tilde{\theta}) = \theta_o  \right\}. $$

Finally, the assumption of affine information costs is less restrictive when the prior is binary. With only two states, the posterior mean fully identifies the posterior distribution. Therefore, a function is affine in posterior means whenever it is affine in the posterior distributions themselves. Indeed, for binary values, the affinity assumption is satisfied by the class of posterior-separable information costs \citep{caplin2021rationally}, which are widely used in the applied literature.

\section{Cost-Compensating Allocations}\label{section:F-ICC}

This section shows it is without loss of optimality to restrict attention to allocations that ensure the buyer's information rents compensate him for the costs of learning. We refer to these allocations as \emph{cost compensating}. Using these allocations, we decompose the seller's problem into two familiar maximization programs: one that involves maximization over allocations, and the other that optimizes over the buyer's information. 

Our definition of cost-compensating allocations builds on \citeapos{dworczak2019simple} duality results. Specifically, $F$ solves \eqref{eq: buyer's problem} if and only if a Lipschitz continuous and convex function $P:\Theta \rightarrow \mathbb{R}$ exists that is affine on any open interval of $F$-pooling types, satisfies $\DMPrice \geq \idu_Q-\icost$, and has
\[
\DMPrice(\theta) = \idu_Q(\theta)-\icost(\theta), \text{ for all } \theta \in \supp F.
\]
\cite{dworczak2019simple} refer to such a $\DMPrice$ as a \emph{price function}. A price function gives the shadow value of producing mean-preserving spreads and contractions. Such spreads never hurt the buyer, so this shadow value is always convex. Moreover if this value is non-affine over some interval, the buyer strictly benefits from producing a local mean-preserving spread. Therefore optimality requires that whenever such spread is feasible, $\DMPrice$ must provide the buyer with no incentives to acquire information; that is, $\DMPrice$ must be affine over $F$-pooling intervals. Given a signal $F$, a convex and Lipschitz continuous function $\DMPrice$ is an $F$-\textbf{price} if it is affine on every $F$-pooling interval.

This duality-based characterization motivates us to restrict our attention to a subset of allocations in which the buyer's net value $\idu_Q - \icost$ coincides with the associated price $\DMPrice$ over the relevant range of types. To achieve such net value, the allocation offered by the seller must provide rents $V_Q$ to the buyer that compensate him for the cost of learning $\icost$ on top of the utility given by the price function. We call such allocations $F$-cost-compensating.

\begin{definition}[Cost-Compensating Allocations] 
An allocation $Q$ is F-\textbf{cost-compensating ($F$-CC)} if an $F$-price exists such that $\idu_Q \leq \DMPrice + \icost$, and
\begin{equation}\label{eq:U-as-P-plus-c}
    V_Q(\theta) \;=\; \DMPrice(\theta) + \icost(\theta)
    \quad\text{for all }\theta\in[\underline{\theta}_Q, \bar{\theta}_Q].
\end{equation}
\end{definition}

Economically, $F$-CC allocations ensure that the buyer's net-information rents, $\idu_Q - \icost$, coincide with the shadow value of generating additional mean-preserving spreads. This coincidence ensures that the buyer is indifferent at the margin between acquiring more or less information over intervals of $F$-pooling types. Moreover, the buyer strictly benefits from further splitting a signal realization of $\theta$ only if she has no more information to learn---that is, only if $\theta$ is $F$-separating. 

While an $F$-price $\DMPrice$ determines the buyer's value from every $F$-CC allocation, it does not pin down the allocation itself. In particular, the allocation is indeterminate at any point in which $\DMPrice$ has a kink.\footnote{Such points are important because $F$ need not absolutely continuous, and so the value of the allocation at Lebesgue-null sets can impact the seller's profits and social surplus.} We therefore parameterize $F$-CC allocations via a novel object: an $F$-marginal price. To introduce this object, note that because $\DMPrice$ is convex and Lipschitz continuous, the usual envelope condition implies that every $F$-CC allocation $Q$ can be rewritten as

\begin{equation}\label{eq:fccallocation}
    Q(\theta) = \DMprice(\theta) + \icost'(\theta), \quad \forall~\theta ~\text{  s.t.  }~\DMprice(\theta) + \icost'(\theta) \in[0, \bar{q}],
\end{equation}
where $p:\Theta \rightarrow \mathbb{R}$ satisfies 

\[
\DMPrice(\theta) = \DMPrice(\thmin) + \int_{\thmin}^{\theta} \DMprice(x)\,dx, \ \theta \in \Theta.
\]

Moreover we can choose $\DMprice$ so that it satisfies two useful properties. First, because the allocation $Q$ must lie between 0 and $\bar{q}$, $\DMprice$ must lie between $-\icost'(\thmin_F)$ and $\bar q - \icost'(\thmax_F)$ over the support of $F$; that is,

    \begin{equation} \label{eq: p-range}
        -\icost'(\thmin_F) \;\leq\; \DMprice(\theta) \;\leq\; \bar q - \icost'(\thmax_F)
        \quad\text{for all }\theta\in[\underline{\theta}_F, \bar{\theta}_F].
    \end{equation}

And second, because every price function is convex and affine over $F$-pooling intervals, we can select $\DMprice$ to be a non-decreasing function that is constant over such intervals. We name any $\DMprice$ satisfying these two properties as an $F$-\textbf{marginal price}, as it is the (almost everywhere) derivative of a price function. 

Unlike price functions, marginal prices completely determine the corresponding cost-compensating allocations. In particular equation~\eqref{eq:fccallocation} shows the quality a cost-compensating allocation delivers to type $\theta$ is pinned down by the sum of the corresponding marginal price $\DMprice$ and the marginal cost of information $\icost'$. Therefore, this equation implies that every $F$-marginal price $\DMprice$ corresponds to a unique $F$-CC allocation, and vice-versa. For every $F$-marginal price $\DMprice$, we let $Q_\DMprice$ denote the induced $F$-CC allocation, and for every $F$-CC allocation $Q$, we let $\DMprice_Q$ be the unique $F$-marginal price which satisfies equation~\eqref{eq:fccallocation}.

Next we demonstrate the above definitions in the context of a binary-prior example.

\paragraph{Example. (Binary types and Variance reduction)} In Figure~\ref{fig: binary FCC}, we illustrate the construction of an $F$-CC allocation from an incentive compatible outcome $(Q,F)$. We assume the prior $F_0$ is binary, and that information costs are quadratic:
\[
\icost (\theta) = \frac{1}{2} (\theta-\theta_0)^2.
\]
Under this specification, the cost of learning a signal is proportional to the reduction in the variance of $F_0$ obtained by observing that signal.

The left panel plots the buyer's net utility $\idu_Q - \icost$ from the allocation $Q$ depicted in dashed black lines in the right panel. This allocation is piece-wise constant, and corresponds to a menu with one cheap, low quality option, and one expensive, high-quality alternative. The high quality alternative is purchased by buyers with high type realizations, while buyers with intermediate realizations purchase the low-quality option. Low $\theta$ buyers do not purchase anything. The result is a gross value function $\idu_Q$ that is piece-wise affine, and a piece-wise concave net value $V_Q - \icost$, as depicted in black on the left panel. The optimal signal $F$ for the buyer is a binary distribution with $\supp \ F = \{\thmin_F, \thmax_F \}$. As \cite{dworczak2019simple} show, one can certify the optimality of this signal using the price function $\DMPrice$, drawn in red. 

Note that $\DMPrice$ must be affine because, for a binary prior, all types in $(\thmin,\thmax)$ are $F$-pooling as long as $F\neq F_0$. Thus, with binary prior and partially revealing signals, price functions are essentially just supergradients of the buyer's net-value function: they correspond to affine functions that support the net-value from above. This observation highlights an alternative method for solving the buyer's problem when the prior is binary: concavify the buyer's net-value $\idu_Q - \icost$ as in \cite{Kamenica2011}, and find the supergradient that is tangent to the concavification's graph at the prior. This method is essentially the Lagrangian lemma from \cite{caplin2021rationally}.

The coincidence between \citeapos{dworczak2019simple} approach and \citeapos{caplin2021rationally} Lagrangian lemma breaks down whenever there are interior separating types. With a binary prior, such a breakdown occurs if and only if the seller induces full learning; that is $F=F_0$. In that case, every type in $\Theta$ is separating, and so every Lipschitz and convex function is an $F_0$-price. Consequently, an allocation $Q$ is $F_0$-CC if and only if $Q-\icost'$ is increasing. 

On the right panel we construct an $F$-CC allocation from the derivative of the price $\DMPrice$, which is a constant $F$-marginal price $\DMprice$. We start by plotting the marginal cost of information, $\icost'$, which is linear. Following equation~\eqref{eq:fccallocation}, we obtain the $F$-CC allocation $Q_\DMprice$ (depicted in light blue) by summing $\DMprice$ and $\icost'$, and then applying a cap of $\bar{q}$ and a floor of $0$. In the depicted example, the sum $\DMprice+\icost'$ exceeds the quality upper-bound $\bar{q}$ for high enough types, and so $Q_\DMprice$ gives those types a quality of $\bar{q}$. Similarly, the depicted allocation $Q_\DMprice$ does not serve the types for which the sum $\DMprice+\icost'$ is negative. Note that $Q_\DMprice(\theta) = Q(\theta)$ for every $\theta \in \supp \ F$.
\hfill$\blacksquare$

\paragraph{ }

\begin{figure} \centering
\begin{subfigure}{.46\textwidth}\centering
    \begin{tikzpicture}

\pgfmathsetmacro{\qHval}{0.5}
\pgfmathsetmacro{\qLval}{0.2}
\pgfmathsetmacro{\tHval}{0.2}
\pgfmathsetmacro{\tLval}{0.03}

\pgfkeys{/pgf/fpu=true}

\pgfmathsetmacro{\thetaH}{ 0.5*( \qHval*\qHval - 2*\qHval*\qLval + \qLval*\qLval + 2*\tHval - 2*\tLval ) / (\qHval-\qLval) }
\pgfmathsetmacro{\thetaL}{ \thetaH - (\qHval-\qLval) }

\pgfmathsetmacro{\yL}{ max(0, max(\thetaL*\qLval-\tLval, \thetaL*\qHval-\tHval)) - 0.5*(\thetaL-0.5)^2 }
\pgfmathsetmacro{\yH}{ max(0, max(\thetaH*\qLval-\tLval, \thetaH*\qHval-\tHval)) - 0.5*(\thetaH-0.5)^2 }

\pgfmathsetmacro{\mP}{ (\yH-\yL)/(\thetaH-\thetaL) }
\pgfmathsetmacro{\bP}{ \yL - \mP*\thetaL }

\pgfkeys{/pgf/fpu=false}

\begin{axis}[
  width=1.2\linewidth,    
  height=1.2\linewidth,
  axis lines=left,
  xmin=0,xmax=1,
  ylabel={$\idu-\icost$},
  xlabel={$\theta$},
  ytick=\empty,
  xtick={\thetaL, .5, \thetaH},
  xticklabels={$\thmin_F$, $\theta_0$ , $\thmax_F$},
  axis x line=middle,
  axis y line = middle, 
  samples=900,
  domain=0:1,
  clip=false,
  legend style={draw=none, fill=none, at={(.4,.6)}, anchor=south east}
]

\addplot[line width=1.5]
({x},{max(0, max(x*\qLval-\tLval, x*\qHval-\tHval)) - .5*(x-0.5)^2});
\addlegendentry{$\idu_Q - \icost$}

\addplot[smooth, red, line width=2]({x},{\mP*x + \bP}) [domain=0:1];
\addlegendentry{$\DMPrice$}

\addplot[only marks, mark=*, mark size=2.2pt]
coordinates {(\thetaL,\yL) (\thetaH,\yH)};

\draw[densely dotted, thick] (axis cs:\thetaL,0) -- (axis cs:\thetaL,\mP*\thetaL + \bP);
\draw[densely dotted, thick] (axis cs:\thetaH,0) -- (axis cs:\thetaH,\mP*\thetaH + \bP);

\end{axis}
\end{tikzpicture}
\end{subfigure} \hfill
\begin{subfigure}{.49\textwidth}\centering
    \begin{tikzpicture}

\pgfmathsetmacro{\qHval}{0.5}
\pgfmathsetmacro{\qLval}{0.2}
\pgfmathsetmacro{\tHval}{0.2}
\pgfmathsetmacro{\tLval}{0.03}

\pgfmathsetmacro{\thetaH}{0.5*(\qHval*\qHval - 2*\qHval*\qLval + \qLval*\qLval + 2*\tHval - 2*\tLval)/(\qHval-\qLval)}
\pgfmathsetmacro{\thetaL}{ \thetaH - (\qHval-\qLval) }

\begin{axis}[
  width=1.2\linewidth,    
  height=1.2\linewidth,
  axis lines=left,
  xmin=0,xmax=1,
  ymax=1,
  xlabel={$\theta$},
  axis x line = middle,
  axis y line = middle,
  ylabel={$Q$},
  xtick={\thetaL, \thetaH},
  xticklabels={$\thmin_F$, $\thmax_F$},
  ytick={0, .7},
  yticklabels={0, $\bar{q}$},
  samples=400,
  domain=0:1,
  clip=false,
  legend style={draw=none, fill=none, at={(0.98,-0.05)}, anchor=south east}
]

\draw[densely dotted, thick] (axis cs:\thetaL,0) -- (axis cs:\thetaL, \qLval);
\draw[densely dotted, thick] (axis cs:\thetaH,0) -- (axis cs:\thetaH, \qHval);

\addplot[line width=1, gray] ({x},{x-0.5});
\addlegendentry{$\icost'$}

\addplot[line width=2, cyan] ({x}, {min(max(0,x-\thetaH+\qHval),.7)});
\addlegendentry{$Q_\DMprice$}

\addplot[dotted, line width=2, blue] ({x}, {x-\thetaH+\qHval});
\addlegendentry{$\DMprice+\icost'$}

\addplot[dashed,
  line width=2
]
({x},{
  ifthenelse( ((x*\qHval-\tHval) >= (x*\qLval-\tLval)) * ((x*\qHval-\tHval) >= 0),
              \qHval,nan)
});

\addplot[dashed,
  line width=2
]
({x},{
  ifthenelse( ((x*\qHval-\tHval) <= (x*\qLval-\tLval)) * ((x*\qLval-\tLval) >= 0),
              \qLval,nan)
});

\addplot[dashed,
  line width=2
]
({x},{
  ifthenelse( 0>= max((x*\qHval-\tHval), (x*\qLval-\tLval)),
              0,nan)
});
\addlegendentry{$Q$}

\draw[densely dotted, thick] (axis cs:0,.7) -- (axis cs:1,.7);

\draw[<->] (axis cs:0.65,.01+.15) -- (axis cs:0.65,.65-\thetaH+\qHval-.01
)
node[midway, left] {$\DMprice$};

\addplot[only marks, mark=*, mark size=2.2pt]
coordinates {(\thetaL,\qLval) (\thetaH,\qHval)};

\end{axis}
\end{tikzpicture}
\end{subfigure}
\caption{From an IC outcome to an $F$-CC allocation} \label{fig: binary FCC}
\end{figure}

The next theorem shows that $F$-CC allocations are without loss of optimality.

\begin{theorem}\label{thm: FCC}
$F$ solves the buyer's information acquisition problem for any $F$-CC  allocation. Moreover, if $F$ solves the buyer's information acquisition problem for allocation $\hat{Q}$, there exists an $F$-CC allocation $Q$ with:
    \begin{enumerate}[(i)]
        \item $ Q(\theta) = \hat Q(\theta)$ \text{ for all } $\theta \in \supp F$;
        \item $V_Q(\theta) \leq V_{\hat{Q}}(\theta) \text{ for all } \theta \in \supp F.   $ 
    \end{enumerate}

\end{theorem}

The theorem's proof proceeds as illustrated by the binary-prior example. We begin by finding a price function $\DMPrice$ that certifies the incentive-compatibility of $(Q,F)$. Then we derive an $F$-marginal price $\DMprice$ from $\DMPrice$, and construct the corresponding $F$-CC allocation $Q_\DMprice$. The result is an $F$-CC allocation $Q_\DMprice$ that coincides with $Q$ in the support of $F$, and gives the buyer a weakly lower gross-value for every signal realization. 

It follows that $F$-CC allocations are without loss for evaluating optimality and efficiency of the resulting allocation: the seller can replace any allocation with an $F$-CC one without foregoing any profits or changing the quality it provides to the buyer's realized types. 

Economically, Theorem~\ref{thm: FCC} shows one can express the buyer's information acquisition incentives as a restriction on the seller's product line. In particular, the speed at which an $F$-CC allocation can increase over $[\thmin_F,\thmax_F]$ must satisfy two conditions. First, to prevent the buyer from pooling different type realizations, the allocation must increase at least as fast as $\icost'$ does. And second, to prevent further learning, the allocation's speed of increase can increase faster than $\icost'$ only at types that cannot be split further---that is, at $F$-separating types. 

An immediate implication of these restrictions is that the seller can induce a signal $F$ only if she provides a minimal degree of quality differentiation. In particular, equation~\eqref{eq:fccallocation} implies
\begin{equation}\label{eq: minimal differentiation}
Q(\thmax_F)-Q(\thmin_F) \geq \icost'(\thmax_F)-\icost'(\thmin_F).
\end{equation}
We use this inequality repeatedly in the sequel. 

Theorem~\ref{thm: FCC} is also useful because it enables us to reason about the seller's problem ``one dimension at a time." 
Specifically, the seller-optimal allocation must solve a familiar-looking allocation design program, whereas the optimal signal must solve a particular mean-measurable information design problem. Using these programs, one can apply familiar techniques to make inferences about the seller's optimal outcome. 

Corollary~\ref{cor: two programs} below outlines the above-mentioned programs. The starting point for both programs is a seller optimal outcome $(Q,F)$. The allocation design problem is based on the observation that, holding $F$ fixed, the allocation $Q$ must maximize the seller's profit across all allocations that make $F$ optimal for the buyer. However, since each such allocation is outcome equivalent (under $F$) to some $F$-CC allocation, there is no loss in focusing solely on $F$-CC allocations. The allocation design program then follows from observing that $F$-CC allocations are indexed by their corresponding $F$-marginal price functions. Denote the set of all $F$-marginal price functions by $\DMpriceset(F)$. 

A similar logic delivers the information design program: holding $Q$ fixed, the signal $F$ must maximize the seller's profit across all signals that are optimal for the buyer given $Q$. In particular, if $Q$ is $F$-CC, the maximizing signal $F$ must be better for the seller than any other signal for which $Q$ is cost-compensating. Given a cost-compensating mechanism $Q$, we let $\ICC(Q)$ be the set of signals for which $Q$ is cost-compensating.\footnote{That is, $\ICC(Q)$ is the set of all signals $F\in \Signals$ such that $\supp(F) \subseteq[\thmin_Q,\thmax_Q]$, and $\int_{\tilde\theta\in[\underline{\theta},\theta]}\left(F_{0}-F\right)(\tilde{\theta})\mathrm{d}\tilde{\theta}> 0$ only if $\DMprice_{Q}$ is constant around $\theta$.} It is straightforward to verify that $\ICC(Q)$ is a compact and convex set.

\begin{corollary}\label{cor: two programs}
Suppose $(Q^*,F^*)$ is seller optimal. For any $F^*$-CC allocation $\tilde Q$, the outcome $(\tilde{Q},F^*)$ is seller optimal if and only if $p_{\tilde Q}$ solves
    \begin{equation}\label{eq: optimal allocation problem}
        \max_{\DMprice \in \DMpriceset(F^*)} \int \pi_{Q_{\DMprice}}(\theta) F^*(\dd \theta).
    \end{equation}
    Moreover, if $Q^*$ is $F^*$-CC, then $(Q^*,\tilde{F})$ is seller optimal if and only if $\tilde{F}$ solves
    \begin{equation}\label{eq: optimal info problem}
        \max_{F \in \ICC(Q^*)} \int \pi_{Q^*}(\theta) F(\dd \theta).
    \end{equation}
\end{corollary}
Thus, once we know the seller-optimal signal, finding the  seller-optimal allocation amounts to finding the $F$-marginal price that solves the program \eqref{eq: optimal allocation problem}. Similarly, once we know the seller-optimal cost-compensating allocation, finding the optimal signal amounts to maximizing the seller's profit across all signals for which that allocation is CC.\footnote{\label{footnote: bipooling} In particular, whenever $F_0$ has a density, it is without loss of optimality to focus on signals that have a bipooling structure \citep{kleiner2021extreme,arieli2023optimal}. The reason is that the set of signals for which the allocation is cost-compensating is equivalent to the set of signals that maximize the expectation of $V_Q - \icost$. Note, however, that the set of maximizers of an affine function over a convex set is a face of that set, and the extreme points of a face are a subset of the extreme points of the original convex set. Therefore, the extreme points of the set of signals that are IC for the buyer for a given allocation $Q$ are a subset of the extreme points of all signals, all of which take a bipooling structure by \cite{kleiner2021extreme} and \cite{arieli2023optimal}. We do not rely on this fact in our analysis.} In the sequel we use these two programs to make inferences about the seller-optimal outcome.

\paragraph{Binary types} We now explain how the programs \eqref{eq: optimal allocation problem} and \eqref{eq: optimal info problem} simplify when the prior is binary. Suppose first  $\supp F^* \subset (\thmin,\thmax)$, and that $Q^*$ is $F^*$-CC with an associated marginal price $p^*$. Start with the optimal allocation problem in \eqref{eq: optimal allocation problem}. As pointed out in the discussion of Figure~\ref{fig: binary FCC}, all intervals $(\theta_1,\theta_2) \subset (\thmin,\thmax)$ are $F^*$-pooling and, as a consequence, any $F^*$-marginal price function must be constant on $(\thmin,\thmax)$. 
Since $\supp F^* \subset (\thmin,\thmax)$, each $F^*$-marginal price $\DMprice$ is equivalent to a constant function, and changing this constant consists of a uniform shift in the allocation $Q_\DMprice (\theta) = \icost'(\theta) + \DMprice$. This observation has two implications. First, over the support of $F^*$, all $F^*$-CC allocations differ from each other only by a level shift. Second, because this level shift determines which types are excluded---i.e., receive zero quality---each constant $\DMprice$ maps into an inclusion threshold, $\thmin_Q$ for $Q=Q_\DMprice$. Thus, by the envelope theorem, $\DMprice$ pins down the rent of every consumer type: $V_Q(\theta) = \int^\theta_{\thmin_Q} Q(x) dx$. Therefore, the allocation problem reduces to the one-dimensional program

\begin{align*}
    \max_{\DMprice} & \int \left( \theta (\icost'(\theta)+\DMprice) - \qcost\left(\icost'(\theta)+\DMprice\right) - \int^\theta_{\thmin_Q} (\icost'(x)+\DMprice)dx \right) F^*(d\theta) 
    \\
    \text{s.t.} & -\icost'(\thmin_{F^*})\leq \DMprice \leq \bar{q} - \icost'(\thmax_{F^*}).
\end{align*}

This problem translates to our setting the usual \citet{mussa1978monopoly} trade-off, with one caveat. In the standard screening program, increasing the quality assigned to a type $\theta$ requires increasing, by incentives, the qualities and rents of all higher types. Here, the same rent-surplus trade-off remains, but the incentives required by information acquisition restrict the seller's feasible perturbations to the uniform level shift $p$. Raising $p$ lowers the exclusion cutoff $\thmin_Q$, raising qualities and rents for all types. Conditional on this restricted class of perturbations, the optimality condition has the familiar form. Namely, when an interior solution exists:

$$ \thmin_Q = \int \qcost'(Q(\theta)) F^*(d\theta). $$

That is, the exclusion cutoff equals the average marginal cost of the qualities induced by the optimal scalar shift.

We now turn to the information problem \eqref{eq: optimal info problem}. Recall that $\mathcal{I}(Q^*)$ is the set of signals for which $Q^*$ is cost-compensating. Because $p^*$ is constant in $(\thmin,\thmax)$, it is constant over any $F$-pooling intervals for any signal $F$. In that case, $\mathcal{I}(Q^*)$ contains any signal $F$ such that $\thmin_{Q^*} \leq \thmin_F$, and $\thmax_{Q^*} \geq \thmax_F$. The information problem can be written as:

\begin{align*}
    \max_{F \in \mathcal{I}} \ & \int \left( \theta Q^*(\theta) - \qcost\left(Q^*(\theta)\right) - \int^\theta_{\thmin_{Q^*}} Q^*(x) dx\right) F(d\theta)
    \\ \text{s.t.} \ & \ \supp \ F \subset [\thmin_{Q^*},\thmax_{Q^*}].
\end{align*}

The information design problem is only constrained by the bounds on the support of $F$. It is easy to see that the usual concavification technique can still be used to find the solution under these restrictions: that is, the seller concavifies the profit function over the interval $[\thmin_Q^*,\thmax_Q^*]$. Consequently, there's always an optimal signal that is at most binary. Moreover, the standard information-design intuition carries over. If the seller's profit function is concave on the relevant interval, the optimal signal is uninformative. If it is convex, the seller implements the most informative signal over that interval.

When $\supp F^* \not\subset (\thmin,\thmax)$---and so $\DMprice^*$ may not be constant over $(\thmin,\thmax)$---the two problems are slightly more involved. First, if the marginal price $\DMprice^*$ associated with the $F^*$-CC allocation $Q^*$ is not constant over $(\thmin,\thmax)$, the information problem is degenerate: the only distribution compatible with $\DMprice^*$ is $F_0$, because that is the only signal that admits an interior separating type. Similarly, if $F^*=F_0$, the allocation problem allows for any increasing marginal price function satisfying the bounds in \eqref{eq: p-range}. Finally, if $F^* \neq F_0$ but has a support that includes $\thmin$ or $\thmax$, the allocation problem must also consider the possibility that the chosen marginal price may jump at $\thmin$ or $\thmax$ (which are always separating), since these jumps now affect the seller's objective.
\hfill $\blacksquare$

\section{Downward Quality Distortions}\label{section:Quality Distortions}

This section studies the inefficiencies in the seller's optimal menu. We prove that the seller distorts quality downward for all of the buyer's realized types. Moreover, we provide sufficient conditions for this distortion to hold strictly for all types, therefore reversing the celebrated ``no distortion at the top'' observation from standard screening with exogenous information.

We begin by stating the main result of this section. 

\begin{theorem}
\label{thm: Inefficiency} 
Every seller-optimal outcome $\left(Q^{*},F^{*}\right)$
admits an allocation $Q$ such that $Q=Q^{*}$ holds $F^{*}$-almost surely,
$\left(Q,F^{*}\right)$ is seller optimal, and every $\theta \in\supp\,F^{*}$ has $\theta\geq \qcost'\left(Q(\theta)\right)$, with equality holding if and only if $\theta =\thmax$.

\end{theorem}

The above theorem states that the seller underprovides quality to all realized types except for $\thmax$, whose allocation is efficient.  An immediate implication is that all realized types are always weakly underprovided, and they are strictly underprovided if and only if $\thmax \notin \supp F$. 

Theorem~\ref{thm: Inefficiency} highlights a fundamental difference between the exogenous and endogenous information cases. When information is exogenous, \citeapos{mussa1978monopoly} analysis shows that quality is distorted downward at all types \emph{except} the highest one. Our results demonstrate that, when information is endogenous, quality is still distorted downwards, but the ``no distortion at the top" observation remains if and only if the highest type under the buyer's chosen signal is the highest possible type when the buyer is fully informed. Later in this section we provide a sufficient condition for the buyer to choose to not learn $\thmax$ and, consequently, for the failure of ``no distortion of the top'' (Proposition~\ref{prop: sufficient distortion}).

Theorem~\ref{thm: Inefficiency}'s logic relies on two economic forces. The first force is familiar from \citeapos{mussa1978monopoly} model of monopolistic screening with exogenous information. In that model, the seller distorts the allocation of lower types downwards in order to reduce the information rents given to higher types. Using the optimal allocation program~\eqref{eq: optimal allocation problem} from Corollary~\ref{cor: two programs}, we show a similar concern arises in the current setting for any type at which the seller's chosen marginal price is strictly increasing. 

The second force comes from the seller steering the buyer's information acquisition. By having the buyer learn different signals, the seller can trade-off the benefit of having different type realizations. At the margin, one can express this trade-off by asking whether the seller benefits from slightly increasing or decreasing each possible type realization. 

The answer to this question turns out to depend on whether or not that type's quality is above or below the efficient level. Roughly speaking, an increase in the buyer's posterior-realization affects the seller's profits in two ways: First, it changes the total surplus generated by the transaction, $\theta Q(\theta) - \qcost(Q(\theta))$, and second, it impacts the information rents the seller cedes to the buyer, $V_Q(\theta)$. When $Q$ is differentiable at $\theta$, the envelope theorem implies that the change in the buyer's information rents is second order, meaning the difference in the available social surplus dominates; that is

$$\pi'_Q(\theta) = \underbrace{Q(\theta) -V'_Q(\theta)}_{=0 \text{ by envelope}} + \left(\theta-\qcost'(Q(\theta))\right) Q'(\theta).$$

Thus, since the provided quality is increasing with the buyer's type, a small increase in $\theta$ raises the seller's profits if and only if quality is under-provided at $\theta$. In other words, $\pi_{Q}$ increases at $\theta$ if and only if $Q(\theta)$ is below the efficient level.\footnote{More precisely, suppose $Q$ is an $F$-CC allocation such that $\DMprice_{Q}$ is constant on $[\theta_1,\theta_2]$. One can show that, in this case, the right derivative $\pi'_{Q+}(\theta_1)$ and the left derivative $\pi'_{Q-}(\theta_2)$ are well defined, and that $\pi'_{Q+}(\theta)$ ($\pi'_{Q-}(\theta)$) is strictly positive (negative) if and only if $Q(\theta)$ is strictly below (above) the efficient level.} 

To see the relevance of this observation to Theorem~\ref{thm: Inefficiency}, take any $\theta_h \in \supp \ F$ with $\theta_h < \thmax$ around which $\DMprice_{Q}$ is constant.  Let $\theta_l$ be the highest type below $\theta_h$ at which $\DMprice_{Q}$ is strictly increasing. Suppose, for the sake of intuition, that both $\theta_h$ and $\theta_l$ occur with positive probability under $F$, and that $\DMprice$ is differentiable at $\theta_l$. Consider the change in profits due to a small mean-preserving contraction that replaces each realization of $\theta_h$ with a slightly lower realization, and each realization of $\theta_l$ with a slightly higher one. By choice of $\theta_l$, $\DMprice_{Q}$ must be constant over $[\theta_l,\theta_h]$, meaning this contraction is incentive compatible for the buyer. Therefore, because $F$ is optimal, the total effect of this contraction must be negative. Recall, however, that because $\DMprice$ is increasing at $\theta_l$, the quality $Q(\theta_l)$ must be below its efficient level. Hence, slightly increasing $\theta_l$ must strictly increase the seller's profit, as explained above. Since this contraction cannot be profitable, we get that the slight reduction in $\theta_h$ must strictly decrease the seller's profits; that is, the seller's profits must be increasing at $\theta_h$, meaning $\theta_h$'s quality must be strictly below its efficient level.

Next we provide sufficient conditions for there to be a downward distortion even at the top of the support of $F$.
In particular, we show that quality is strictly under-provided to the highest signal realization whenever the scale of information costs is sufficiently large. We also establish a converse, namely that when the scale of information costs is low, the highest type in the support is $\thmax$, and so there is no distortion at the top. To show this converse we require an additional condition. This condition also plays a role in the comparative statics of Section~\ref{sec: Comparative Statics}.

\begin{condition}[Finiteness and Regularity] \label{as: finitesupport}
 The type distribution $F_0$ is \textbf{Finite and Regular} (FAR) if it has finite support---$\supp F_0 = \{\theta_1, ..., \theta_L \}$, where $\theta_1 < \theta_2 < ...< \theta_L$---, and its virtual values,\footnote{In this expression, we take $F_0(\theta_{1-1}):=0$.}
   \begin{equation}\label{eq: virtual values}
   v_i := \theta_i - \frac{1-F_0(\theta_i)}{F_0(\theta_i) - F_0(\theta_{i-1})} (\theta_{i+1} -\theta_{i}), \ \ i\leq L,
   \end{equation}
are strictly increasing in $i$, and are strictly larger than $\qcost'(0)$ for all $i\geq 2$.
\end{condition}

The assumption on virtual values is equivalent to the seller finding it optimal to fully separate the buyer's types when the buyer's information is fixed at full learning. Moreover, one can show Condition~\ref{as: finitesupport} implies that, when information is endogenous and learning is free, the seller finds it uniquely optimal to induce the buyer to acquire full information. 
For an intuitive explanation, note that with free learning, every incentive-compatible signal must result in an ex-post optimal selection from the seller's menu. Therefore, the seller's profit from any menu must be the profit she gets under perfect learning, and so the best the seller can do is to offer the optimal full-learning menu. Moreover, because virtual-values are strictly increasing, this menu is fully-separating, and so full-learning uniquely maximizes the seller's profits. 

Hence, Condition~\ref{as: finitesupport} implies that full-learning is the uniquely seller optimal signal when information is free. In fact, the condition implies something stronger: because the value distribution is finite, one can show that perfect learning is also optimal for the seller when the scale of information costs is sufficiently low---a result we prove in section~\ref{sec: Comparative Statics}.

The next proposition provides two sets of sufficient conditions, one for strict distortion everywhere, and one for no distortion at the top. 

\begin{proposition} \label{prop: sufficient distortion}
    Suppose information costs satisfy $C(F) = \gamma \int c(\theta) F(d\theta)$, for $\gamma>0$. Then:
\begin{enumerate}[(i)]
    \item \label{prop: sufficient distortion - high costs} 
    When $\gamma$ is sufficiently large, $\thmax \notin \supp F^*$, and so $\thmax_{F^*} > \qcost'(Q(\thmax_{F^*}))$. 
    
    \item \label{prop: sufficient distortion - low costs}
    If $F_0$ is FAR and $\gamma$ is sufficiently small, $\thmax \in \supp F^*$, and so $\thmax_{F^*} = \qcost'(Q(\thmax_{F^*}))$.
\end{enumerate}

\end{proposition}

Part~\eqref{prop: sufficient distortion - high costs} of the proposition provides sufficient conditions for existence of ``distortion at the top.''
When $\gamma$ parametrizes the scale of marginal information costs, the result implies that all qualities are strictly below efficient for high $\gamma$. Specifically, letting $\bar{q}_e$ be the efficient quality level for type $\thmax$ (i.e., $\bar{q}_e$ solves $\qcost'(\bar{q}_e)=\thmax$), one has that $\thmax \notin \supp \ F^*$ whenever
\[
\gamma \geq \frac{\bar{q}_e}{\icost'(\thmax) - \icost'(\theta_0)}.
\]
The reason is that the above inequality implies the seller can make a signal $F$ with $\thmax_F=\thmax$ incentive compatible only if she upward distorts the quality she provides $\thmax$, which is suboptimal. For an explanation, recall that by Theorem~\ref{thm: FCC}, the seller can make $F$ incentive compatible only if she provides a minimal amount of quality differentiation, described by equation~\eqref{eq: minimal differentiation}. Specializing this equation to the case where costs are parameterized by $\costcoef$ and observing that $\thmax_F=\thmax$ only if $\thmin_F < \theta_0$ gives the following inequality for any $Q$ that makes $F$ incentive compatible: 
\[
Q(\thmax_F) \geq  Q(\thmax_F) - Q(\thmin_F) \geq \costcoef\left(\icost'(\thmax_F) - \icost'(\thmin_F)\right) > \costcoef\left(\icost'(\thmax_F) - \icost'(\theta_0)\right).
\]
When $\costcoef$ surpasses the above-mentioned threshold, the inequality's right-hand side is larger than $\bar{q}_e$, meaning that $Q$ makes $F$ IC only if $Q$ gives $\thmax_F$ too high of a quality. Since quality must be weakly under-provided to all realized types at the optimum, such an $F$ cannot be part of a profit maximizing outcome.

The proposition's part~\eqref{prop: sufficient distortion - low costs} provides sufficient conditions for our model to exhibit the classic ``efficiency at the top'' result. These conditions require the true type distribution $F_0$ to be FAR, and for learning costs to be sufficiently small. This part follows from the proof of Theorem~\ref{thm: low cost comparative statics}. As explained earlier, that result shows that when $F_0$ is finite and regular, the seller finds full-learning optimal for all sufficiently low $\costcoef$. It then follows from Theorem~\ref{thm: Inefficiency} that the seller optimal outcome induces the buyer to obtain the efficient quality whenever he learns that his type is the highest possible---i.e., when $\theta=\thmax.$

\paragraph{Example. (Binary types and Variance reduction)}

\begin{figure}[t] 
\centering

\begin{minipage}[t]{0.49\textwidth}
    \centering
    \resizebox{\linewidth}{!}{\input{quadraticqualities0}}
    \subcaption{Optimal and Efficient Allocations}
\end{minipage}
\hfill
\begin{minipage}[t]{0.49\textwidth}
    \centering
    \resizebox{\linewidth}{!}{\input{quadraticsignal0}}
    \subcaption{Optimal Signal}
\end{minipage}

\caption{Variance-Reduction Information Costs with Quadratic Production Costs} \label{fig: quadraticproductioncosts}
\end{figure}

We now illustrate Theorem~\ref{thm: FCC} and Proposition~\ref{prop: sufficient distortion} using two examples where the true type is binary and uniformly distributed over $\supp\ F_0=\{0.1,0.9\}$. Information costs are quadratic and scaled by a constant $\gamma$: $C(F) = \frac{\gamma}{2} \int \left(\theta-\theta_0 \right)^2 F(\dd \theta).$  When $F_0$ is binary, an extreme point argument delivers an optimal signal $F$ that has at most two points in its support.\footnote{This argument follows a similar logic as outlined in footnote~\ref{footnote: bipooling}, combined with the observation that the extreme points of $\Signals$ have binary support whenever $\supp \ F_0$ is binary.} 
Because the mean of $F$ must be $\theta_0$, $F$ is completely determined by these points, which we denote by $\theta_L$ and $\theta_H$, where $\theta_L\leq\theta_0\leq\theta_H$. We let $Q_L$ and $Q_H$ be the qualities each of these types receives in the optimal allocation, and take $(Q_L^e,Q_H^e)$ and $(Q_L^s, Q_H^s)$ respectively to be the qualities they receive under the efficient allocation and under standard screening when information is exogenous as in \cite{mussa1978monopoly}. Whenever only one quality is offered, we denote it by $Q_L$ in the figures.

We illustrate the seller's optimal outcome using two examples of production costs. We start by describing the right panel in Figure~\ref{fig: quadraticproductioncosts}, which depicts the optimal signal $F$ when production costs are quadratic: $\qcost(q) = \frac{q^2}{2}$. In this case, when the level of information costs $\costcoef$ is below some $\costcoef_l$, the buyer chooses to be fully informed. When $\costcoef$ is above $\costcoef_l$ but below another cutoff $\costcoef_h$, one of the signal's realizations fully reveals the high type, $\theta_H=\thmax$. The other realization of the signal pools some realizations of the high type with realizations of the low type, leading to a posterior $\theta_L \in (\thmin,\theta_0]$. As $\gamma$ grows, the  signal's informativeness deteriorates monotonically, with both the low posterior's value $\theta_L$ and its probability increasing until $\theta_L$ reaches $\theta_0$ at $\gamma=\gamma_h$. At that point, the optimal signal becomes uninformative---$\supp F=\{\theta_0\}$---and remains so for any higher $\gamma$. The next section shows that the existence of full learning at low $\costcoef$ and no learning at high $\costcoef$ extends beyond the current example---See Theorem~\ref{thm: large cost comparative statics} and Theorem~\ref{thm: low cost comparative statics}.

The left panel in Figure~\ref{fig: quadraticproductioncosts} depicts the optimal qualities for quadratic production costs as $\gamma$ varies. Because the high signal is fully revealing, the high type receives an efficient quality by Theorem~\ref{thm: Inefficiency}, and all the distortions in this example come from the allocation of the low type. In the Figure, we plot the allocation for that type with endogenous information, compared to the one that would hold with exogenous information, $Q_L^s$, and with efficiency, $Q_L^e$. The distance between $Q_L^e$ and $Q_L^s$ is the standard screening distortion: it stems from asymmetric information exclusively and so vanishes when information becomes symmetric at the optimum (i.e., when $\gamma\geq \gamma_h$). The difference between $Q_L^s$ and $Q_L$, in turn, is the distortion given by information acquisition: the result of the agent's ability to acquire information that is not prescribed by the principal. This difference does not disappear when the signal becomes uninformative: the distortion persists and only vanishes asymptotically, as $\gamma$ approaches infinity.

Figure~\ref{fig: cubicproductioncosts} considers the same example but with cubic production cost; that is $\qcost(q) = \frac{q^3}{3}$. Like the previous example, the optimal signal is fully informative when $\costcoef$ is below a cutoff $\costcoef'_l$, and completely uninformative when $\costcoef$ exceeds a high cutoff $\costcoef'_h$. However, unlike the quadratic $\qcost$ case, the optimal signal only reveals the high type with positive probability when $\costcoef$ is below a cutoff $\costcoef'_m$. For $\costcoef \in (\costcoef'_m, \costcoef'_h)$ the high type is not fully disclosed, and the optimal quality for the high posterior is strictly lower than the efficient one $Q_H<Q_H^e$. Indeed, the qualities given to both types are strictly below their efficient levels, and neither type is fully revealed. Moreover, both signal realizations fall as $\gamma$ rises within this range, meaning the optimal signal becomes neither more nor less informative as $\costcoef$ increases. \hfill $\blacksquare$

\begin{figure}[t] 
\centering

\begin{minipage}[t]{0.49\textwidth}
    \centering
    \resizebox{\linewidth}{!}{\input{cubicqualities0}}
    \subcaption{Optimal and Efficient Allocations}
\end{minipage}
\hfill
\begin{minipage}[t]{0.49\textwidth}
    \centering
    \resizebox{\linewidth}{!}{\input{cubicsignal0}}
    \subcaption{Optimal Signal}
\end{minipage}

\caption{Variance-Reduction Information Costs with Cubic Production Costs} \label{fig: cubicproductioncosts}
\end{figure}

\section{Comparative Statics} \label{sec: Comparative Statics}

This section studies how the scale of information costs affects consumer surplus and seller's profits. Several mechanisms influence the interplay between costs and surplus. First, higher information costs mechanically reduce surplus for any fixed signal. Second, because information allows the seller to better tailor her offers to the buyer, it has a productive role and affects total surplus. Finally, the value of different deviations for the buyer depends on information costs. Therefore, obtaining global comparative statics predictions that hold regardless of the shape of information and production costs is difficult. Nevertheless, the results below show that the third mechanism, how the value of deviations varies with information costs, dominates when information costs are very small or very large. 

Formally, let information costs be
\[
C(F) = \costcoef \int \icost(\theta) F(d\theta),
\]
where $\costcoef>0$ parametrizes the scale of information costs. Take $(Q_\costcoef,F_\costcoef)$ to be any cost-compensating seller optimal outcome when the scale is $\costcoef$.
The main results of this section describe how profits $U_S$ and consumer surplus $U_B$ vary for small and large $\costcoef$. We begin by showing that when information costs are large, the seller's profits increase with $\costcoef$, whereas the consumer's surplus moves in opposite direction.

\begin{theorem} \label{thm: large cost comparative statics}
    There exists $\costcoef_h$ such that:
\begin{enumerate}[(i)]
    \item \label{thm: large cost comparative statics - uninformative F} $F_\costcoef$  is uninformative for $\costcoef \geq \costcoef_h$ (i.e., $F_\costcoef = \mathbf{1}_{[\theta_0,\infty)}$).
    \item \label{thm: large cost comparative statics - seller payoff} $U_S$ strictly increases in $\costcoef$ over $[\costcoef_h, \infty)$.
    \item \label{thm: large cost comparative statics - buyer payoff} $U_B$ strictly decreases in $\costcoef$ over $[\costcoef_h, \infty)$.
\end{enumerate}
\end{theorem}

Part~\eqref{thm: large cost comparative statics - uninformative F} shows that when costs are sufficiently high, the seller finds it optimal to completely disincentivize  learning. For a rough explanation, note that, by Theorem~\ref{thm: FCC}, inducing the buyer to obtain two different type realizations $\theta_1<\theta_2$ requires the seller to provide the two types with a quality differential $Q(\theta_2)-Q(\theta_1)$ of at least $\costcoef(\icost'(\theta_2) - \icost'(\theta_1))$ (see equation~\eqref{eq: minimal differentiation}). Because the seller's production costs $\qcost$ are strictly convex, providing this differential reduces the seller's profits from the buyer's information acquisition. As $\costcoef$ increases, this profit reduction increases as well, resulting in the seller preferring to keep the buyer uninformed whenever $\costcoef$ is sufficiently high.

This intuition implies that  the buyer remains uninformed for high $\costcoef$, and only one level of quality is chosen with positive probability. However, note that the seller provides the buyer with positive rents because she must dissuade him from acquiring information. As the buyer's costs increase, information acquisition becomes less attractive, and dissuading learning becomes easier. The rents required to prevent learning change accordingly: the seller benefits, and the buyer loses.

Next we establish comparative statics for low information costs. For this result, we need $F_0$ to be FAR---that is, to satisfy Condition~\ref{as: finitesupport}. Fixing such an $F_0$, we let $\supp F_0 = \{\theta_1, ..., \theta_L \}$ with $\theta_1 < \theta_2 < ...< \theta_L$, and define the function $\tilde{\icost}: \Theta \rightarrow \mathbb{R}$ via
\[
\tilde{\icost}(\theta) = \icost(\thmin) + \sum_{j<J_\theta} \icost'(\theta_j) (\theta_{j+1}-\theta_j)+\icost'(\theta_{J_\theta}) (\theta - \theta_{J_\theta}), \quad 
\]
where $J_\theta$ satisfies $\theta \in [\theta_{J_\theta},\theta_{J_\theta+1}).$ The function $\tilde{\icost}$ is a linear-by-parts approximation of the information cost, $\icost$, taking the slope of $\icost$ at the points in the support of $F_0$. Because information costs are convex, $\tilde{\icost}$ always lies below $\icost$. The finer the support of $F_0$, the more precise this approximation is. Figure~\ref{fig:tildec} illustrates these observations by plotting information costs against two approximations for different numbers of points in the support of $F_0$. In the figure, $\icost$ is quadratic, and we consider two distributions of types. The first one gives rise to $\tilde\icost_2$, with $\supp F_0 = \{0,1\}$. The second approximation $\tilde \icost_3$ is generated by a distribution $F_0$ with three points in the support, 0, $.3$ and 1. Our next result shows that, for low levels of information costs,  changes in consumer surplus are connected to $\tilde{c}$.

\begin{samepage}
\begin{theorem} \label{thm: low cost comparative statics}
   Suppose $F_0$ is FAR. A $\costcoef_l>0$ exists such that:
   \begin{enumerate}[(i)]
       \item \label{thm: low cost comparative statics - fully informative F} For all $\costcoef \in [0,\costcoef_l], F_\costcoef = F_0$, and $Q_\costcoef(\theta) = Q_0(\theta)$ for all $\theta \in \supp F_0$.
       \item \label{thm: low cost comparative statics - seller payoff} $U_S$ is strictly decreasing in $\costcoef$ over $[0, \costcoef_l]$.
       \item \label{thm: low cost comparative statics - buyer payoff} $U_B$ is strictly increasing (decreasing) in $\costcoef$ over $[0,\costcoef_l]$ whenever
       \[
       \int \tilde{\icost}(\theta) F_0(d\theta) < (>) 0.
       \]
   \end{enumerate}
\end{theorem}
\end{samepage}

\begin{figure}
    \centering
    \begin{tikzpicture}
\begin{axis}[
    width=13cm,
    height=8cm,
    xmin=0, xmax=1,
    ymin=-0.4, ymax=0.2,
    axis lines=middle,
    axis line style={->},
    xlabel={$\theta$},
    ylabel={},
    xtick={0,0.3,0.5,1},
    xticklabels={$0$,$0.3$,$0.5$,$1$},
    ytick=\empty,
    domain=0:1,
    samples=200,
    legend style={
        at={(0.3,.25)},
        anchor=north,
        legend columns=-1,
        /tikz/every even column/.append style={column sep=0.5cm}
    }
]

\addplot[thick,blue] {(x-0.5)^2/2};

\addplot[thick,red,dashed,domain=0:0.3,forget plot] {0.125 - 0.5*x};

\addplot[thick,red,dashed,domain=0.3:1] {0.035 - 0.2*x};

\addplot[thick,gray,dotted] {0.125 - 0.5*x};

\legend{$c$, $\tilde c_3$, $\tilde c_2$}

\addplot[
    only marks,
    mark=*,
    mark options={fill=red},
    red,
    forget plot
] coordinates {
    (0,   0.125)                     (0.3, 0.125 - 0.5*0.3)           (1,   0.035 - 0.2*1)             };

\addplot[
    only marks,
    mark=*,
    mark options={fill=gray},
    gray,
    forget plot
] coordinates {
    (1, 0.125 - 0.5*1)               };

\end{axis}
\end{tikzpicture}
\caption{Illustration of $\tilde{c}$}
\label{fig:tildec}
\end{figure}

Part~\eqref{thm: low cost comparative statics - fully informative F} of Theorem~\ref{thm: low cost comparative statics} says that when information costs are small, the seller finds it optimal to induce full learning. Moreover, the buyer's realized types receive the same qualities they would get if the buyer's information was exogenous (though $Q(\theta)\neq Q_0(\theta)$ for $\theta \notin\supp \ F_0$). The proof of this part is technical and relies heavily on our $F$-CC characterization. Broadly speaking, the result follows from noting that when $F_0$ is FAR and learning is free, the seller has a strict incentive to induce the optimal perfect-learning outcome, and so prefers to keep this outcome for small learning costs.

Parts~\eqref{thm: low cost comparative statics - seller payoff} and \eqref{thm: low cost comparative statics - buyer payoff} establish comparative statics for profits and consumer surplus, respectively: profits are decreasing for small costs, whereas consumer surplus could move in either direction. For a rough explanation, recall that Theorem~\ref{thm: FCC} implies that incentivizing the buyer to acquire a particular signal constrains the variety of products the seller can offer. In other words, the allocation $Q$ must change at just the right rate: it cannot increase too slowly nor too quickly. This constraint in turn restricts the way in which the seller can allocate information rents across buyer types. In particular, if the seller wishes to incentivize full-learning and have $Q$ match $Q_0$ on $\supp \ F_0$, she must---on top of the information rents provided by the allocation $Q_0$---provide each type $\theta_i$ with a rent of at least $\costcoef(\icost(\theta_i) - \tilde{\icost}(\theta_i))$. Since the seller wishes to minimize the buyer's rents, at the optimum a $\theta_i$-buyer's gross utility equals
\[
\idu_Q(\theta_i) = \int_{\thmin}^{\theta_i}(\DMprice(\theta) + \costcoef\icost'(\theta))\dd \theta = \sum_{j<i}Q_0(\theta_j)(\theta_{j+1}-\theta_j) + \costcoef(\icost(\theta_i) - \tilde{\icost}(\theta_i)).
\]
The proof of parts~\eqref{thm: low cost comparative statics - seller payoff} and \eqref{thm: low cost comparative statics - buyer payoff} is now straightforward. As $\costcoef$ increases over $[0,\costcoef_l]$, the buyer's gross surplus increases, because $\tilde{\icost}$ underestimates $\icost$. Since the seller induces the same joint distribution of qualities and types for all $\costcoef$ in $[0,\costcoef_l]$, the expected trade-surplus remains constant, and so expected profits must decrease. As for the buyer's expected net-payoff, the only term it contains that depends on $\costcoef$ is $-\costcoef\int\tilde{\icost}(\theta)F_0(\dd \theta)$, immediately delivering part~\eqref{thm: low cost comparative statics - buyer payoff} of the theorem.

At a higher level, the reason seller's profits decrease in $\costcoef$ over $[0,\costcoef_l]$  goes as follows. Because the seller offers the same quality and induces full information for $\costcoef<\costcoef_l$, increasing $\costcoef$ keeps production surplus fixed. Hence, aggregate surplus decreases, as the buyer acquires the same signal, but at a higher cost. 
Moreover, to incentivize the acquisition of this signal, the seller must increase the buyer's  information rents, since increasing $\costcoef$ makes acquiring less information more tempting. Thus, as $\costcoef$ increases, the seller's profits must go down, because aggregate surplus decreases and the buyer's gross surplus increases. 

The change in buyer's net surplus, however, depends on the way the buyer's optimal learning constraint restricts the way the seller can allocate information rents across buyer types. If the seller must allocate these rents in a manner that is more generous on average than the increase in costs, the buyer's net-value increases in $\costcoef$. Otherwise, his net-value decreases.

One prominent case in which the buyer prefers information to be cheap rather than free is when $F_0$ is binary. With a binary prior and free learning, the optimal menu leaves the high-type buyer indifferent between his product and the one targeted to the low type. Therefore, with that same menu the buyer is not willing to pay to learn whether his type is low or high. To induce full-learning when information is costly but cheap, the seller must change the menu so that it provides the buyer with information rents that are at least equal to the cost of acquiring full information. Providing these rents, however, creates additional profitable deviations for the buyer, which in turn necessitates providing the buyer even more rents. Consequently, the buyer is better off when learning is cheap than she is when learning is free. We document this result in the following corollary.

\begin{corollary} \label{cor: consumer surplus increasing for binary}
    Suppose $F_0$ has binary support. Then $U_B$ is strictly increasing in $\costcoef$ for all sufficiently small $\costcoef$.
\end{corollary}

The corollary's proof is straightforward: when $F_0$ is binary, $\tilde{\icost}$ is just a subgradient of $\icost$ at $\thmin$. Therefore, we obtain the following inequality chain:
\[
\int \tilde{\icost}(\theta) F_0(\dd \theta) = \tilde{\icost}(\theta_0) < \icost(\theta_0) =0,
\]
where the first equality follows from $\tilde{\icost}$ being linear, and the strict inequality from $\icost$ being strictly convex. Thus, when $F_0$ is binary and learning is cheap, higher learning costs increase the buyer's net surplus in addition to reducing the seller's profits.

To conclude this section, we note that our comparative statics apply either to very high costs (i.e., $\costcoef > \costcoef_h$), or very low costs (namely, $\costcoef < \costcoef_l$). Our results do not apply to the intermediate values of $\costcoef$. In particular, our results do not imply that profits are quasiconvex in $\costcoef$ or that the buyer's net surplus is quasiconcave in $\costcoef$. And, indeed, one can find examples where both of these properties are violated. 

\section{Steep Learning Costs and Simple Menus}
\label{section:Optimal Info}

If buyers are fully informed and $F_0$ has a continuous distribution, \citeapos{mussa1978monopoly} analysis shows the optimal menu is usually quite complicated, containing many alternatives. This section proves that the same is not generally true under costly learning. In particular, we show the optimal menu contains at most two purchasing options whenever learning costs are steep. In this case, we also strengthen our inefficiency result: on average, the wedge between the optimal allocation and the efficient one is larger in our model, compared to standard screening. 

We refer to an outcome $(Q,F)$ that is implementable with at most two purchasing options as a \textbf{binary-menu} outcome.  Notice that $(Q,F)$ is a binary-menu outcome whenever $F$ has binary support. The reason is that restricting the seller's menu to include only the alternatives the buyer chooses with positive probability has no impact on the buyer's decisions as to what to learn and what to buy. Consequently, the seller can always maximize profits with a menu that contains weakly less options than the number of realizations in the support of the buyer's signal. More generally, $(Q,F)$ is a binary-menu outcome whenever $Q(\supp \ F)\setminus\{0\}$ has two or less elements.

Our next result requires a few definitions. A signal $F\in \Signals$ is a \textbf{cutoff signal} if it has two realizations in its support and admits at least one interior $F$-separating type $\theta^c \in (\thmin,\thmax)$. Informally, a cutoff signal is induced by the buyer learning whether or not $\theta$ is above $\theta^c$. We say learning costs satisfy the \textbf{steep-slope condition} if every cutoff signal $F$ has $\icost'(\thmax_F) - \icost'(\thmin_F) > \bar{q}_e$ (recall that $\bar{q}_e$ is the efficient quality level for type $\thmax$).\footnote{
We remark here that the steep-slope condition implies the condition used to establish strict distortion at the top in Proposition~\ref{prop: sufficient distortion}, namely $\icost'(\thmax) - \icost'(\theta_0) \geq \bar{q}_e$. For an explanation, note that any cutoff signal $F$ with cutoff $\theta^c$ has $\thmin_F \leq \theta_0$ and $\theta^c \leq \thmax_F$. Hence, one has $\thmax_{F_n}\rightarrow \thmax$ for any sequence of cutoff signals $F_n$ with $\theta^c_n\rightarrow \thmax$. Therefore, if $\icost$ has steep slopes, one must have $\bar{q}_e \geq \icost'(\thmax_{F_n}) - \icost'(\thmin_{F_n}) \geq \icost'(\thmax_{F_n}) - \icost'(\theta_0)\rightarrow \icost'(\thmax) - \icost'(\theta_0)$. In other words, the steep slope condition is sufficient for existence of strict distortion at the top.  
} 

Proposition~\ref{prop: binary signals} below shows that when learning costs are steep, the seller implements a binary-menu outcome.

\begin{proposition}   
    \label{prop: binary signals}
    Suppose learning costs satisfy the steep-slope condition. Then some binary-menu outcome is seller optimal. 
\end{proposition}

The proposition again stems from the constraints that the buyer's optimal learning incentives impose on product variety. When learning costs are steep, these constraints require the qualities of some realized types to be inefficiently high for any signal that contains interior separating types. Since providing such qualities is suboptimal by Theorem~\ref{thm: Inefficiency}, it follows the seller finds inducing such signals suboptimal. Restricted to signals without interior separating types, one can employ standard arguments from the information design literature to show the program~\eqref{eq: optimal info problem} admits a binary-support solution.

Of particular interest are outcomes that can be implemented via a menu with a single-purchasing option. We refer to such outcomes as \textbf{posted-price outcome}, because they can be implemented with a single quality sold at a fixed price. Note that a sufficient condition for $(Q,F)$ to be a posted-price outcome is for $Q(\supp(F))$ to have only one non-zero element---that is, for $\supp(F)\setminus \{\thmin_Q\}$ to be a singleton. 

Corollary~\ref{cor: singleton menus} below provides conditions for optimality of a posted-price outcome. This condition relies on the following definition. For every $\thmin^* \in [\thmin, \theta_0]$, let $\hat{\pi}_{\thmin^*}:[\thmin^*,\thmax]\rightarrow \mathbb{R}$ be the function given by 
\[
\hat{\pi}_{\thmin^*}(\theta) = \theta(\icost'(\theta) - \icost'(\thmin^*)) - \qcost(\icost'(\theta) - \icost'(\thmin^*)) - \left[\icost(\theta)-\icost(\thmin^*)\right].
\]
To understand the motivation for this definition, suppose $Q$ is an $F$-CC allocation whose induced marginal price $\DMprice_{Q}$ is constant. In this case, $\pi_{Q}$ equals to $\hat{\pi}_{\thmin_Q}+\icost(\thmin_Q)$ on $[\thmin_{Q},\thmax_{Q}]$, and so has the same curvature $\hat{\pi}$. Corollary~\ref{cor: singleton menus} below utilizes this relationship. 

\begin{corollary}\label{cor: singleton menus}
    Suppose learning costs satisfy the steep slope condition.
    \begin{enumerate}[(i)]
    \item \label{cor: singleton menus - concave} If $\hat{\pi}_{\thmin^*}$ is concave for every $\thmin^*\in [\thmin, \theta_0]$, then a seller optimal outcome $(Q,F)$ exists such that $F$ is uninformative, and so $(Q,F)$ is a posted price outcome.
    \item \label{cor: singleton menus - convex} If $\hat{\pi}_{\thmin^*}$ is convex for  every $\thmin^*\in [\thmin, \theta_0]$, then a seller optimal outcome $(Q,F)$ exists such that $\supp(F)=\{\thmin_{Q},\thmax_{Q}\}$, and so $(Q,F)$ is a posted price outcome.
    \end{enumerate}
\end{corollary}

Next we explain the corollary's proof. As discussed above, when learning costs are steep, all seller-optimal outcomes involve signals that do not separate any interior types. Moreover, the support of every optimal $F$ must be interior, $\supp \ F \subseteq (\thmin,\thmax)$. Thus, the induced marginal price of any optimal CC allocation $Q$ is equivalent to a constant, meaning the seller's interim profit $\pi_{Q}$ must be equal to $\hat{\pi}_{\thmin_{Q}}+\icost(\thmin_Q)$ over the interval $[\thmin_{Q},\thmax_{Q}]$. The desired result then follows from the information design problem in Corollary~\ref{cor: two programs}. Specifically, whenever the conditions of  part~\eqref{cor: singleton menus - concave} hold, the program \eqref{eq: optimal info problem} admits an uninformative solution, whereas the conditions of part~\eqref{cor: singleton menus - convex} imply \eqref{eq: optimal info problem} admits a binary solution supported on $\{\thmin_{Q},\thmax_{Q}\}$. 
In both cases, one gets a posted-price seller-optimal outcome.\footnote{The logic underlying Corollary~\ref{cor: singleton menus} is reminiscent of the proof of Theorem~3 in \cite{mensch2022screening}. Within our setting, one can interpret that theorem as providing sufficient conditions for a single-quality menu to be optimal when $\qcost(q)=0$ for all $q$ and the state is binary. And, indeed, the last step of our argument is very similar to the proof of \citeapos{mensch2022screening} Theorem~3.
Thus, one can interpret Corollary~\ref{cor: singleton menus}  as an extension of \citeapos{mensch2022screening} Theorem~3 to case where costs are mean-measurable and have a steep-slope. Note that unlike \citeapos{mensch2022screening} Theorem~3, our result does not require $F_0$ to have binary support.   
}

Comparing the corollary to the standard analysis with a fully-informed buyer is instructive. In that setup, the seller typically finds it optimal to offer a single-purchasing option in one of two fairly restrictive cases: either marginal quality costs $\qcost'$ are constant, \citep[e.g.,][]{myerson1981optimal,riley1983optimal}, or the marginal profit of serving a positive quality to any type other than the highest is negative, in which case the seller chooses to only serve the highest type. 
Corollary~\ref{cor: singleton menus} highlights that costly information acquisition can also lead single-item menus. In particular, single-item menus arise when learning costs are steep, and either the seller wants to induce as little learning or as much learning as possible.

In addition to delivering simple menus, steep learning costs also enable us to compare the monopolistic distortions between the cases of acquired and exogenously given information. Towards this comparison, first note that type $\theta$'s quality is efficient if and only if the marginal benefit of a quality increase equals its marginal costs, $\theta = \qcost'(Q(\theta))$. Whenever $\theta > \qcost'(Q(\theta))$, the marginal benefit is strictly larger than the marginal costs, so the marginal surplus is positive and quality is underprovided. Following this logic, for a type distribution $F$ and an allocation $Q$, we define the average distortion as the wedge between expected marginal benefits and marginal costs of production: $\avgdist\left(Q,F\right) := \int\left[\theta -  \qcost'\left(Q(\theta) \right)\right] F(d\theta)$. Intuitively, a higher wedge implies that qualities are, in some sense, more distorted on average. Our next result compares the wedge $\avgdist$ under the seller-optimal allocation $Q$ with the allocation $Q^s_F$ obtained under standard screening for type distribution $F$.

\begin{proposition} \label{prop: average efficiency}
    Under the steep-slope condition, every optimal binary-menu outcome $(Q,F)$ satisfies $\avgdist (Q,F) > \avgdist(Q^s_F,F)$.
\end{proposition}

For intuition, suppose first that $\thmin_F$ is excluded under costly learning. In this case, since $\thmin_F$ is excluded, its allocated quality is weakly below the quality it would get under standard screening. Since $F$ is binary, the support of $F$ includes at most one additional type $\thmax_F$, and under the steep slope condition this type is strictly below $\thmax$.  It follows by Theorem~\ref{thm: Inefficiency} that $\thmax_F$ receives an inefficiently low quality. Since the standard screening allocation gives that type an efficient allocation, we get that the costly-learning allocation is pointwise lower than its  standard screening counterpart. It follows the average inefficiency is higher under the former than under the latter.

What if $\thmin_F$ is not excluded under costly learning? In this case one must consider the seller's marginal profits from changing the allocation. But not all changes are allowed, as some would trigger learning deviations. One change that would not trigger such a deviation, however, is decreasing the quality of all non-excluded types. Such a change must therefore reduce profits: the resulting marginal loss in revenue must be below the expected marginal cost savings; that is, $\thmin_Q \geq \int \qcost'(Q(\theta))F(\dd \theta)$. By contrast, under standard screening the seller can change the allocation in any way that preserves the buyer's reporting incentives. In particular, the seller can increase the quality of only the types weakly above $\thmin_F$. At the optimal standard screening allocation, this change must be unprofitable at the margin, meaning the expected marginal cost of production must be weakly above the lowest realized type, $\thmin_F$. Collecting these observations we get
\[
\int \qcost' (Q(\theta))F(\dd \theta) \leq \thmin_Q < \thmin_F \leq \int \qcost' (Q^s_F(\theta))F(\dd \theta),
\]
where the strict inequality follows from $\thmin_F$ receiving a strictly positive quality. Thus, the average distortion is larger when information is acquired than in standard screening.

\section{Concluding Remarks}\label{sec: Discussion}

We conclude our paper with a few brief remarks regarding our assumptions
and results.

\bigskip

\textbf{Support vs. positive probability.} 
Our results show whenever learning costs are not too flat, the seller strictly distorts downward the quality she provides to \emph{all} buyer types. This result stands in contrast to the conclusion one obtains when information is exogenous, where the highest buyer type is allocated the efficient quality. As such, our paper suggests that an analyst who examines the market under the assumption that information
is exogenous may come to erroneous conclusions regarding the efficiency
of the market's allocation. However, one might wonder whether this
error actually occurs: since the buyer's type distribution is endogenous,
the buyer may choose an $F$ that assigns zero probability
to the top of its support. It turns out, however, that it is without loss for $F$ to put positive probability on 
$\thmax_{F}$ whenever $\thmax_F <\thmax$. 
For an explanation, consider a seller optimal outcome $(Q,F)$ in which $\thmax_F <\thmax$ and where $F$ generates $\thmax_F$ with zero probability. We now sketch the construction of an alternative signal $G$ that attains the same value as $F$.\footnote{When $F_0$ admits a density, one can get this result by observing that restricting attention to bipooling signals \citep{kleiner2021extreme,arieli2023optimal} is without loss of optimality---see footnote~\ref{footnote: bipooling}.} Because $\thmax_F<\thmax$, $\thmax_F$ must be a pooling type, and so $F$ must have a pooling interval $(\theta_1,\theta_2)$ that contains $\thmax_F$ at the top of its support. Let $\tilde\theta$ be the mean of $F$ conditional on its realization being in that interval. Since $\thmax_F$ has zero probability, $F$ must generate multiple signal realizations inside  $(\theta_1,\theta_2)$, some of which are below $\tilde\theta$ and some are above. Denote the highest realization below $\tilde\theta$ as $l$, the lowest realization above $\tilde\theta$ as $h$, and take $G$ to be the signal that is equal to $F$ outside of $(\theta_1,\theta_2)$, but generates only $l$ and $h$ within  $(\theta_1,\theta_2)$. One can show such a $G$ exists, and attains the same value as $F$ for the program \eqref{eq: optimal info problem}. Moreover, $G$ is feasible in that program. Thus$(Q,G)$ must seller optimal, because $(Q,F)$ is. Moreover, $G$ generates $\thmax_G = h$ with positive probability. Hence, whenever our model yields downward distortion at top, it also enables this distortion to occur with strictly positive probability. 

\bigskip

\textbf{Inefficiency of serving some types.} Throughout, we assumed that $\qcost'(0) < \thmin$, meaning it is efficient to serve all types. We use this assumption to prove  Theorem~\ref{thm: Inefficiency} and Theorem~\ref{thm: large cost comparative statics}. In Theorem~\ref{thm: Inefficiency}, we use the assumption that $\qcost'(0) < \thmin$ to show that excluding $\thmin_F$ is inefficient. Thus, the theorem continues to hold whenever it is efficient to serve all types that can arise in a signal that is incentive compatible for some menu. More specifically, suppose $C(F)=\costcoef\int \icost(\theta)F(\dd\theta)$, and there is some $\hat\thmin_\costcoef$ such that $\icost'(\hat\thmin_\costcoef) = \icost'(\theta_0) - \tfrac{1}{\costcoef}\bar{q}$. Then Theorem~\ref{thm: Inefficiency} holds as stated so long  $\qcost'(0) < \hat\thmin_\costcoef$. The reason is simple: by Theorem~\ref{thm: FCC}, a signal $F$ is IC for some allocation only if  $\thmin_F \geq \hat\thmin_\costcoef$.\footnote{Formally, if $F$ is IC for some allocation, it is also IC for some $F$-CC allocation $Q$. Consequently, $\bar{q} \geq Q(\thmax_F) - Q(\thmin) \geq \costcoef\icost'(\thmax_F) - \costcoef\icost'(\thmin_F) \geq \costcoef\icost'(\theta_0) - \costcoef\icost'(\thmin_F).$ That $\thmin_F \geq \hat\thmin_\costcoef$ then follows from $\icost'$ being strictly increasing.} Therefore, it is always optimal to serve the lowest realized buyer type. The rest of Theorem~\ref{thm: Inefficiency} goes through as before. 

The assumption that $\qcost'(0) < \hat\thmin_\costcoef$ is also sufficient for extending Theorem~\ref{thm: large cost comparative statics}. In fact, Theorem~\ref{thm: large cost comparative statics}, requires an even weaker condition: $\qcost'(0) < \hat\thmin_\costcoef$ must hold for \emph{some} $\costcoef$. For an explanation, note that this weaker condition is equivalent to the assumption that $\qcost'(0) < \theta_0$, which in turn implies the seller can induce no learning while still making a positive profit. This implication is sufficient for Theorem~\ref{thm: large cost comparative statics}: whenever the implication holds, the theorem is true regardless of whether it is efficient to serve the lowest possible type $\thmin$. 

Without these alternative assumptions, Theorem~\ref{thm: large cost comparative statics}, may fail, because it may be optimal for the seller to not serve anyone at the optimum. Theorem~\ref{thm: Inefficiency}, however, remains true so long as it is efficient to give a strictly positive quality to $\thmax$. To prove this, one can show  there is at most one signal realization $\theta^*\in\supp (F)$ for which $\qcost'(Q(\theta^*))>\theta^*$. Moreover, this realization must be excluded, meaning that $\theta^*=\thmin_Q$. So, in effect, the seller never provides an inefficiently high, positive quality to any realized $\theta$. We provide a proof sketch in the online appendix.

\bigskip

\textbf{Bounded Derivatives of $\icost$.}  Throughout the analysis we assumed that $\icost$ has a bounded second derivative, which in turn implies a bounded first derivative. Most of our results do not depend on this assumption---their proof applies without change even without it. The only results whose proof does not go through are the results for the case with low learning costs.\footnote{Our proof Theorem~\ref{thm: FCC} does not change if $\icost'$ is unbounded, since it applies \citeapos{dizdar2020simple} generalization of \cite{dworczak2019simple}. See the Appendix for precise formulation.} These are Part~\eqref{prop: sufficient distortion - low costs} of Proposition~\ref{prop: sufficient distortion} and Theorem~\ref{thm: low cost comparative statics}. These results both rely on the fact that the seller induces full learning when costs are low. This fact fails when the first derivative of $\icost$ is unbounded: in that case, the seller cannot create a sufficient degree of product differentiation to make full-learning incentive compatible, even when $\costcoef$ is very low (see equation~\ref{eq: minimal differentiation}).

Our arguments also fail when the first derivative $\icost'$ happens to be bounded but the second derivative $\icost''$ is not. The reason is the curvature of the seller's profits as a function of the buyer's realized type. When $\icost''$ is bounded, one can use Condition~\ref{as: finitesupport} to show that, when $\costcoef$ is low enough, the seller strictly benefits from having the buyer acquire additional information for all cost-compensating allocations. With unbounded $\icost''$, this argument could fail. In particular, the seller could prefer to avoid revealing to the buyer that his true type is $\thmin$ for all values of $\costcoef$. Such a preference would violate Theorem~\ref{thm: low cost comparative statics}, which in turn invalidates our argument for Part~\eqref{prop: sufficient distortion - low costs} of Proposition~\ref{prop: sufficient distortion}. 

It turns out, however, that Proposition~\ref{prop: sufficient distortion}'s Part~\eqref{prop: sufficient distortion - low costs} holds as long as $\icost'$ is bounded---even if $\icost''$ is not. While our current proof fails (since it relies on 
Theorem~\ref{thm: low cost comparative statics}), one can make an alternative argument, sketched as follows. If $F_0$ is finite and regular, one can show that for any optimal cost compensating $(Q,F)$, the associated $F$-marginal price $\DMprice$ must be constant around any $\theta$ that is not in $F_0$'s support. Using this observation, one can find a $\hat\theta \in (\thmin,\thmax)$ that satisfies the following three properties for all sufficiently small $\costcoef$. First, the highest realization of any seller optimal $F$ must be in $(\hat\theta,\thmax]$. Second, every optimal $F$-CC allocation must be essentially flat on $(\hat\theta,\thmax)$. And third, every optimal $F$-CC allocation has a strictly positive upward jump at $\thmax$ whose size is bounded away from zero. Together, these properties imply that $\pi_Q$ must have a similar shape on $(\hat\theta,\thmax_F]$ as $Q$: $\pi_Q$ must be nearly flat on $(\hat\theta,\thmax_F)$ and have a strictly positive upward jump at $\thmax$. This shape, in turn, implies that $\thmax_F$ must equal $\thmax$---if not, the seller would strictly benefit from spreading $\thmax_F$ between $\thmax_F-\epsilon$ and $\thmax$ for some sufficiently small $\epsilon>0$.

\bigskip

\textbf{Affine Learning Costs.} We assume the buyer's costs of learning to be affine in the distribution of her posterior estimate. As explained earlier, the assumption that costs depend only on the distribution of the buyer's posterior estimate is without loss. However, the assumption that costs are affine in this distribution is substantive. Next, we explain how our analysis changes if we instead assume the buyer's learning costs can be locally approximated by an affine function. Specifically, say $C$ is \textbf{Gateaux differentiable} if every $F \in \Signals$ admits some twice differentiable strictly convex function $\icost_{F}:\Theta \rightarrow \mathbb{R}$ such that for every $G \in \Signals$,
\[
\lim_{\epsilon \searrow 0}\frac{1}{\epsilon}\left[C(F+\epsilon(G-F)) - C(F)\right] = \int \icost_{F}(\theta)(G-F)(\dd\theta).
\]
Assumptions of this type were first introduced into information acquisition models by \cite{ravid2022learning}. \cite{lipnowski2022predicting} introduce a generalization of this class of cost functions that are not mean-measurable---i.e., costs that depend on the distribution of the buyer's posterior \emph{belief}. 

Under Gateaux differentiability, our reduction to $F$-CC allocations is still without loss, though one must adjust the definition of CC allocations so that $\icost^{'}_{F}$ replaces $\icost'$. To get this result, one first applies Lemma 1 from \cite{georgiadis2022flexible} to get that $F$ solves the buyers problem if and only if it solves the buyer's problem when her costs are given by their affine approximation at $F$.\footnote{The lemma is stated for the case where the agent can induce any distribution, but applies as stated to any convex constraint set.} Said differently, a given $F$ is incentive compatible for the buyer if and only if it is buyer optimal when the buyer's cost function is given by $G \mapsto \int c_{F}(\theta) G(\dd \theta)$. Theorem~\ref{thm: FCC} therefore implies it is without loss to focus on (the properly defined) $F$-CC allocations. 

The appropriate adjustment of Corollary~\ref{cor: two programs} also continues to hold. In particular, it turns out the adjusted mechanism design problem~\eqref{eq: optimal allocation problem} is amenable to the same techniques as the problem presented in this paper. Consequently, one can show that the seller optimal outcome $(Q,F)$ distorts quality downward for any $F$-separating $\theta$.  The reason is that this observation relies only on perturbations of the $F$-marginal price. 

The rest of our results, however, cannot be proven as is. The reason is that the affine approximation $\icost_F$ depends on $F$, making the analogue of the information design problem~\eqref{eq: optimal info problem} much less tractable. For an explanation, note that we prove our other results by fixing a CC allocation $Q$ and varying $F$ within the set of signals for which $Q$ is CC. When learning costs are affine, this set contains any signal that separates any $\theta$ at which $\DMprice_Q$ is strictly increasing. The same is not true when $C$ is Gateaux differentiable: in that case, changing $F$ typically involves changing $\icost^{\prime}_{F}$, and therefore the allocation. Hence, the arguments for the rest of our results do not apply when costs are merely Gateaux differentiable. Developing a method for analyzing the case with Gateaux differentiable costs remains an open question.

\bibliographystyle{jpe.bst}
\bibliography{QualityLearning}

\pagebreak{}

\appendix

\section{Online Proofs Appendix}

We introduce some notation and definitions that we use throughout the appendix. For every CDF $F$, define the function $I_{F}:\Theta \rightarrow \mathbb{R}$ as
\begin{align*}
I_{F}(\theta):=\int_{\tilde\theta\in[\underline{\theta},\theta]}\left(F_{0}-F\right)(\tilde{\theta})\mathrm{d}\tilde{\theta}.
\end{align*}
Note a CDF $F$ is a signal---that is, $F \in \Signals$---if and only if $I_F(\thmax)=0$ and $I_F(\theta)\geq 0$ holds for all $\theta$. Moreover, $\theta$ is $F$-separating if and only if $I_F(\theta)=0$.

If $(Q,F)$ is IC, we say that $Q$ is $F$\textbf{-incentive compatible } ($F$-IC). For every $\theta \in (\thmin,\thmax]$, let $\varphi_{-}(\theta) = \sup_{\theta'<\theta}\varphi(\theta')$ be the left limit of $\varphi$ at $\theta$, and set $\varphi_{-}(\thmin) = x$. Similarly, for $\theta \in [\thmin,\thmax)$, define the right limit of $\varphi$ and $\theta$ by $\varphi_{+}(\theta) = \inf_{\theta' > \theta}\varphi(\theta')$, while setting $\varphi_{+}(\thmax) = y$. Relatedly, for a convex function $\phi:\Theta \rightarrow [x,y]$, we let we let $\phi'_{-}$ and $\phi'_{+}$ denote its left
and right derivatives, respectively, whenever those exist (which is the case for every $\theta \in (\thmin,\thmax)$). 

Armed with the definitions above, we say that an allocation $Q$ \textbf{jumps towards efficiency} if 
\begin{equation}\label{eq: efficient jumps}
    Q(\theta) \in \argmax_{q \in [Q_{-}(\theta),Q_{+}(\theta)]} \left[\theta q - \qcost(q) \right].
\end{equation}
The following observations, proved next, simplify the analysis. First, one can take any allocation and replace it with an allocation that jumps towards efficiency without reducing the seller's profits. Consequently, focusing on allocations that jump towards efficiency is without loss of optimality. Second,  allocations that jump towards efficiency convey a technical benefit: if $Q$ jumps towards efficiency, then $\pi_{Q}$ is an upper-semicontinuous function. Therefore, we restrict attention to allocations that jump towards efficiency in all the proofs.

\subsection{Jumps Towards Efficiency}
In this section we prove that jumps towards efficiency are without loss of optimality. Moreover, we show that whenever $Q$ jumps towards efficiency, $\pi_{Q}$ is upper-semicontinuous.

We begin with proving that jumping towards efficiency is without loss of optimality. 

\begin{lemma}\label{lem: jumps towards efficiency are wlog}    
    Suppose $(Q,F)$ is IC. Then there is an allocation $Q^*$ that jumps towards efficiency such that $(Q^*,F)$ is IC, and $\int \pi_{Q^*}(\theta)F(\dd \theta) \geq \int \pi_{Q}(\theta)F(\dd \theta)$. Moreover, if $(Q,F)$ is seller optimal, $Q^*$ equals $Q$ $F$-almost surely. 
\end{lemma}
\begin{proof}[Proof of Lemma~\ref{lem: jumps towards efficiency are wlog}]
Suppose $Q$ is $F$-IC. Define the allocation $Q^*$ via
\[
Q^*(\theta):=\min\left\{\tilde{q}: \tilde{q} \in \argmax_{q\in\left[Q_{-}(\theta),Q_{+}(\theta)\right]}[\theta q-\qcost\left(q\right)]\right\},
\]
which is well-defined because $\qcost$ is convex. Note that $Q^*$ equals $Q$ at any $\theta$ where $Q$ is continuous. Since $Q$ is discontinuous in at most a countable set of points, we get that $Q^{*}_{+}=Q_{+}$ and $Q^{*}_{-}=Q_{-}$. It follows $Q$ jumps towards efficiency. In addition, note that $Q$ and $Q^*$ differ on at most a countable set of points, which has Lebesgue measure zero, and so $\idu_{Q} = \idu_{Q^*}$. It follows $(Q,F)$ is IC if and only if $(Q^*,F)$ is IC as well. 

We now turn to showing the sellers profit under $(Q^*,F)$ is no lower than it is under $(Q,F)$. To see this, note that 
\begin{equation}\label{eq: 08-14-23b}
\begin{split}
\int [\pi_{Q}(\theta) - \pi_{Q^*}(\theta)]F(\dd \theta) 
= \int\left[\left(\theta Q(\theta)-\qcost\circ Q(\theta) \right)-\left(\theta Q^*(\theta)-\qcost\circ Q^*(\theta)\right)\right]F(\dd\theta)\leq 0,
\end{split}
\end{equation}
where the first equality follows from $\idu_{Q}=\idu_{Q^*}$.

We now conclude the proof by arguing that if $(Q,F)$ is seller optimal, then $Q^*=Q$ $F$-almost surely. 
For this purpose, note that strict convexity of $\qcost$ means that $Q^*$ is the unique maximizer of $q\mapsto \theta [q - \qcost (q)]$ over $[Q_{-}(\theta), Q_{+}(\theta)]$. Consequently, \eqref{eq: 08-14-23b} must hold with strict inequality whenever $Q^* \neq Q$ over a positive $F$-measure set. But $Q^*$ is $F$-IC, and so $(Q,F)$ being seller optimal means that the inequality in \eqref{eq: 08-14-23b} must hold with equality. It follows that $Q^*=Q$ $F$-almost surely. 
\end{proof}

Next, we show that $Q$ jumping towards efficiency has a useful technical benefit: it makes $\pi_Q$ upper semicontinuous.

\begin{lemma}\label{lem: jumping towards effiency yields usc objective}
    If an allocation $Q$ jumps towards efficiency, $\pi_{Q}$ is upper semicontinuous.
\end{lemma}

\begin{proof}[Proof of Lemma~\ref{lem: jumping towards effiency yields usc objective}]
Suppose $Q$ jumps towards efficiency. In what follows, define
\[
Q^{e}_{-}(\theta) = \min \{q: q \in \argmax_{\tilde q \in [0,\bar{q}]} [\theta \tilde q - \qcost(\tilde q)]\}, 
\quad \text{and} \quad 
Q^{e}_{+}(\theta) = \max \{q: q \in \argmax_{\tilde q \in [0,\bar{q}]} [\theta \tilde q - \qcost(\tilde q)]\}.
\]
It is easy to verify that both $Q^e_{-}$ and $Q^e_{+}$ are increasing. Berge's Maximum Theorem then delivers that $Q^e_{-}$ and $Q^e_{+}$ are respectively lower and upper semicontinuous, which (combined with monotonicity of the two functions) delivers that the functions are respectively left and right continuous. Moreover, Berge's Maximum Theorem also delivers that $\lim_{\theta_n\nearrow \theta} Q^{e}_{+}(\theta_n) = Q^{e}_{-}(\theta)$ and $\lim_{\theta_n\searrow \theta} Q^{e}_{-}(\theta_n) = Q^{e}_{+}(\theta)$. Finally, $\argmax_{q \in [0,\bar q]} [\theta q - \qcost(q)] = [Q^{e}_{-}(\theta),Q^{e}_{+}(\theta)]$ holds because $q \mapsto (\theta q - \qcost(q))$ is a concave function. 

Let $(\theta_n)_{n\in \mathbb{N}}$ be some convergent sequence, and take $\theta_{\infty}$ to be its limit. Our goal is to show that $\limsup_{n} \pi_{Q}(\theta_n) \leq \pi_{Q}(\theta_\infty)$. Since $V_{Q}$ is continuous, it is sufficient to show that
\begin{equation}\label{eq: 8-14-23a}
\limsup_{n}\left[\theta_n Q(\theta_n) - \qcost\circ Q (\theta_n) \right] \leq \theta_\infty Q(\theta_\infty) - \qcost\circ Q(\theta_\infty). 
\end{equation}
Obviously, equation~(\ref{eq: 8-14-23a}) holds if $Q$ is continuous at $\theta_\infty$. Thus, from now on we assume a discontinuity in $Q$ at $\theta_\infty$, so $Q_{-}(\theta_\infty)< Q_{+}(\theta_\infty)$. Moreover, since every sequence admits a monotone subsequence, to show the above inequality it is sufficient to show that it holds when $(\theta_n)_{n\in\mathbb{N}}$ is monotone. Without loss of generality, suppose $\theta_n$ is monotone increasing.

We proceed in cases. 
\begin{itemize}
    \item [Case 1.] Suppose that $Q(\theta_\infty) \in [Q^{e}_{-}(\theta_\infty),Q^{e}_{+}(\theta_\infty)]$ (i.e., $Q(\theta_\infty)$ is efficient). Then for every $n$, 
\[
\begin{split}   
\theta_n Q(\theta_n) - \qcost\circ Q (\theta_n)
\leq 
\theta_n Q^{e}_{+}(\theta_n) - \qcost\circ Q^{e}_{+} (\theta_n) 
& \rightarrow \theta_\infty Q^{e}_{-}(\theta_\infty) - \qcost\circ Q^{e}_{-} (\theta_\infty)
\\ & = \theta_\infty Q(\theta_\infty) - \qcost\circ Q (\theta_\infty),
\end{split}
\]
where convergence follows from Berge's Maximum Theorem. 

\item [Case 2.] Suppose $Q(\theta_\infty) < Q^{e}_{-}(\theta_\infty)$. Because $q \mapsto \theta q - \qcost(q)$ is concave and $Q$ jumps towards efficiency, $Q(\theta_\infty) = Q_{+}(\theta_\infty)<Q^{e}_{-}(\theta_\infty)$. In the next paragraph we argue that $Q(\theta_n) < Q^{e}_{-}(\theta_n)$ holds for all sufficiently large $n$. Taking this inequality as given, note that, because $\theta_n \nearrow \theta_\infty$, $Q^{e}_{-}(\theta_n)\rightarrow Q^{e}_{-}(\theta_\infty)$, and so for all sufficiently large $n$, $Q(\theta_n) \leq Q(\theta_\infty) < Q^{e}_{-}(\theta_n).$ Since $q \mapsto \theta q - \qcost(q)$ is concave, we get that
\[
\theta_n Q(\theta_n) - \qcost\circ Q (\theta_n) 
\leq \theta_n Q(\theta_\infty) - \qcost\circ Q (\theta_\infty) 
\rightarrow \theta_\infty Q(\theta_\infty) - \qcost\circ Q (\theta_\infty),
\]
as required. 

Thus, to complete the proof of this case, it remains to show that $Q(\theta_n) < Q^{e}_{-}(\theta_n)$ holds for all sufficiently large $n$. Suppose otherwise. Then one can find a subsequence $(Q_n)_{n\in N}$ for some infinite $N \subseteq \mathbb{N}$ such that $Q(\theta_n) \geq Q^{e}_{-}(\theta_n)$ for all $n \in N$. Since $Q^{e}_{-}$ is left continuous and $\theta_n \nearrow \theta_\infty$, one gets that
\[
Q(\theta_\infty) \geq Q(\theta_n) \geq Q^{e}_{-}(\theta_n) \rightarrow Q^{e}_{-}(\theta_\infty),
\]
meaning $Q(\theta_\infty) \geq Q^{e}_{-}(\theta_\infty)$, a contradiction. The proof of this case is therefore complete. 

\item [Case 3.] Suppose $Q(\theta_\infty) > Q^{e}_{+}(\theta_\infty)$. Then concavity of $q \mapsto \theta q - \qcost(q)$ and $Q$ jumping towards efficiency means that $Q(\theta_\infty) = Q_{-}(\theta_\infty)$. Since $\theta_n \nearrow \theta_\infty$ and $Q$ is monotone, $Q(\theta_n)\rightarrow Q_{-}(\theta_\infty)=Q(\theta_{\infty})$. It follows \eqref{eq: 8-14-23a} holds.
\end{itemize}
\end{proof}

\subsection{Proof of Lemma \ref{Existence}}

We begin by formally defining the buyer's maximization problem holding
the seller's menu fixed. By standard arguments we only need to consider $t\geq 0$. Let $X=\left[0,\bar{q}\right]\times\mathbb{R}_{+}$,
and endow the set of Borel measures over $X\times\Theta$, $\Delta\left(X\times\Theta\right)$,
with the weak{*} topology. Given a menu $M$, the buyer's program
can be written as 
\begin{align*}
\max_{\xi\in\Delta\left(X\times\Theta\right)} & \int\left(\theta q-t\right)\xi\left(\mathrm{d}(q,t,\theta)\right)-C\left(\mathrm{marg}_{\Theta}\,\xi\right)\\
\text{s.t. } & \supp\,\xi\subseteq M\times\Theta,\\
 & \text{marg}_{\Theta}\,\xi\preceq F_{0}.
\end{align*}
Observe the above program involves the maximization of a continuous
objective over a compact constraint set, and so the set of solution,
$\Xi\left(M\right)$, is non-empty for every compact $M$. Letting
$\mathcal{M}$ be the collection of compact subsets of $X$ that contain the non-participation option
$\left(0,0\right)$, the seller's program can be written as
\begin{align*}
\max_{\left(M,\xi\right)\in\mathcal{M}\times\Delta\left(X\times\Theta\right)} & \int\left(t-\qcost\left(q\right)\right)\xi\left(\dd(q,t,\theta)\right)\\
\mathrm{s.t.}\, & \xi\in\Xi\left(M\right).
\end{align*}
Notice it is without loss to assume $M\subseteq\bar{X}=\left[0,\bar{q}\right]\times\left[0,\bar{\theta}\bar{q}\right]$, because
the buyer strictly prefers $\left(0,0\right)$ to any menu item
that includes a transfer strictly above $\bar{\theta}\bar{q}$. Let
$\mathcal{K}\left(\bar{X}\right)$ be the set of all compact non-empty
subsets of $\bar{X}$ endowed with the Hausdorff metric, 
\[
d\left(A,B\right)=\max\left\{ \max_{b\in B}\min_{a\in A}d\left(b,a\right),\max_{a\in A}\min_{b\in B}d\left(a,b\right)\right\} ,
\]
and take $\bar{\mathcal{M}}$ to be the elements of $\mathcal{K}\left(\bar{X}\right)$
that contain $\left(0,0\right)$. Taking $\bar{\Xi}$ to be the restriction
of $\Xi$ to $\bar{\mathcal{M}}$, and letting 
\[
\mathrm{Gr} \ \bar{\Xi} = \left\{(M,\xi)\in \bar{\mathcal{M}}\times \Delta(X\times \Theta): \xi \in \bar{\Xi}(M) \right\}
\] 
denote the restriction's graph, we get that the seller's problem
can be rewritten as
\begin{equation}
\max_{\left(M,\xi\right)\in\mathrm{Gr}\,\bar{\Xi}}\int\left(t-\qcost\left(q\right)\right)\xi\left(\dd(q,t,\theta)\right).\label{eq:ExistencePrincipal}
\end{equation}
Observe $\bar{\mathcal{M}}$ is a closed subset of $\mathcal{K}\left(\bar{X}\right)$,
and so because $\mathcal{K}\left(\bar{X}\right)$ is compact (\citet{aliprantis2006infinite},
Theorem 3.85), $\bar{\mathcal{M}}$ must be compact as well. It follows,
by Berge's theorem of the maximum, that $\bar{\Xi}$ is
upper-hemicontinuous and has a closed graph (\citet{aliprantis2006infinite},
Theorem 17.10). Hence, this graph must be compact because it is a
subset of $\bar{\mathcal{M}}\times\Delta\left(\bar{X}\times\Theta\right)$,
which is compact. That (\ref{eq:ExistencePrincipal}) admits a solution
follows. \qedhere

\subsection{Proofs from Section \ref{section:F-ICC}}
We begin with stating \citeapos{dizdar2020simple} generalization of \citeapos{dworczak2019simple} duality result. 

\begin{theorem}[\citeauthor{dizdar2020simple}, \citeyear{dizdar2020simple}]\label{thm:DM-Theorem} Let $\Phi:\Theta \rightarrow \mathbb{R}$ be a bounded, upper semicontinuous function that admits some $\epsilon>0$ and $\bar L\in \mathbb{R}_{+}$ such that $\Phi(\theta) - \Phi(\theta') \geq \bar{L}(\theta' - \theta)$ holds for 
$\thmin\leq \theta' \leq \theta \leq \thmin+\epsilon$, and
$\Phi(\theta) - \Phi(\theta') \leq \bar{L}(\theta - \theta')$ holds for 
$\thmax-\epsilon\leq \theta' \leq \theta \leq \thmax$. 
Then,
\[
F \in \argmax \int \Phi (\theta) F(\dd\theta) 
\]
if and only if an $F$-price $\DMPrice$ exists such that $\DMPrice(\theta) \geq \Phi(\theta)$  for $\theta$, where the inequality holds with equality for $\theta \in \supp \ F$.\footnote{
The statement of the theorem here is slightly more general than the one stated by \cite{dizdar2020simple}. However, the exact same steps as in their proof hold, with $\bar{L}$ replacing $L$ in equation (8) of their paper.
}    
\end{theorem}

We now use this theorem to prove the following specialization of \citeapos{dworczak2019simple} duality result to our setting.

\begin{lemma}\label{lem:dwm}
    Fix an allocation $Q$. Then, $F$ solves \eqref{eq: buyer's problem} if and only if a Lipschitz continuous and convex function $P:\Theta \rightarrow \mathbb{R}$ exists such that $P \geq V_Q-\icost$,  
    $$ P(\theta) = \idu_Q(\theta)-\icost(\theta), \text{ for all } \theta \in \supp F,$$
    and $\DMPrice$ is affine on any interval $(\theta_1,\theta_2)$ of $F$-pooling types.
\end{lemma}
\begin{proof}[\textbf{Proof of Lemma \ref{lem:dwm}}]
    We begin by arguing that $\idu_Q-\icost$ satisfies the prerequisites of Theorem~\ref{thm:DM-Theorem}. By definition, $\idu_Q-\icost$ is continuous, and therefore bounded and upper-semicontinuous. Now pick any $\epsilon>0$. Then for any $\theta'<\theta$ such that $(\theta',\theta) \in [\thmin,\thmin+\epsilon]$ we have
    \[
    \idu_Q(\theta)-\icost(\theta) - \idu_Q(\theta')-\icost(\theta') = \int_{\tilde\theta \in (\theta',\theta)} \left[Q(\tilde\theta) - \icost'(\tilde(\theta)) \right]\dd \tilde{\theta}
    \geq -\icost'(\thmin+\epsilon)(\theta - \theta') = \icost'(\thmin+\epsilon)(\theta' - \theta),
    \]
    where the inequality follows from $Q\geq 0$ and $\icost'$ being increasing. Similarly, for every $\theta'<\theta$ such that $(\theta',\theta) \in [\thmax-\epsilon,\thmax]$,
    \[
    \idu_Q(\theta)-\icost(\theta) - \idu_Q(\theta')-\icost(\theta') = \int_{\tilde\theta \in (\theta',\theta)} \left[Q(\tilde\theta) - \icost'(\tilde(\theta)) \right]\dd \tilde{\theta}
    \leq (\bar{q} - \icost'(\thmax-\epsilon))(\theta - \theta'),
    \]
    where the inequality follows from $Q\leq \bar{q}$ and $\icost'$ being increasing. It follows that, by taking $\Phi=\idu_Q-\icost$, the assumptions of Theorem~\ref{thm:DM-Theorem} are satisfied. Because the lemma's conditions imply $\DMPrice$ is an $F$-price, Theorem~\ref{thm:DM-Theorem} delivers that $F$ is buyer optimal. 

    For the ``only if" part, suppose $F$ solves the buyer's problem. Let $\DMPrice$ be the $F$-price delivered by Theorem~\ref{thm:DM-Theorem}. We then have $\DMPrice\geq \idu_Q-\icost(Q)$, with the inequality holding with equality over the support of $F$.

\end{proof}

Next we prove a lemma that recovers an $F$-marginal price from an $F$-price that certifies $F$ as buyer optimal given $Q$. Moreover, we show one can choose $\DMprice$ such that $\DMprice+\icost'$ coincide with $Q$ on the support of $F$.

\begin{lemma}\label{lem: marginal price from price}
Fix an outcome $(Q,F)$ and let $\DMPrice$ be an $F$-price with $\DMPrice \geq \idu_Q - \icost$, with equality in $\supp F$. Then, there exists an $F$-marginal price $\DMprice$ with $Q = \DMprice + \icost'$ in $\supp F$, and:

$$\DMPrice(\theta) = \DMPrice(\thmin)+ \int^\theta_{\thmin} \DMprice(x) dx, \ \theta \in [\thmin,\thmax].  $$
\end{lemma}
\begin{proof}[Proof of Lemma~\ref{lem: marginal price from price}]
Since $\DMPrice$ is Lipschitz, there is a function $\DMprice:\Theta \rightarrow \mathbb{R}$ such that $\DMPrice(\theta) = \DMPrice(\thmin) + \int_{\thmin}^{\theta}\DMprice(\tilde\theta) \dd \tilde\theta$ for all $\theta$. 

    Next, we claim one can take $\DMprice$ such that $\DMprice(\theta) = Q(\theta) - \icost'(\theta)$ for all $\theta \in \supp \ F$. To do so, we show below that for each such $\theta$, 
    \begin{equation}\label{eq: 2024-02-10a}
    \DMprice_{+}(\theta) \geq Q_{+}(\theta) - \icost'(\theta)\geq Q_{-}(\theta) - \icost'(\theta) \geq \DMprice_{-}(\theta).
    \end{equation}
    Equation~\ref{eq: 2024-02-10a} immediately implies the desired equality whenever $\DMprice_{+}(\theta) = \DMprice_{-}(\theta)$. Moreover, since $\DMPrice$ is convex, we can take $\DMprice$ to be increasing, and so $\DMprice_{+}(\theta) > \DMprice_{-}(\theta)$ holds over a Lebesgue-null set, meaning one can edit $\DMprice$ so that it satisfies the desired equality without impacting its integral. Hence, to prove that one can take $\DMprice$ such that $\DMprice(\theta) = Q(\theta) - \icost'(\theta)$ for all $\theta \in \supp \ F$, it is sufficient to show that \eqref{eq: 2024-02-10a} holds for all such $\theta$. To show this inequality, notice that for any $\epsilon >0$,
    \[
    \begin{split}
    0 
    & \leq \frac{1}{\epsilon}\left[\DMPrice(\theta-\epsilon) - \left(\idu_Q - \icost\right)(\theta-\epsilon)\right] 
    \\ & = \frac{1}{\epsilon}\left[\DMPrice(\theta-\epsilon) - \DMPrice(\theta) + \left(\idu_Q-\icost\right)(\theta) - \left(\idu_Q-\icost\right)(\theta-\epsilon)\right] 
    \\ & \xrightarrow[]{\epsilon\searrow 0}-\DMprice_{-}(\theta) + Q_{-}(\theta) - \icost'(\theta),
    \end{split}
    \]
    where the first inequality follows from $\DMPrice \geq \idu_Q - \icost$, and the second equality from $\DMPrice$ being equal to $\idu_Q-\icost$ over the support of $F$ (and $\theta$ being in that support). It follows $Q_{-}(\theta) - \icost'(\theta) \geq \DMprice_{-}(\theta)$. An analogous argument delivers that $\DMprice_{+}(\theta) \geq Q_{+}(\theta) - \icost'(\theta)$. 
    
    Thus, all that remains is to argue that we can take $\DMprice$ so that it satisfies the conditions of an $F$-marginal-price. Towards this goal, recall we already argued we can take $\DMprice$ to be increasing. Moreover, because $\DMPrice$ is affine on any interval over which $I_{F}$ is strictly positive, $\DMprice$ is constant over any such interval. Continuity of $I_F$ then implies that if $I_F(\theta)$ is strictly positive, $I_F$ must be strictly positive over a neighborhood of $\theta$, and so $\DMprice$ must be constant around $\theta$. 
    
    Finally, we need to argue that $\DMprice \in [-\icost'(\thmin_F),\bar{q}-\icost'(\thmax_F)]$. To do so, notice first it is without loss to choose $\DMprice$ so that $\DMprice(\thmax) = \DMprice_{-}(\thmax_F)$: either $\thmax_F = \thmax$, or $\thmax_F < \thmax$, in which case $I_F$ must be strictly positive over $(\thmax_F-\epsilon,\thmax)$ for some small $\epsilon>0$, meaning $\DMprice$ must be constant over the same interval, and so setting $\DMprice(\thmax) = \DMprice(\thmax_F)=\DMprice_{-}(\thmax_F)$ is without loss. An analogous argument delivers it is without loss to have $\DMprice(\thmin) = \DMprice_{+}(\thmin_F)$. Combined with equation~\ref{eq: 2024-02-10a}, we get that 
    \[
    \DMprice(\thmax) = \DMprice(\thmax_F) = Q(\thmax_F) - \icost(\thmax_F) \leq \bar{q} - \icost(\thmax_F).
    \]
    An analogous argument delivers $\DMprice(\thmin)=\DMprice(\thmin_F) \geq -\icost'(\thmin_{F})$. Since $\DMprice$ is increasing, we get $\DMprice \in [-\icost'(\thmin_F),\bar{q}-\icost'(\thmax_F)]$ as required.  

\end{proof}

Finally, we prove that focusing on F-CC allocations is without loss of generality. 

\begin{proof}[\textbf{Proof of Theorem \ref{thm: FCC}}]

As a preliminary step, suppose $F$
is buyer-optimal for some mechanism, and that we have some $F$-CC allocation
$Q$. Let $\DMprice = \DMprice_{Q}$ be the $F$-marginal price associated with $Q$.

We first show $F$ is buyer-optimal for every $F$-CC allocation. 
Let $Q$ be an $F$-CC allocation, with associated $F$-price  $\DMPrice_Q$. To show that $F$ is buyer-optimal given $Q$, by Lemma~\ref{lem:dwm}, it suffices to show that $\DMPrice_Q(\theta)\geq \idu_Q(\theta)-\icost(\theta)$ holds for all $\theta$, and that $\DMPrice_Q(\theta)=\idu_Q(\theta)-\icost(\theta)$ for all $\theta \in \supp \ F$. But by the definition of $F$-CC allocations, $\DMPrice_Q \geq \idu_Q - \icost$, with $\DMPrice_Q=\idu_Q - \icost$ in $[\thmin_Q,\thmax_Q]$. Because $[\thmin_F,\thmax_F] \subset [\thmin_Q,\thmax_Q]$, the result is proved.

Next, we argue every allocation $\tilde{Q}$ for which $F$ is buyer-optimal admits an equivalent $F$-CC allocation. Let $F$ be buyer-optimal for an allocation $\tilde{Q}$. Let $\tilde{\DMPrice}$ be an $F$-price that certifies buyer-optimality of $\tilde{Q}$. We shall prove there is an $F$-CC allocation, $Q$ such that, in the support of $F$, $Q=\tilde{Q}$ , and $V_{Q} \leq V_{\tilde{Q}}$. By Lemma~\ref{lem: marginal price from price}, we can find an $F$-marginal price $\tilde{\DMPrice}$ such that: 

\[
\tilde{\DMPrice}(\theta) = \tilde{\DMPrice}(\thmin) +\int^\theta_{\thmin} \tilde{\DMprice} (x) dx, \ \ \theta\in[\thmin,\thmax], 
\]
and $\tilde{\DMPrice}=\tilde{Q}-\icost'$ in the support of $F$. Define:

$$ Q(\theta) = \min\{\max\{ \tilde{\DMprice} (\theta)+\icost'(\theta),0 \},\bar{q}\}.$$

Also define $\thmin_Q = \inf[\{\thmax\} \cup Q^{-1}(0,\bar{q}]]$ and $\thmax_Q = \sup[\{\thmin\} \cup Q^{-1}[0,\bar{q})]$. Note that, because $p \in [-\icost'(\thmin_F),\bar{q}-\icost'(\thmax_F)]$, one has $\thmin_Q \leq \thmin_F$ and $\thmax_F \leq \thmax_Q$, and so $Q=\tilde{\DMprice}(\theta) +\icost'(\theta)=\tilde{Q}(\theta)$ for all $\theta \in \supp \ F$.

We now find an $F$-price that shows that $Q$ is an $F$-CC allocation and that certifies $F$ as optimal for $Q$. For $\theta \in [\thmin_Q,\thmax_Q]$: 

$$V_Q(\theta) = \int^\theta_{\thmin} Q(x) dx = \int^{\theta}_{\thmin_Q} \left(\tilde{\DMprice} (x)+\icost'(x)\right) dx = \tilde{\DMPrice} (\theta) - \tilde{\DMPrice}(\thmin_Q) + \icost(\theta) - \icost(\thmin_Q).$$  Then, letting $\eta = -\tilde{\DMPrice}(\thmin_Q) - \icost(\thmin_Q),$ every $\theta \in [\thmin_Q,\thmax_Q]$ has

\[
V_Q(\theta) - \icost(\theta) = \tilde{\DMPrice} (\theta) + \eta.
\]

We show now that $\DMPrice := \tilde{\DMPrice} + \eta$ is an $F$-price and, therefore, $Q$ is $F$-CC. Because $\DMPrice$ differs from $\tilde{\DMPrice}$ only by a constant, it inherits convexity and Lipschitz continuity. It must also be affine for any $F$-pooling intervals. The condition above shows that $\DMPrice = V_Q - \icost$ for $\theta \in [\thmin_Q,\thmax_Q]$. All that is left is to show $Q$ is $F$-CC  with $F$-price $\DMPrice$ is that $\DMPrice \geq V_Q - \icost$ for all $\theta$. We show this fact next.

For $\theta \leq \thmin_Q$, we have

$$ 0 \geq \int^{\thmin_Q}_{\theta} \left( \tilde{\DMprice} (x) + \icost'(x) \right)dx = \tilde{\DMPrice}(\thmin_Q)-\tilde{\DMPrice}(\theta)+\icost(\thmin_Q)-\icost(\theta), $$
where the first inequality holds because, by definition of $\thmin_Q$, $\tilde{\DMprice} (x) + \icost'(x)\leq0$ for $x \leq \thmin_Q$. Reorganizing that expression, we obtain $V_Q (\theta) - \icost(\theta) = -\icost(\theta) \leq \tilde{\DMPrice}(\theta) + \eta$. 

For $\theta \geq \thmax_Q$,
\begin{equation*}
    \begin{split}
        \idu_Q(\theta) - \icost(\theta) = \idu_Q (\thmax_Q) + \bar{q} (\theta - \thmax_Q) - \icost(\theta) = \tilde{\DMPrice}(\thmax_Q) + \eta + \icost(\thmax_Q) + \bar{q} (\theta - \thmax_Q) - \icost(\theta)\\  = \tilde{\DMPrice}(\thmax_Q) + \int^\theta_{\thmax_Q}\left(\bar{q} - \icost'(x) \right) dx + \eta \leq \tilde{\DMPrice} (\theta) + \eta,
    \end{split}
\end{equation*}
where the first equality follows from envelope and the definition of $\thmax_Q$, the second equality from $\idu_Q = \tilde{\DMPrice} + \eta$ at $\thmax_Q$, and the inequality follows from $\tilde{\DMPrice} (x) + \icost'(x)\geq\bar{q}$ for $x \geq \thmax_Q$, by definition of $\thmax_Q$. We then conclude $\DMPrice = \tilde{\DMPrice} + \eta$ is an $F$-price certifying optimality of $F$-CC allocation $Q$. 

What is left to prove is $\idu_Q \leq \idu_{\tilde{Q}}$ for $\theta \in \supp F$. Because, in the support of $F$, $\tilde{\DMPrice}+\eta = \DMPrice = \idu_Q + \icost$, and $\tilde{\DMPrice} = \idu_{\tilde{Q}} + \icost$, it suffices to show $\eta \leq 0$. But this follows from:

\[
0=\idu_Q(\thmin_Q)=\DMPrice(\thmin_Q)+\icost(\thmin_Q)=\tilde{\DMPrice}(\thmin_Q)+\icost(\thmin_Q)+\eta \geq \idu_{\tilde{Q}}(\thmin_Q)+\eta,
\]
where the inequality follows from $\tilde{\DMPrice}$ certifying optimality of $\tilde{Q}$ as in Lemma~\ref{lem:dwm}.
It follows that $-\eta \geq V_{\tilde{Q}}(\thmin_Q) \geq 0$---that is, $\eta \leq 0$, as required.
\end{proof}

\subsection{Proof of Theorem~\ref{thm: Inefficiency}}
The ultimate goal of this section is to prove Theorem~\ref{thm: Inefficiency}. En route, we prove several auxiliary results about the programs~\eqref{eq: optimal allocation problem} and \eqref{eq: optimal info problem}.

\subsubsection{Allocation Pertubations}
We first consider the program~\eqref{eq: optimal allocation problem}. We begin with two lemmas. The first lemma notes the set of allocations that are $F$-IC is convex. The second lemma uses this convexity to derive a necessary first order condition for an allocation to be part of a seller optimal allocation. 

\begin{lemma}
Suppose $Q$ and $\tilde{Q}$ are both $F$-IC. Then, $\left(1-\beta\right)Q+\beta\tilde{Q}$
is also $F$-IC for all $\beta\in\left[0,1\right]$. 
\end{lemma}
\begin{proof}
Note that for any two allocations $Q$, $\tilde{Q}$, and any $\beta\in\left[0,1\right],$
\begin{align*}
\idu_{\left(1-\beta\right)Q+\beta\tilde{Q}}(\theta) & =\int_{\thmin}^{\theta}\left(\left(1-\beta\right)Q\left(\tilde{\theta}\right)+\beta\tilde{Q}\left(\tilde{\theta}\right)\right)\dd\tilde{\theta}\\
 & =\left(1-\beta\right)\int_{\thmin}^{\theta}Q\left(\tilde{\theta}\right)\dd\tilde{\theta}+\beta\int_{\thmin}^{\theta}\tilde{Q}\left(\tilde{\theta}\right)\dd\tilde{\theta}=\left(1-\beta\right)\idu_{Q}(\theta)+\beta\idu_{\tilde{Q}}(\theta).
\end{align*}
Therefore, if both $Q$, $\tilde{Q}$ are $F$-IC, one obtains the
following inequality for all $\tilde{F}$:
\begin{align*}
\int\left(\idu_{\left(1-\beta\right)Q+\beta\tilde{Q}}-\icost\right)(\theta)F\left(\dd \theta\right) & =\left(1-\beta\right)\int\left(\idu_{Q}-\icost\right)(\theta)F\left(\dd \theta\right)+\beta\int\left(\idu_{\tilde{Q}}-\icost\right)(\theta)F\left(\dd \theta\right)\\
 & \geq\left(1-\beta\right)\int\left(\idu_{Q}-\icost\right)(\theta)\dd\tilde{F}(\theta)+\beta\int\left(\idu_{\tilde{Q}}-\icost\right)(\theta)\dd\tilde{F}(\theta)
 \\
 & =\int\left(\idu_{\left(1-\beta\right)Q+\beta\tilde{Q}}-\icost\right)(\theta)\dd\tilde{F}(\theta),
\end{align*}
meaning $\left(1-\beta\right)Q+\beta\tilde{Q}$ is also $F$-IC. 
\end{proof}
Next, we obtain a necessary first-order condition for the seller's optimal
outcome by perturbing the allocation while keeping the buyer's information
fixed.

\begin{lemma}
\label{AllocationPerturbation}Let $\left(Q^{*},F^{*}\right)$ be
seller optimal. Suppose $Q$ also incentivizes $F^{*}$. Then,
\[
\int\left[\left(\theta-\qcost'\left(Q^{*}(\theta)\right)\right)\left(Q-Q^{*}\right)(\theta)-\left(\idu_{Q}-\idu_{Q^{*}}\right)(\theta)\right]F^{*}(\dd \theta)\leq0.
\]
\end{lemma}
\begin{proof}
Suppose $\left(Q^{*},F^{*}\right)$ is seller optimal, and let
$Q$ be any other $F^*$-IC allocation. Defining the allocation $Q_{\varepsilon}:=Q^{*}+\epsilon\left(Q-Q^{*}\right)$
for every $\epsilon\in\left(0,1\right)$, it follows from the previous
lemma that $Q_{\epsilon}$ is also $F^*$-IC. Therefore, it must be
that $\left(Q_{\epsilon},F^{*}\right)$ is weakly worse for the
seller than $\left(Q^{*},F^{*}\right)$ . In other words, we must
have 
\[
\int\left(\pi_{Q_{\epsilon}}(\theta)-\pi_{Q}(\theta)\right) F^{*}(\dd \theta) \leq 0
\]
for all $\epsilon$. Dividing this inequality by $\epsilon>0$, and
taking the limit as $\epsilon\searrow0$, gives 
\begin{align*}
0 
& \geq
\frac{1}{\epsilon}\int\left(\pi_{Q_{\epsilon}}(\theta)-\pi_{Q}(\theta)\right)F^{*}(\dd\theta)
\\ & =\int\theta\left(Q-Q^{*}\right)(\theta)-\left(\idu_{Q}-\idu_{Q^{*}}\right)(\theta)F^{*}(\dd\theta)
\\ & \quad \quad -\int\frac{1}{\epsilon}\left(\kappa\left(Q^{*}(\theta)+\epsilon\left(Q-Q^{*}\right)(\theta)\right)-\kappa\left(Q^{*}(\theta)\right)\right)F^{*}(\dd\theta)
\\ & \rightarrow \int\theta\left(Q-Q^{*}\right)(\theta)-\left(\idu_{Q}-\idu_{Q^{*}}\right)(\theta)F^{*}(\dd\theta)
\\ & \quad \quad -\int\kappa'\left(Q^{*}(\theta)\right)\left(Q-Q^{*}\right)(\theta)F^{*}(\dd\theta),
\end{align*}
where convergence follows from Beppo Levi's Theorem (e.g., \citet{aliprantis2006infinite}
Theorem 11.18).\footnote{Because $\qcost$ is convex, the function $\epsilon\mapsto\frac{1}{\epsilon}\left(\qcost\left(q+\epsilon\left(\tilde{q}-q\right)\right)-\qcost\left(q\right)\right)$
is decreasing in $\epsilon$ for all $\tilde{q}$ and $q$.} The lemma follows.
\end{proof}

We now work towards the main result of this section. This result establishes conditions under-which the following inequalities hold:
\begin{align}
    \int_{\theta\geq\theta^{*}}\qcost'\left(Q(\theta)\right)F\left(\dd\theta\right) & \geq  \theta^{*}(1-F_{-}(\theta^*)),  \label{eq: avg MC above threshold type} \\
    \int_{\theta\geq\theta^{*}}\qcost'\left(Q(\theta)\right)F\left(\dd\theta\right) & \leq  \theta^{*}(1-F_{-}(\theta^*)).\label{eq: avg MC below threshold type}
\end{align}
These inequalities compare the average marginal cost of the allocation provided to every $\theta$ above $\theta^*$ to $\theta^*$. \cite{mussa1978monopoly} provide conditions under which $\theta^*$ is larger or smaller than this average marginal production costs when information is exogenous. 

Before discussing these inequalities for the endogenous information case, we first show that the following inequality implies \eqref{eq: avg MC below threshold type}:
\begin{equation}\label{eq: avg MC below threshold type strong}
    \int_{\theta>\theta^{*}}\qcost'\circ Q(\theta)F\left(\dd \theta\right)  \leq\left(1-F\left(\theta^{*}\right)\right)\theta^{*}.
\end{equation}
This inequality is the same as \eqref{eq: avg MC below threshold type}, except that the integral on the left hand side excludes $\theta^*$, and the right hand side has $1-F(\theta^*)$ instead of $1-F_{-}(\theta^*)$. 
\begin{lemma}
\label{lem: strong FOC implies weak FOC}If \eqref{eq: avg MC below threshold type strong} holds for $\theta^* \in [\thmin_F,\thmax_F)$, then \eqref{eq: avg MC below threshold type} also holds
\end{lemma}
\begin{proof}
We first argue \eqref{eq: avg MC below threshold type strong} implies $\kappa'\circ Q\left(\theta^{*}\right)<\theta^{*}$. To do so, assume (\ref{eq: avg MC below threshold type strong}) holds, and suppose
$\kappa'\circ Q\left(\theta^{*}\right)\geq\theta^{*}$ for a contradiction.
Because $Q$ is $F$-CC, the allocation $Q$ is strictly increasing on $\left[\underline{\theta}_{F},\bar{\theta}_{F}\right]\supseteq\left[\theta^{*},\bar{\theta}_{F}\right]$,
and so $\kappa'\circ Q(\theta)>\qcost'\circ Q\left(\theta^{*}\right)$
for all $\theta>\theta^{*}$, because $\qcost'$ is strictly increasing.
Therefore, 
\begin{equation*}
\theta^{*}<\int\qcost'\circ Q(\theta) F\left(\dd\theta|\theta\in(\theta^{*},\bar{\theta}_{F}]\right)=\int\qcost'\circ Q(\theta) F\left(\dd\theta|\theta>\theta^{*}\right)=\frac{\int_{\theta>\theta^{*}}\qcost'\circ Q(\theta)F\left(\dd \theta\right)}{1-F\left(\theta^{*}\right)},
\end{equation*}
contradicting \eqref{eq: avg MC below threshold type strong}. Thus, we have shown that $\qcost'\circ Q(\theta^*) < \theta^*$. Using this inequality, we now show that \eqref{eq: avg MC below threshold type strong} implies \eqref{eq: avg MC below threshold type}. Specifically, \eqref{eq: avg MC below threshold type} follows from the inequality chain
\begin{align*}
\int_{\theta\geq\theta^{*}}\qcost'\circ Q(\theta)F\left(\dd \theta\right) & =\int_{\theta>\theta^{*}}\qcost'\circ Q(\theta)F\left(\dd \theta\right)+\left(F\left(\theta^{*}\right)-F_{-}\left(\theta^{*}\right)\right)\kappa'\circ Q\left(\theta^{*}\right)\\
 & \leq\left(1-F\left(\theta^{*}\right)\right)\theta^{*}+\left(F\left(\theta^{*}\right)-F_{-}\left(\theta^{*}\right)\right)\theta^{*}=\left(1-F_{-}\left(\theta^{*}\right)\right)\theta^{*}.
\end{align*}
The proof is now complete.
\end{proof}

We now state Lemma~\ref{lem: allocation perturbations summary}, which is the main result of this section. 

\begin{samepage}
\begin{lemma}\label{lem: allocation perturbations summary}
    Suppose $(Q,F)$ is seller optimal, and that $Q$ is $F$-CC. Fix any $\hat\theta \in \Theta$, and let 
    \[
    \theta^*= \max\{\hat\theta,\thmin_{Q}\}.
    \]
    Then:
    \begin{enumerate}[(i)]
    \item\label{lem: allocation perturbations summary - high mc} If $I_{F}(\hat\theta) = 0$ and $Q(\thmax_F)<\bar q$, the inequality \eqref{eq: avg MC above threshold type} holds for $\theta^*$.     
    \item\label{lem: allocation perturbations summary - low mc and min-theta} If $\hat\theta = \thmin$ and $Q(\thmin_{F}) > 0$, the inequality \eqref{eq: avg MC below threshold type} holds for $\theta^*$.
    \item\label{lem: allocation perturbations summary - low mc and strictly increasing} If $\DMprice_{Q}$ strictly increases at $\hat\theta$ and $\hat\theta > \thmin_{Q}$, the  inequality \eqref{eq: avg MC below threshold type} holds for $\theta^*$.
    \end{enumerate}
\end{lemma}
\end{samepage}

To understand our proof, revisiting \cite{mussa1978monopoly} is helpful. To show \eqref{eq: avg MC above threshold type}, \cite{mussa1978monopoly} consider the change in the seller's profit due to a slight increase in the quality given to all types weakly above $\theta^*$. If one starts from an optimal allocation, the change in the seller's revenues---represented by equation \eqref{eq: avg MC above threshold type}'s right hand side---must be below the change in the seller's production costs, which are given by the left hand side of \eqref{eq: avg MC above threshold type}. The opposite inequality \eqref{eq: avg MC below threshold type} is derived in a similar fashion by noting the seller cannot benefit from slightly \emph{reducing} the quality given to all types above $\theta^*$---provided that the allocation $Q$ is strictly increasing at $\theta^*$. The reason for this caveat is that, if $Q$ is constant around $\theta^*$, reducing the quality given to $\theta^*$ would result in non-monotone allocation, which would violate the buyer's incentive constraint.

It turns out that, by moving $\DMprice$, one can apply \citeapos{mussa1978monopoly} perturbation arguments when reasoning about the solution to \eqref{eq: optimal allocation problem}, subject to two caveats. First, one needs to take care to perturb $\DMprice$ in a way that results in a new $F$-marginal price. In particular, one cannot increase $\DMprice$ at $F$-pooling points, and one must make sure $Q_{\DMprice}$ remains in $[0,\bar q]$ over $F$'s support. Second, shifting $\DMprice$ only affects the induced allocation $Q_{\DMprice}$ for types at which $\DMprice + \icost'$ is in the interval $[0,\bar q]$. Thus, a marginal increase in $\DMprice$ for all types above $\theta$ only changes the allocation for types above $\thmin_{Q_{\DMprice}}$. Taking these caveats into account and applying the ideas of \cite{mussa1978monopoly} delivers the result. 

\begin{proof}[\textbf{Proof of Lemma~\ref{lem: allocation perturbations summary}-\eqref{lem: allocation perturbations summary - high mc}}]

Suppose $(Q,F)$ is seller optimal, and that $Q(\thmax_F)<\bar{q}$. We want to show that the following inequality holds for every $\theta^*$ such that $I_{F}(\theta^*)=0$:
    \begin{equation}\label{eq: avg_mc geq threshold}
    \int_{\theta \geq \theta^*} \qcost'\circ Q(\theta) F(\dd \theta) \geq (1-F_{-}(\theta^*))(\theta^*\vee\thmin_{Q}).
    \end{equation}
    Let $\DMprice$ be the $F$-marginal price associated with the allocation $Q$. For any $\epsilon>0$ such that $\DMprice(\thmax_F)+\epsilon<\bar{q}$, define
    \[
    \DMprice_{\epsilon}(\theta) = \begin{cases}
        \DMprice(\theta) & \text{if }\theta < \hat{\theta}, \\
        \DMprice(\theta)+\epsilon & \text{otherwise.}
    \end{cases}
    \]
    Because $I_{F}(\hat\theta)=0$ and $\DMprice$ is an $F$-marginal price, $\DMprice_{\epsilon}$ is also an $F$-marginal price. Note that $Q_{\DMprice^\epsilon}\geq Q$ because $\DMprice_\epsilon \geq \DMprice$, meaning that $\thmin_{Q_{\DMprice^{\epsilon}}} \leq \thmin_{Q} \leq \thmin_{F}$. Therefore, every $\theta \geq \thmin_{F}$ has $Q_{\DMprice^\epsilon}(\theta) = Q(\theta) + \epsilon\mathbf{1}_{\theta \geq \hat{\theta}}$. Below we prove that 
    \begin{equation}\label{eq: 09-02-2023a}
    \lim_{\epsilon\rightarrow 0} \frac{1}{\epsilon}\left[\idu_{Q^{\DMprice^\epsilon}}(\theta) - \idu_{Q}(\theta)\right] = \left[ \theta - (\hat\theta\vee\thmin_{Q})\right]_{+}.
    \end{equation}
    Using this convergence, we can appeal to Lemma~\ref{AllocationPerturbation} to get
    \[
    \begin{split}
    0 
    & \geq \frac{1}{\epsilon}\left\{ 
    \int \left[ (\theta - \qcost'\circ Q(\theta))(Q_{\DMprice^{\epsilon}}(\theta) - Q(\theta)) - (\idu_{Q_{\DMprice^\epsilon}}(\theta) - \idu_{Q}(\theta))\right] F(\dd\theta) 
    \right\}
    \\ & = \frac{1}{\epsilon}\left\{ 
    \int \left[ (\theta - \qcost'\circ Q(\theta))\epsilon\mathbf{1}_{\theta \geq \hat\theta} - (\idu_{Q_{\DMprice^\epsilon}}(\theta) - \idu_{Q}(\theta))\right] F(\dd\theta)\right
    \}
    \\ & = \int_{\theta \geq \hat{\theta}} (\theta - \qcost'\circ Q(\theta)) F(\dd \theta) - \frac{1}{\epsilon}\int (\idu_{Q_{\DMprice^\epsilon}}(\theta) - \idu_{Q}(\theta)) F(\dd\theta) 
    \\ & \rightarrow \int_{\theta \geq \hat{\theta}} (\theta - \qcost'\circ Q(\theta)) F(\dd \theta) 
    -  \int \left[ \theta - (\hat{\theta}\vee\thmin_{Q})\right]_{+} F(\dd \theta) 
    \\ & = \int_{\theta \geq \hat\theta} \left[(\hat{\theta}\vee\thmin_{Q}) - \qcost'\circ Q(\theta)\right] F(\dd \theta),
    \end{split}
    \]
    where the convergence follows from Beppo Levi's Theorem \citep[e.g.,][Theorem 11.18]{aliprantis2006infinite}, and the last equality follows from observing that $\thmin_{Q} \leq \thmin_{F}$ implies that $(\hat{\theta}\vee\thmin_{Q}) > \thmin_{F}$ if and only if $\hat\theta> \thmin_{F}$. Rearranging the above inequality then delivers part~\eqref{lem: allocation perturbations summary - high mc} of the lemma.

    Thus, to complete the proof of this part, arguing that \eqref{eq: 09-02-2023a} holds suffices. We divide our argument into two cases. 
    \begin{itemize}
        \item[Case 1:] Suppose $\hat{\theta} \geq \thmin_{Q}$. Direct computation reveals 
        \[
        \idu_{Q^{\DMprice^\epsilon}}(\theta) - \idu_{Q_{\DMprice}}(\theta) 
        = \int_{\hat\theta}^{\theta \vee \hat\theta} \epsilon \dd \tilde\theta 
        = \epsilon [\theta \vee \hat{\theta} - \hat{\theta}] 
        = \epsilon [\theta - \hat{\theta}]_{+}
        = \epsilon\left[\theta - \hat{\theta}\vee\thmin_{\DMprice}\right]_{+}.
        \]
       
        The equality \eqref{eq: 09-02-2023a} immediately follows. 

        \item[Case 2:] 
        Suppose $\hat{\theta} < \thmin_Q$. First, fix any  $\theta< \thmin_Q$, and note that there exists $\epsilon_\theta$ such that $\thmin_{Q_{\DMprice^\epsilon}} \in [\theta,\thmin_Q]$ for all $\epsilon<\epsilon_\theta$. Therefore, $Q_{\DMprice^\epsilon} (\theta) = Q(\theta) = 0$, and $\idu_{Q_{\DMprice^\epsilon}}(\theta) - \idu_{Q}(\theta) = 0 = [\theta - \hat{\theta}\vee\thmin_{Q}]_{+}$ for all small enough $\epsilon$, as required. We have then proved the result for $\theta < \thmin_Q$.

        We now consider $\theta \geq \thmin_Q$. Note that for $\theta\in[\thmin_{Q_{\DMprice^\epsilon}},\thmin_Q]$,
        $Q_{\DMprice^\epsilon}(\theta)=p(\theta) + \icost'(\theta)+\epsilon \in [0,\epsilon]$. So, for any $\theta\geq\thmin_Q$, 
        \[
        \begin{split}
        \idu_{Q_{\DMprice^\epsilon}}(\theta) - \idu_{Q}(\theta) &= \int_{\thmin_Q}^\theta \epsilon d\tilde{\theta}+\int_{\thmin_{Q_{\DMprice^\epsilon}}}^{\thmin_Q} Q_{\DMprice^\epsilon}(\theta)d\tilde{\theta}\\
        &=\epsilon[\theta-\thmin_Q] +\int_{\thmin_{Q_{\DMprice^\epsilon}}}^{\thmin_Q} Q_{\DMprice^\epsilon}(\theta)d\tilde{\theta}\\
        \end{split}
        \]
        Therefore,
        \[
        \begin{split}
        \theta-\thmin_Q \leq \frac{1}{\epsilon}\left[\idu_{Q^{\DMprice^\epsilon}}(\theta) - \idu_{Q}(\theta)\right] 
        & =[\theta-\thmin_Q] + \frac{1}{\epsilon} \int_{\thmin_{Q_{\DMprice^\epsilon}}}^{\thmin_Q} Q_{\DMprice^\epsilon}(\theta)d\tilde{\theta}
        \\ & \leq [\theta-\thmin_Q] +  \int_{\thmin_{Q_{\DMprice^\epsilon}}}^{\thmin_Q} d\tilde{\theta}
        \\ & \xrightarrow{\epsilon\rightarrow 0} \theta-\thmin_Q=\left[\theta - \hat{\theta}\vee\thmin_{Q}\right]_{+},
        \end{split}
        \] 
        where the first and second inequalities follow from $Q_{\DMprice^\epsilon} (\theta) \in [0,\epsilon]$ for $\theta \in [\thmin_{Q_{\DMprice^\epsilon}}, \thmin_Q]$, convergence follows from $Q_{\DMprice^{\epsilon}}$ uniformly converging to $Q$ and the continuity of $Q$, and the last equality from the current case's assumption that $\hat\theta < \thmin_{Q}$.

\end{itemize} 
\end{proof}

\begin{proof}[\textbf{Proof of Lemma~\ref{lem: allocation perturbations summary}-\eqref{lem: allocation perturbations summary - low mc and min-theta}}]
For any $\varepsilon\in\left(0,Q\left(\underline{\theta}_{F}\right)\right)$,
let $\theta_{\varepsilon}=\inf\left\{ \theta:Q(\theta)\geq\varepsilon\right\} ,$
and define $\DMprice_{\varepsilon}\left(\cdot\right):=\DMprice\left(\cdot\right)-\varepsilon$.
Observe $\DMprice_{\varepsilon}$ is an $F$-marginal price because $\DMprice$ is an $F$-marginal price, and $\epsilon<Q(\thmin_F) = \DMprice(\thmin_F) + \icost'(\thmin_F)$, meaning $\DMprice_{\varepsilon}(\thmin_{F}) > -\icost'(\thmin_F)$ (which is sufficient for $\DMprice_{\varepsilon}\geq -\icost'(\thmin_F)$ to hold). 

Let $Q_{\varepsilon}:=Q_{\DMprice_{\varepsilon}}$ be the $F$-ICC allocation
induced by $\DMprice_{\varepsilon}$.
Obviously, $\theta_{\varepsilon}\leq\underline{\theta}_{F}$, and
\[
Q_{\varepsilon}(\theta)=\DMprice_{Q}(\theta)+c'(\theta)-\varepsilon=Q(\theta)-\epsilon
\]
for all $\theta\in\left[\theta_{\varepsilon},\thmax_{F}\right]$.
Noting that for all $\theta \in [\thmin_{F},\thmax_{F}]$, 
\[
V_{Q_{\varepsilon}}(\theta)-V_{Q}(\theta)=\int_{\theta_{\varepsilon}}^{\theta}-\varepsilon\ddd\tilde{\theta}+\int_{\underline{\theta}}^{\theta_{\varepsilon}}-Q\left(\tilde{\theta}\right)\dd\tilde{\theta}=\varepsilon\left(\theta_{\varepsilon}-\theta\right)-V_{Q}\left(\theta_{\varepsilon}\right),
\]
and so by Lemma \ref{AllocationPerturbation}, we have 
\begin{align*}
0 & \geq\int\left[\left(\theta-\qcost'\circ Q(\theta)\right)\left(-\varepsilon\right)-\varepsilon\left(\theta_{\varepsilon}-\theta\right)+V_{Q}\left(\theta_{\varepsilon}\right)\right]F\left(\dd\theta\right)\\
 & =\int\left[\left(\qcost'\circ Q(\theta)-\theta_{\varepsilon}\right)\varepsilon+V_{Q}\left(\theta_{\varepsilon}\right)\right]F\left(\dd\theta\right).
\end{align*}
This inequality, however, implies that 
\begin{align*}
\varepsilon\int_{\theta\geq\underline{\theta}_{Q}}\qcost'\circ Q(\theta)F\left(\dd \theta\right)
=\varepsilon\int\qcost'\circ Q(\theta)F\left(\dd \theta\right) 
\leq\int\left[\varepsilon\theta_{\varepsilon}-V_{Q}\left(\theta_{\varepsilon}\right)\right]F(\dd\theta)\leq\varepsilon\theta_{\varepsilon}.
\end{align*}
Dividing both sides by $\epsilon$, taking $\epsilon \searrow 0$, and noting that $\theta_{\epsilon} \searrow \thmin_{Q}$ completes the proof. 
\end{proof}

\paragraph*{\textbf{Proof of Lemma~\ref{lem: allocation perturbations summary}-\eqref{lem: allocation perturbations summary - low mc and strictly increasing}}}
The proof of this part proceeds in 3 cases: 
\begin{enumerate}
    \item $\DMprice_{Q}$ has an upward jump at $\theta^{*}$. 
    \item $\DMprice_{Q}$ strictly increases immediately below $\theta^{*}$.
    \item $\DMprice_{Q}$ strictly increases immediately above $\theta^{*}$.
\end{enumerate}
We begin with the case in which $\DMprice_{Q}$ jumps at $\theta^{*}$. 
\begin{proof}[\textbf{Case 1: $\DMprice_{Q-}\left(\theta^{*}\right)<\DMprice_{Q+}\left(\theta^{*}\right)$.}]
Observe $Q$ being $F$-ICC and $\DMprice_{Q-}\left(\theta^{*}\right)<\DMprice_{Q+}\left(\theta^{*}\right)$
means $I_{F}\left(\theta^{*}\right)=0$, and so $\theta^{*}>\underline{\theta}_{F}$.
For any $\varepsilon\in\left(0,\DMprice_{Q+}\left(\theta^{*}\right)-\DMprice_{Q-}\left(\theta^{*}\right)\right),$
define
\[
\DMprice_{\varepsilon}(\theta)=\begin{cases}
\DMprice_{Q}(\theta) & \text{if }\theta<\theta^{*},\\
\DMprice_{Q}\left(\theta^{*}\right)\wedge\left(\DMprice_{Q+}\left(\theta^{*}\right)-\varepsilon\right) & \text{if }\theta=\theta^{*},\\
\DMprice_{Q}(\theta)-\varepsilon & \text{if }\theta>\theta^{*}.
\end{cases}
\]
It is easy to verify that $\DMprice_{\varepsilon}$ is an $F$-marginal
price.
Let $Q_{\varepsilon}:=Q_{\DMprice_{\varepsilon}}$ be the $F$-ICC mechanism associated with $\DMprice_{\varepsilon}$.
It follows $Q_{\varepsilon}$ is $F$-IC, and thus one can apply Lemma \ref{AllocationPerturbation}
to get the following inequality for every $\varepsilon\in\left(0,\DMprice_{Q+}\left(\theta^{*}\right)-\DMprice_{Q-}\left(\theta^{*}\right)\right),$
\begin{align*}
0 & \geq\int_{\theta>\theta^{*}}\left[\left(\theta-\qcost'\circ Q(\theta)\right)\left(-\varepsilon\right)-\varepsilon\left(\theta^{*}-\theta\right)\right]F\left(\dd\theta\right)\\
 & \,\,\,\,+\left(F\left(\theta^{*}\right)-F_{-}\left(\theta^{*}\right)\right)\left(\theta^{*}-\qcost'\circ Q\left(\theta^{*}\right)\right)\left(\DMprice_{\varepsilon}\left(\theta^{*}\right)-\DMprice_{Q}\left(\theta^{*}\right)\right)\\
 & =\int_{\theta>\theta^{*}}\left(\qcost'\circ Q(\theta)-\theta^{*}\right)\varepsilon F\left(\dd\theta\right)\\
 & \,\,\,\,+\left(F\left(\theta^{*}\right)-F_{-}\left(\theta^{*}\right)\right)\left(\theta^{*}-\qcost'\circ Q\left(\theta^{*}\right)\right)\left(\DMprice_{\varepsilon}\left(\theta^{*}\right)-\DMprice_{Q}\left(\theta^{*}\right)\right).
\end{align*}
Rearranging gives
\begin{align}
\int_{\theta>\theta^{*}}\qcost'\circ Q(\theta)F\left(\dd\theta\right)\leq & \left(1-F\left(\theta^{*}\right)\right)\theta^{*}\label{eq:1043}\\
 & +\left(F\left(\theta^{*}\right)-F_{-}\left(\theta^{*}\right)\right)\left(\theta^{*}-\qcost'\circ Q\left(\theta^{*}\right)\right)\left(\frac{\DMprice_{\varepsilon}\left(\theta^{*}\right)-\DMprice_{Q}\left(\theta^{*}\right)}{\varepsilon}\right).\nonumber 
\end{align}
We now distinguish between two cases. Suppose first $\DMprice_{Q}\left(\theta^{*}\right)<\DMprice_{Q+}\left(\theta^{*}\right)$.
Then for all small enough $\varepsilon>0$, $\DMprice_{\varepsilon}\left(\theta^{*}\right)-\DMprice_{Q}\left(\theta^{*}\right)=0$,
and so equation (\ref{eq:1043}) is equivalent to (\ref{eq: avg MC below threshold type}).

Suppose then $\DMprice_{Q}\left(\theta^{*}\right)=\DMprice_{Q+}\left(\theta^{*}\right)$.
Then $\DMprice_{Q}\left(\theta^{*}\right)-\DMprice_{\varepsilon}\left(\theta^{*}\right)=\varepsilon$.
Substituting into (\ref{eq:1043}) and rearranging gives (\ref{eq: avg MC below threshold type}), as desired.
\end{proof}
\begin{proof}[\textbf{Case 2: $\DMprice_{Q-}\left(\theta^{*}\right)=\DMprice_{Q+}\left(\theta^{*}\right)$
and $\DMprice_{Q-}\left(\theta^{*}\right)>\DMprice_{Q}(\theta)$
for all $\theta<\theta^{*}$:}] 

We begin by arguing
that we can find a sequence $(\theta_{n}) _{n\in\mathbb{N}}$
in $\left[\underline{\theta}_{F},\bar{\theta}_{F}\right]$ such that
$\theta_{n}\nearrow\theta^{*}$, $I_{F}\left(\theta_{n}\right)=0$
for all $n$, and $\DMprice_{Q}\left(\theta_{n}\right)<\DMprice_{Q}\left(\theta_{n+1}\right)$
for all $n$. We then use this sequence to construct a sequence of
allocations that keep $F$ incentive compatible. This allocation
sequence, combined with Lemma \ref{AllocationPerturbation}, delivers
a sequence of first-order conditions whose limit delivers \eqref{eq: avg MC below threshold type}.

Let us find the sequence $(\theta_{n}) _{n\in\mathbb{N}}$.
For every $\delta>0$, $\DMprice_{Q}$ is non-constant on $\left[\theta^{*}-\delta,\theta^{*}\right]$,
because if it were, $\DMprice_{Q-}\left(\theta^{*}\right)=\DMprice_{Q}\left(\theta^{*}-\delta\right)<\DMprice_{Q-}\left(\theta^{*}\right)$.
It follows we can find a sequence $\left\{ \tilde{\theta}_{n}\right\} _{n\in\mathbb{N}}$
in $\left(\underline{\theta}_{F},\theta^{*}\right)$ with $\tilde{\theta}_{n}\nearrow\theta^{*}$
such that $\DMprice_{Q}\left(\tilde{\theta}_{n}\right)<\DMprice_{Q}\left(\tilde{\theta}_{n+1}\right)$
for all $n$. It follows $\DMprice_{Q}$ is non-constant on $\left[\tilde{\theta}_{m},\tilde{\theta}_{n}\right]$
for any $m<n$, and so every $m<n$ admits some $\theta_{m,n}\in\left[\tilde{\theta}_{m},\tilde{\theta}_{n}\right]$
for which $I_{F}\left(\theta_{m,n}\right)=0$. Choosing $\theta_{n}:=\theta_{2n,2n+1}$,
we have $\theta_{n}\nearrow\theta^{*}$, and 
\[
\DMprice_{Q}\left(\theta_{n}\right)=\DMprice_{Q}\left(\theta_{2n,2n+1}\right)\leq\DMprice_{Q}\left(\tilde{\theta}_{2n+1}\right)<\DMprice_{Q}\left(\tilde{\theta}_{2n+2}\right)\leq\DMprice_{Q}\left(\theta_{2\left(n+1\right),2\left(n+1\right)+1}\right)=\DMprice_{Q}\left(\theta_{n+1}\right),
\]
meaning $(\theta_{n}) _{n\in\mathbb{N}}$ is as desired.

We now construct an $F$-CC allocation for every $\theta_{n}$ in
the above sequence. For this purpose, let $\delta_{n}=\DMprice_{Q}\left(\theta^{*}\right)-\DMprice_{Q}\left(\theta_{n}\right)>0$,
\[
\DMprice_{n}(\theta)=\begin{cases}
\DMprice_{Q}(\theta) & \text{if }\theta\leq\theta_{n}\\
\DMprice_{Q}(\theta)-\delta_{n} & \text{if }\theta\geq\theta^{*}\\
\DMprice_{Q}\left(\theta_{n}\right) & \text{if }\theta\in\left[\theta_{n},\theta^{*}\right].
\end{cases}
\]
Since $\DMprice_{Q}$ is an $F$-marginal price, and because $I_F(\theta^*) = 0$ (because $\DMprice_{Q}$ is not constant around $\theta^*$), the function $\DMprice_{n}$ is an $F$-marginal price for every $n$, and so the allocation  $Q_{n}:=Q_{\DMprice_n}$ is an $F$-CC allocation.  Thus, $Q_{n}$
is $F$-CC.

Next, we apply Lemma \ref{AllocationPerturbation} to get
a first-order condition indexed by $n$. For this purpose, observe that for all $\theta \in [\thmin_{F},\thmax_{F}]$,
\[
\idu_{Q_{n}}(\theta)-\idu_{Q}(\theta)=\int_{\theta_{n}\wedge\theta}^{\theta^{*}\wedge\theta}\left(\DMprice_{Q}\left(\theta_{n}\right)-\DMprice_{Q}\left(\tilde{\theta}\right)\right)\dd\tilde{\theta}-\delta_{n}\left(\theta-\theta^{*}\wedge\theta\right).
\]
Therefore, Lemma \ref{AllocationPerturbation} delivers the following
inequality for all $n$,
\begin{align*}
0\geq & \int_{\theta\geq\theta^{*}}\left(\theta-\qcost'\circ Q(\theta)\right)\left(-\delta_{n}\right)F\left(\dd \theta\right)+\int_{\theta\in[\theta_{n},\theta^{*})}\left(\theta-\qcost'\circ Q(\theta)\right)\left(\DMprice_{Q}\left(\theta_{n}\right)-\DMprice_{Q}(\theta)\right)F\left(\dd \theta\right)\\
 & -\int_{\theta\geq\theta_{n}}\int_{\theta_{n}}^{\theta^{*}\wedge\theta}\left(\DMprice_{Q}\left(\theta_{n}\right)-\DMprice_{Q}(\tilde{\theta})\right)\dd\tilde{\theta}F\left(\dd \theta\right)-\int_{\theta\geq\theta^{*}}-\delta_{n}\left(\theta-\theta^{*}\right)F\left(\dd \theta\right)\\
= & \int_{\theta\geq\theta^{*}}\left(\qcost'\circ Q(\theta)-\theta^{*}\right)\delta_{n}F\left(\dd \theta\right)+\int_{\theta\in[\theta_{n},\theta^{*})}\left(\theta-\qcost'\circ Q(\theta)\right)\left(\DMprice_{Q}\left(\theta_{n}\right)-\DMprice_{Q}(\theta)\right)F\left(\dd \theta\right)\\
 & -\int_{\theta\geq\theta_{n}}\int_{\theta_{n}}^{\theta^{*}\wedge\theta}\left(\DMprice_{Q}\left(\theta_{n}\right)-\DMprice_{Q}(\tilde{\theta})\right)\dd\tilde{\theta}F(\dd \theta).
\end{align*}
Rearranging and noting that $\DMprice_{Q}\left(\theta_{n}\right)\le\DMprice_{Q}(\theta)$
for all $\theta\geq\theta_{n}$ delivers
\begin{align}
\int_{\theta\geq\theta^{*}}\qcost'\circ Q(\theta)F\left(\dd \theta\right)
\leq & \int_{\theta\geq\theta^{*}}\theta^{*}F\left(\dd \theta\right)-\int_{\theta\in[\theta_{n},\theta^{*})}\left(\theta-\qcost'\circ Q(\theta)\right)\frac{1}{\delta_{n}}\left(\DMprice_{Q}\left(\theta_{n}\right)-\DMprice_{Q}(\theta)\right)F\left(\dd \theta\right)\nonumber 
\\ & +\int_{\theta\geq\theta_{n}}\int_{\theta_{n}}^{\theta^{*}\wedge\theta}\frac{1}{\delta_{n}}\left(\DMprice_{Q}\left(\theta_{n}\right)-\DMprice_{Q}(\tilde{\theta})\right)\dd\tilde{\theta}F\left(\dd \theta\right)\nonumber 
\\ \leq & \int_{\theta\geq\theta^{*}}\theta^{*}F\left(\dd \theta\right)-\int_{\theta\in[\theta_{n},\theta^{*})}\left(\theta-\qcost'\circ Q(\theta)\right)\frac{1}{\delta_{n}}\left(\DMprice_{Q}\left(\theta_{n}\right)-\DMprice_{Q}(\theta)\right)F\left(\dd \theta\right)\nonumber 
\\ \leq & \int_{\theta\geq\theta^{*}}\theta^{*}F\left(\dd \theta\right)+\int_{\theta\in[\theta_{n},\theta^{*})}\left|\theta-\qcost'\circ Q(\theta)\right|\left|\frac{1}{\delta_{n}}\left(\DMprice_{Q}\left(\theta_{n}\right)-\DMprice_{Q}(\theta)\right)\right|F\left(\dd \theta\right).\label{eq:835am}
\end{align}
We now show taking the limit of equation (\ref{eq:835am}) as $n\rightarrow\infty$
delivers equation (\ref{eq: avg MC below threshold type}). To do so, observe
$\left|\DMprice_{Q}\left(\theta_{n}\right)-\DMprice_{Q}(\theta)\right|\leq\delta_{n}$
for all $\theta\in[\theta_{n},\theta^{*})$, and that $\theta-\qcost'\circ Q(\theta)\leq\bar{\theta}_{F}+\qcost'\circ Q\left(\bar{\theta}_{F}\right)$
for all $\theta\in[\theta_{n},\theta^{*})$. Therefore, an $M$ exists
such that $\left|\theta-\qcost'\circ Q(\theta)\right|\left|\frac{1}{\delta_{n}}\left(\DMprice_{Q}\left(\theta_{n}\right)-\DMprice_{Q}(\theta)\right)\right|\leq M$
for all $n$. Substituting back into (\ref{eq:835am}) and taking
limit with $n$ delivers
\[
\int_{\theta\geq\theta^{*}}\qcost'\circ Q(\theta)F\left(\dd \theta\right)\leq\int_{\theta\geq\theta^{*}}\theta^{*}F\left(\dd \theta\right)+M\left(F_{-}\left(\theta^{*}\right)-F\left(\theta_{n}\right)\right)\rightarrow\int_{\theta\geq\theta^{*}}\theta^{*}F\left(\dd \theta\right).
\]
Hence (\ref{eq: avg MC below threshold type}) holds at $\theta^{*}$.
\end{proof}
\begin{proof}[\textbf{Case 3:  $\DMprice_{Q-}\left(\theta^{*}\right)=\DMprice_{Q+}\left(\theta^{*}\right)$
and $\DMprice_{Q+}\left(\theta^{*}\right)<\DMprice_{Q}(\theta)$
holds for all $\theta>\theta^{*}$.}]
We begin by finding a sequence $(\theta_{n}) _{n\in\mathbb{N}}$
in $\left[\underline{\theta}_{F},\bar{\theta}_{F}\right]$ such that
$\theta_{n}\searrow\theta^{*}$, $I_{F}\left(\theta_{n}\right)=0$
for all $n$, and $\DMprice_{Q}\left(\theta_{n}\right)>\DMprice_{Q}\left(\theta_{n+1}\right)$
for all $n$. We then construct a corresponding sequence of allocations
that keep $F$ incentive compatible for the buyer. This allocation
sequence, combined with Lemma \ref{AllocationPerturbation}, delivers
a sequence of first-order conditions whose limit delivers \eqref{eq: avg MC below threshold type}.

Let us find the sequence $(\theta_{n}) _{n\in\mathbb{N}}$.
Observe that for every $\delta>0$, $\DMprice_{Q}$ is non-constant
on $\left[\theta^{*},\theta^{*}+\delta\right]$, because if it were constant,
$\DMprice_{Q+}\left(\theta^{*}\right)=\DMprice_{Q}\left(\theta^{*}+\delta\right)>\DMprice_{Q+}\left(\theta^{*}\right)$.
It follows we can find a sequence $(\tilde{\theta}_{n})_{n\in\mathbb{N}}$
in $\left(\theta^{*},\bar{\theta}\right)$ with $\tilde{\theta}_{n}\searrow\theta^{*}$
such that $\DMprice_{Q}(\tilde{\theta}_{n})>\DMprice_{Q}(\tilde{\theta}_{n+1})$
for all $n$. To define $(\theta_{n})_{n\in\mathbb{N}}$,
observe that $\DMprice_{Q}$ is non-constant on $[\tilde{\theta}_{m},\tilde{\theta}_{n}]$
for any $m<n$, and so every $m<n$ admits some $\theta_{m,n}\in\left[\tilde{\theta}_{m},\tilde{\theta}_{n}\right]$
for which $I_{F}\left(\theta_{m,n}\right)=0$. Choosing $\theta_{n}:=\theta_{2n,2n+1}$,
we have $\theta_{n}\searrow\theta^{*}$, and 
\[
\DMprice_{Q}\left(\theta_{n}\right)=\DMprice_{Q}\left(\theta_{2n,2n+1}\right)\geq\DMprice_{Q}\left(\tilde{\theta}_{2n+1}\right)>\DMprice_{Q}\left(\tilde{\theta}_{2n+2}\right)\geq\DMprice_{Q}\left(\theta_{2\left(n+1\right),2\left(n+1\right)+1}\right)=\DMprice_{Q}\left(\theta_{n+1}\right).
\]
Finally, observe $I_{F}(\theta)=0$ and $\theta<\bar{\theta}$
implies $\theta<\bar{\theta}_{F}$. Hence, because $(\theta_{n})_{n \in \mathbb{N}}$ is
a strictly decreasing sequence, it has at most one element weakly above $\bar{\theta}_{F}$,
and so it is without loss to take $(\theta_{n}) _{n\in\mathbb{N}}$
to be strictly below $\bar{\theta}_{F}$, as desired.

We now construct an $F$-CC mechanism for every $\theta_{n}$ in
the above sequence. Let $\delta_{n}:=\DMprice_{Q}\left(\theta_{n}\right)-\DMprice_{Q}\left(\theta^{*}\right)>0$.
Define 
\[
\DMprice_{n}(\theta)=\begin{cases}
\DMprice_{Q}(\theta) & \text{if }\theta\leq\theta^{*}\\
\DMprice_{Q}(\theta)-\delta_{n} & \text{if }\theta\geq\theta_{n}\\
\DMprice_{Q}\left(\theta^{*}\right) & \text{if }\theta\in\left[\theta^{*},\theta_{n}\right],
\end{cases}
\]
Note $I_{F}(\theta^*)=0$ because $\DMprice$ is not constant around $\theta^*$. Using this fact and the fact that $\DMprice$ is an $F$-marginal price, it is straightforward to verify that $\DMprice_n$ is an $F$-marginal price as well for every $n$. We let $Q_n:=Q_{\DMprice_n}$ be the $F$-CC allocation induced by $\DMprice_n$. 

Our next goal is to apply Lemma \ref{AllocationPerturbation} to get
a first-order condition indexed by $n$. For this purpose, observe that for $\theta \in [\thmin_{F},\thmax_{F}]$,
\begin{align*}
\idu_{Q_{n}}(\theta)-\idu_{Q}(\theta) & =\int_{\theta^{*}\wedge\theta}^{\theta_{n}\wedge\theta}\left(\DMprice_{Q}\left(\theta^{*}\right)-\DMprice_{Q}(\tilde{\theta})\right)\dd\tilde{\theta}-\delta_{n}\left(\theta-\theta_{n}\wedge\theta\right).
\end{align*}
Therefore, Lemma \ref{AllocationPerturbation} delivers the following
inequality for all $n$:
\begin{align*}
0\geq & \int_{\theta\geq\theta_{n}}\left(\theta-\qcost'\circ Q(\theta)\right)\left(-\delta_{n}\right)F\left(\dd \theta\right)+\int_{\theta\in[\theta^{*},\theta_{n})}\left(\theta-\qcost'\circ Q(\theta)\right)\left(\DMprice_{Q}\left(\theta^{*}\right)-\DMprice_{Q}(\theta)\right)F\left(\dd \theta\right)\\
 & -\int_{\theta\geq\theta^{*}}\int_{\theta^{*}}^{\theta_{n}\wedge\theta}\left(\DMprice_{Q}\left(\theta^{*}\right)-\DMprice_{Q}(\tilde{\theta})\right)\dd\tilde{\theta}F\left(\dd \theta\right)-\int_{\theta\geq\theta_{n}}-\delta_{n}\left(\theta-\theta_{n}\right)F\left(\dd \theta\right)\\
= & \int_{\theta\geq\theta_{n}}\left(\qcost'\circ Q(\theta)-\theta_{n}\right)\delta_{n}F\left(\dd \theta\right)+\int_{\theta\in[\theta^{*},\theta_{n})}\left(\theta-\qcost'\circ Q(\theta)\right)\left(\DMprice_{Q}\left(\theta^{*}\right)-\DMprice_{Q}(\theta)\right)F\left(\dd \theta\right)\\
 & -\int_{\theta\geq\theta^{*}}\int_{\theta^{*}}^{\theta_{n}\wedge\theta}\left(\DMprice_{Q}\left(\theta^{*}\right)-\DMprice_{Q}(\tilde{\theta})\right)\dd\tilde{\theta}F\left(\dd \theta\right).
\end{align*}
Dividing both sides by $\delta_{n}$ and noting that $p_{Q}\left(\theta^{*}\right)\leq p_{Q}(\theta)$
for all $\theta\geq\theta^{*}$ delivers
\begin{align}
0 & \geq\int_{\theta\geq\theta_{n}}\left(\qcost'\circ Q(\theta)-\theta_{n}\right)F\left(\dd \theta\right)-\int_{\theta\in[\theta^{*},\theta_{n})}\left(\qcost'\circ Q(\theta)-\theta\right)\left(\frac{\DMprice_{Q}\left(\theta^{*}\right)-\DMprice_{Q}(\theta)}{\delta_{n}}\right)F\left(\dd \theta\right)\nonumber \\
 & \geq\int_{\theta\geq\theta_{n}}\left(\qcost'\circ Q(\theta)-\theta_{n}\right)F\left(\dd \theta\right)-\int_{\theta\in[\theta^{*},\theta_{n})}\left|\qcost'\circ Q(\theta)-\theta\right|\left|\frac{\DMprice_{Q}\left(\theta^{*}\right)-\DMprice_{Q}(\theta)}{\delta_{n}}\right|F\left(\dd \theta\right).\label{eq:216pm}
\end{align}
We now show taking the limit of equation (\ref{eq:216pm}) as $n\rightarrow\infty$
delivers equation \eqref{eq: avg MC below threshold type}. To do so, observe
first $\mathbf{1}_{[\theta_{n},\infty)}(\theta)\left(\qcost'\circ Q(\theta)-\theta_{n}\right)$
converges pointwise to $\mathbf{1}_{(\theta^{*},\infty)}(\theta)\left(\qcost'\circ Q(\theta)-\theta^{*}\right)$.
Second, notice $\left|\DMprice_{Q}\left(\theta^{*}\right)-\DMprice_{Q}(\theta)\right|\leq\delta_{n}$
for all $\theta\in[\theta_{n},\theta^{*})$, and that $\theta-\qcost'\circ Q(\theta)\leq\bar{\theta}_{F}+\qcost'\circ Q\left(\bar{\theta}_{F}\right)$
for all $\theta\in[\theta_{n},\theta^{*})$. Therefore, an $M$ exists
such that 
\[\left|\theta-\qcost'\circ Q(\theta)\right|\left|\frac{1}{\delta_{n}}\left(\DMprice_{Q}\left(\theta_{n}\right)-\DMprice_{Q}(\theta)\right)\right|\leq M
\]
for all $n$. Substituting these facts back into (\ref{eq:216pm})
gives
\begin{align*}
0 & \geq\int_{\theta\geq\theta_{n}}\left(\qcost'\circ Q(\theta)-\theta_{n}\right)F\left(\dd \theta\right)-\int_{\theta\in[\theta^{*},\theta_{n})}\left|\qcost'\circ Q(\theta)-\theta\right|\left|\frac{\DMprice_{Q}\left(\theta^{*}\right)-\DMprice_{Q}(\theta)}{\delta_{n}}\right|F\left(\dd \theta\right)\\
 & \geq\int\mathbf{1}_{[\theta_{n},\infty)}(\theta)\left(\qcost'\circ Q(\theta)-\theta_{n}\right)F\left(\dd \theta\right)-M\left(F_{-}\left(\theta_{n}\right)-F\left(\theta^{*}\right)\right)\\
 & \rightarrow\int\mathbf{1}_{(\theta^{*},\infty)}(\theta)\left(\qcost'\circ Q(\theta)-\theta^{*}\right)F\left(\dd \theta\right)=\int_{\theta>\theta^{*}}\qcost'\circ Q(\theta)F\left(\dd\theta\right)-\theta^{*}\left(1-F\left(\theta^{*}\right)\right),
\end{align*}
where convergence follows from right continuity of $F$ and the Lebesgue
dominated convergence theorem. Hence, we've shown \eqref{eq: avg MC below threshold type strong} holds for $\theta^*$. Lemma~\ref{lem: strong FOC implies weak FOC} then implies \eqref{eq: avg MC below threshold type} also holds, as desired.
\end{proof}
We have now completed the proof of Lemma~\ref{lem: allocation perturbations summary}.

\subsubsection{Information Perturbations}

In this section we discuss the consequences of applying a perturbation based approach for analyzing the information design program~\eqref{eq: optimal info problem}. Such perturbations must satisfy two broad restrictions. The first restriction is that the perturbation must result in a signal---that is, a mean-preserving contraction of $F_0$. This restriction is satisfied whenever the perturbation creates a mean-preserving contraction of the original distribution, or when the perturbation creates a (small) mean-preserving spread over a set of $F$-pooling types. The second restriction is that the original allocation must be an $\hat{F}$-CC allocation with respect to the perturbed signal $\hat{F}$. This requirement means one can only alter $F$ in regions where $\DMprice$ is constant. 

Lemma~\ref{lem: ConvCav} outlines the consequence of two perturbation satisfying the above-mentioned restrictions. The lemma's first part identifies situations in which one can conduct mean-preserving contractions in the buyer's signal. Since such contractions cannot be profitable for the seller, they imply $\pi_{Q}$ must satisfy the concave inequality \eqref{lem: ConvCav:CavEq}. The lemma's second part identifies situations where one can spread the buyer's signal in mean-preserving manner. Profit maximization then delivers that $\pi_{Q}$ must satisfy the convex inequality \eqref{lem: ConvCav:ConvEq}.

\begin{lemma}
\label{lem: ConvCav}Let $\left(Q^{*},F^{*}\right)$ be seller optimal, and suppose $Q^{*}$ is $F^{*}$-CC. Suppose $\DMprice_{Q^{*}}$ is constant
over $\left(\theta_{*},\theta^{*}\right)\subseteq\left[\thmin_{Q},\thmax_{Q}\right]$,
and $F(\theta^*) > F_{-}(\theta_*)  $. 
Then,
\begin{enumerate}[(i)]
\item \label{lem: ConvCav:Conv}For every  $\theta_{1},\theta_{2}\in\supp\,F^{*}\left(\cdot|\theta\in\left[\theta_{*},\theta^{*}\right]\right)$ and every $\alpha \in [0,1]$,
\begin{equation}
\pi_{Q^{*}}\left(\alpha\theta_{1}+\left(1-\alpha\right)\theta_{2}\right)\leq\alpha\pi_{Q^{*}}\left(\theta_{1}\right)+\left(1-\alpha\right)\pi_{Q^{*}}\left(\theta_{2}\right)\label{lem: ConvCav:ConvEq}
\end{equation}
\item \label{lem: ConvCav:Cav} If $\theta_1<\theta_2$ are such that $\theta_1$ satisfies either $I_{F^*}(\theta_1)>0$ or $F_0(\theta_1) > F^*(\theta_1)$, $\theta_2$ satisfies either $I_{F^*}(\theta_2)>0$ or $F^*_{-}(\theta_2) > F_{0-}(\theta_2)$, and $I_{F^*}(\theta)>0$ for all $\theta\in\left(\theta_{1},\theta_{2}\right)\subseteq\left[\theta_{*},\theta^{*}\right]$, 
then
\begin{equation}
\pi_{Q^{*}}\left(\alpha\theta_{1}+\left(1-\alpha\right)\theta_{2}\right)\geq\alpha\pi_{Q^{*}}\left(\theta_{1}\right)+\left(1-\alpha\right)\pi_{Q^{*}}\left(\theta_{2}\right)\label{lem: ConvCav:CavEq}
\end{equation}
for all $\alpha\in\left[0,1\right]$ such that $\alpha\theta_{1}+\left(1-\alpha\right)\theta_{2}\in\mathrm{supp}\,F^{*}\left(\cdot|\theta\in\left[\theta_{*},\theta^{*}\right]\right)$.
\end{enumerate}
\end{lemma}

The lemma's first part says that, for appropriately chosen $\theta_1$ and $\theta_2$, the seller cannot benefit from contracting the mass on $\theta_1$ and $\theta_2$ into $\alpha \theta_1 + (1-\alpha)\theta_2$. Similarly, the lemma's second condition says the seller cannot benefit from the spreading mass on $\alpha \theta_1 + (1-\alpha)\theta_2$ across $\theta_1$ and $\theta_2$. For a rough proof sketch, consider the lemma's part \eqref{lem: ConvCav:Conv}, and suppose that $F^{*}$ has atoms at $\theta_{1}$ and $\theta_{2}$. 
As explained above, that $\DMprice_{Q^{*}}$ is constant on $\left[\theta_{*},\theta^{*}\right]$
means one can pool together some mass from $\theta_{1}$ and $\theta_{2}$
without violating the buyer's incentive constraints. It follows
that such pooling cannot benefit the seller; that is, (\ref{lem: ConvCav:ConvEq})
must hold. To prove the result without atoms, we approximate 
$\theta_{1}$ and $\theta_{2}$ with a shrinking neighborhood. The intuition for part \eqref{lem: ConvCav:Cav}
of the lemma is similar: if equation (\ref{lem: ConvCav:CavEq}) did not
hold, the seller would strictly benefit from having the buyer
spread the mass he puts on (a small neighborhood around) $\alpha\theta_{1}+\left(1-\alpha\right)\theta_{2}$
across $\theta_{1}$ and $\theta_{2}$, thereby violating optimality of $F^{*}$.

Before discussing the formal proof, observe first that both parts of the lemma trivially hold when $\theta_{1}=\theta_{2}$ or when $\alpha\in\left\{ 0,1\right\} $. Therefore, suppose (without
loss of generality) that $\theta_{1}<\theta_{2}$. 

Broadly speaking, the formal proof of the lemma proceeds as follows. Using that $\DMprice_{Q^*}$
is constant on $\left[\theta_{*},\theta^{*}\right]$, we construct
a family of informational deviations which are incentive compatible
for the buyer and that are indexed by $\epsilon>0$. As $\epsilon$
vanishes, the difference between these deviations and $F^{*}$ converges
to the difference between an atom at $\alpha\theta_{1}+\left(1-\alpha\right)\theta_{2}$
and a split of that atom's mass between an atom on $\theta_{1}$ and
an atom on $\theta_{2}$ for the first part, and vice-versa for the
second part. Then, we show the desired inequality using optimality
of $\left(Q^{*},F^{*}\right)$ and continuity of $\pi_{Q^{*}}|_{\left[\theta_{*},\theta^{*}\right]}$
(where the latter is implied by continuity of $c'$ and $\DMprice_{Q^*}$
being constant over $\left[\theta_{*},\theta^{*}\right]$).

We now proceed with the actual proof. 
As a preliminary step, let $G=F^{*}\left(\cdot|\theta\in\left[\theta_{*},\theta^{*}\right]\right)$,
$H=F^{*}\left(\cdot|\theta\notin\left[\theta_{*},\theta^{*}\right]\right)$,
and $\beta=F^{*}\left(\theta^{*}\right)-F_{-}^{*}\left(\theta_{*}\right)$,
and observe $F^{*}=\beta G+\left(1-\beta\right)H$. In addition, notice
that $\idu_{Q^*} - \icost$ is affine on $\left[\theta_{*},\theta^{*}\right]$,
because for any $\theta\in\left[\theta_{*},\theta^{*}\right]\subseteq\left[\thmin_{Q^*},\thmax_{Q^*}\right]$,
\begin{align*}
\idu_{Q^*}(\theta) - \icost(\theta)
=\idu_{Q^*}\left(\theta_{*}\right) - \icost\left(\theta_{*}\right)+\int_{\theta_{*}}^{\theta}\left(Q^{*}-\icost'\right)(\theta)\mathrm{d}\theta
& 
 =\idu_{Q^*}\left(\theta_{*}\right) - \icost\left(\theta_{*}\right)+\int_{\theta_{*}}^{\theta}\DMprice_{Q^{*}}(\theta)\mathrm{d}\theta
 \\ &
 =\idu_{Q^*}\left(\theta_{*}\right) - \icost\left(\theta_{*}\right)+\DMprice_{Q^*}\left(\theta_{*}\right)\left(\theta-\theta_{*}\right),
\end{align*}
where the last equality follows from $\DMprice_{Q^*}$ being constant
on $\left[\theta_{*},\theta^{*}\right]\subseteq\left[\thmin_{Q^*},\thmax_{Q^*}\right]$.

\textbf{Proof of Part \eqref{lem: ConvCav:Conv}. }We begin by constructing
the above-mentioned class of informational deviations. Take any $\epsilon\in\left(0,\frac{1}{2}\left(\theta_{2}-\theta_{1}\right)\right)$
(which implies $\left[\theta_{1}-\epsilon,\theta_{1}+\epsilon\right]\cap\left[\theta_{2}-\epsilon,\theta_{2}+\epsilon\right]=\varnothing$),
and define the following objects: 
\[
\begin{split}G_{0,\epsilon} & =G\left(\cdot|\theta\notin\left[\theta_{1}-\epsilon,\theta_{1}+\epsilon\right]\cup\left[\theta_{2}-\epsilon,\theta_{2}+\epsilon\right]\right),\\
G_{1,\epsilon} & =G\left(\cdot|\theta\in\left[\theta_{1}-\epsilon,\theta_{1}+\epsilon\right]\right),\\
G_{2,\epsilon} & =G\left(\cdot|\theta\in\left[\theta_{2}-\epsilon,\theta_{2}+\epsilon\right]\right),\\
\gamma_{1,\epsilon} & =G(\theta_{1}+\epsilon)-G_{-}(\theta_{1}-\epsilon),\\
\gamma_{2,\epsilon} & =G(\theta_{2}+\epsilon)-G_{-}(\theta_{2}-\epsilon),\\
\gamma_{0,\epsilon} & =1-\gamma_{1,\epsilon}-\gamma_{2,\epsilon}.
\end{split}
\]
Clearly, $G=\sum_{i=0}^{2}\gamma_{i,\epsilon}G_{i,\epsilon}$. Moreover,
since $\theta_{1},\theta_{2}\in\supp\ G$, both $\gamma_{1,\epsilon}$
and $\gamma_{2,\epsilon}$ are strictly positive for all $\epsilon>0$.
For any $\epsilon\in\left(0,\frac{1}{2}\left(\theta_{2}-\theta_{1}\right)\right)$,
define 
\[
\begin{split}\theta_{\epsilon} & =\int\theta\ddd\left(\alpha G_{1,\epsilon}+(1-\alpha)G_{2,\epsilon}\right),\\
\tilde{\gamma}_{\epsilon} & =\min\{\gamma_{1,\epsilon},\gamma_{2,\epsilon}\}>0,\text{ and}\\
G_{\epsilon} & =\gamma_{0,\epsilon}G_{0,\epsilon}+\tilde{\gamma}_{\epsilon}\mathbf{1}_{[\theta_{\epsilon},\infty)}+(\gamma_{1,\epsilon}-\alpha\tilde{\gamma}_{\epsilon})G_{1,\epsilon}+(\gamma_{2,\epsilon}-(1-\alpha)\tilde{\gamma}_{\epsilon})G_{2,\epsilon}.
\end{split}
\]
In words, $G_{\epsilon}$ alters $G$ by pooling $\alpha\tilde{\gamma}_{\epsilon}$
mass from the $\epsilon$-ball around $\theta_{1}$ and $(1-\alpha)\tilde{\gamma}_{\epsilon}$
mass from the $\epsilon$-ball around $\theta_{2}$ and pooling them
to create an $\tilde{\gamma}_{\epsilon}>0$ mass on $\theta_{\epsilon}$; that
is, 
\[
G_{\epsilon}-G=\tilde{\gamma}_{\epsilon}\left(\mathbf{1}_{[\theta_{\epsilon},\infty)}-\left(\alpha G_{1,\epsilon}+(1-\alpha)G_{2,\epsilon}\right)\right).
\]
With the above in hand, we can finally define our informational pertubation: specifically,
take $F_{\epsilon}=\beta G_{\epsilon}+\left(1-\beta\right)H$.

Next, we argue $F_{\epsilon}\in\Signals$ and that $Q^{*}$ is $F_{\epsilon}-$IC.
For the first claim, observe that because $\alpha G_{1,\epsilon}+(1-\alpha)G_{2,\epsilon}\succ\mathbf{1}_{[\theta_{\epsilon},\infty)}$,
$G_{\epsilon}$ is less informative than $G$, and so $F_{\epsilon}\preceq F^{*}\preceq F_{0}$.
That $F_{\epsilon}\in\Signals$ follows from $\preceq$ being transitive.
To see $Q^{*}$ is $F_{\epsilon}$-IC for all $\epsilon\in\left(0,\frac{1}{2}\left(\theta_{2}-\theta_{1}\right)\right)$,
observe that
\begin{align*}
\int[\idu_{Q^*}(\theta) - \icost(\theta)] \left(F_{\epsilon}-F^{*}\right)(\dd\theta) & =\beta\int[\idu_{Q^*}(\theta) - \icost(\theta)]\left(G_{\epsilon}-G\right)(\dd \theta)\\
 & =\beta\tilde{\gamma}_{\epsilon}\int[\idu_{Q^*}(\theta) - \icost(\theta)]\left[\mathbf{1}_{[\theta_{\epsilon},\infty)}-\left(\alpha G_{1,\epsilon}+(1-\alpha)G_{2,\epsilon}\right)\right](\dd\theta)=0,
\end{align*}
where the last equality follows from $\alpha G_{1,\epsilon}+(1-\alpha)G_{2,\epsilon}\succ\mathbf{1}_{[\theta_{\epsilon},\infty)}$,
the support of $\alpha G_{1,\epsilon}+(1-\alpha)G_{2,\epsilon}$ being
contained in $\left[\theta_{*},\theta^{*}\right]\subseteq\left[\thmin_{Q},\thmax_{Q}\right]$,
and $\idu_{Q^*} - \icost$ being affine on $\left[\theta_{*},\theta^{*}\right]$.

Now, because $\left(Q^{*},F^{*}\right)$ is seller-optimal, that
$Q^{*}$ is $F_{\epsilon}-$IC all small $\epsilon>0$ means that
$\int\pi_{Q^{*}}\mathrm{d}F_{\epsilon}\leq\int\pi_{Q^{*}}\mathrm{d}F$.
Rearranging this inequality, dividing by $\beta\tilde{\gamma}_{\epsilon}$,
and taking $\epsilon$ to zero delivers 
\[
\begin{split}
0\leq
\frac{1}{\beta\tilde{\gamma}_{\epsilon}}\int\pi_{Q^{*}}(\theta)(F^{*}-F_{\epsilon})(\dd\theta) 
& 
=\frac{1}{\tilde{\gamma}_{\epsilon}}\int\pi_{Q^{*}}(\theta)\left(G-G_{\epsilon}\right)(\dd\theta)
\\ & 
=\left[\alpha\int\pi_{Q^{*}}(\theta)G_{1,\epsilon}(\dd\theta)+\left(1-\alpha\right)\int\pi_{Q^{*}}(\theta)G_{2,\epsilon}(\dd\theta)\right]-\pi_{Q^{*}}(\theta_{\epsilon})
\\ & 
\rightarrow\left(\alpha\pi_{Q^{*}}(\theta_{1})+(1-\alpha)\pi_{Q^{*}}(\theta_{2})\right)-\pi_{Q^{*}}\left(\alpha\theta_{1}+(1-\alpha)\theta_{2}\right),
\end{split}
\]
where convergence follows from continuity of $\pi_{Q^{*}}|_{\left[\theta_{*},\theta^{*}\right]}$,
convergence of $G_{1,\epsilon}$ and $G_{2,\epsilon}$ to $\mathbf{1}_{[\theta_{1},\infty)}$
and $\mathbf{1}_{[\theta_{2},\infty)}$ respectively, and $\theta_{\epsilon}\rightarrow\alpha\theta_{1}+(1-\alpha)\theta_{2}$.
\qedhere

\textbf{Proof of Part \eqref{lem: ConvCav:Cav}.} Suppose now $\theta_1<\theta_2$ are such that $\theta_1$ satisfies either $I_F(\theta_1)>0$ or $F_0(\theta_1) > F(\theta_1)$, $\theta_2$ satisfies $I_F(\theta_2)>0$ or $F_{-}(\theta_2) > F_{0-}(\theta_2)$, and $I_{F}(\theta)>0$ for all $\theta\in\left(\theta_{1},\theta_{2}\right)\subseteq\left[\theta_{*},\theta^{*}\right]$. Let $\alpha\in\left(0,1\right)$ is such that $\theta_{\alpha}:=\alpha\theta_{1}+(1-\alpha)\theta_{2}\in\supp\ G$.
We begin by defining the following family of deviations. For
any strictly positive $\epsilon<\min\left\{ \theta_*-\theta_{1},\theta_{2}-\theta^*\right\} $,
define 
\[
\begin{split}G_{0,\epsilon}(\cdot) & :=G(\cdot|\theta\notin[\theta_{\alpha}-\epsilon,\theta_{\alpha}+\epsilon]),\\
G_{1,\epsilon}(\cdot) & :=G(\cdot|\theta\in[\theta_{\alpha}-\epsilon,\theta_{\alpha}+\epsilon]),\\
\theta_{\epsilon} & :=\int\theta G_{1,\epsilon}\left(\dd\theta\right)\\
\gamma_{\epsilon} & :=G(\theta_{\alpha}+\epsilon)-G_{-}(\theta_{\alpha}-\epsilon).
\end{split}
\]
Clearly, $G=(1-\gamma_{\epsilon})G_{0,\epsilon}+\gamma_{\epsilon}G_{1,\epsilon}$.
Observe $\gamma_{\epsilon}>0$, because $\theta_{\alpha}\in\supp\ G$,
and that an $\alpha_{\epsilon}\in\left(0,1\right)$ exists such that
\[
\theta_{\epsilon}=\alpha_{\epsilon}\theta_{1}+(1-\alpha_{\epsilon})\theta_{2},
\]
by our choice of $\epsilon$. Obviously, $\theta_{\epsilon}\rightarrow\theta_{\alpha}$
and $\alpha_{\epsilon}\rightarrow\alpha$ as $\epsilon \rightarrow 0$. For a given $\tilde{\gamma}\in\left(0,\gamma_{\epsilon}\right)$,
define 
\[
G_{\tilde{\gamma},\epsilon}=(1-\gamma_{\epsilon})G_{0,\epsilon}+(\gamma_{\epsilon}-\tilde{\gamma})G_{1,\epsilon}+\tilde{\gamma}\left(\alpha_{\epsilon}\mathbf{1}_{[\theta_{1},\infty)}+(1-\alpha_{\epsilon})\mathbf{1}_{[\theta_{2},\infty)}\right).
\]
Clearly, $G_{\tilde{\gamma},\epsilon}$ is a CDF.

We now construct our informational deviation: set $F_{\tilde{\gamma},\epsilon}:=\beta G_{\tilde{\gamma},\epsilon}+\left(1-\beta\right)H$
for all $\epsilon$ and $\tilde{\gamma}$ satisfying the above conditions.
We begin by arguing that this deviation is a signal---that is, $F_{\tilde{\gamma},\epsilon}\in\Signals$---whenever
$\tilde{\gamma}$ is sufficiently small (holding $\epsilon$ fixed).
To do so, observe that the function $F\mapsto I_{F}(\theta)$
is affine for all $\theta$, meaning that 
\begin{equation}
I_{F_{\tilde{\gamma},\epsilon}} - I_{F^{*}}=\tilde{\gamma}\beta\left(\alpha_{\epsilon}I_{\mathbf{1}_{[\theta_{1},\infty)}}+(1-\alpha_{\epsilon})I_{\mathbf{1}_{[\theta_{2},\infty)}}-I_{G_{1,\epsilon}}\right)\leq 0,\label{eq:lem: ConvCav1}
\end{equation}
where the inequality follows from $G_{1,\epsilon}$ being a mean preserving contraction of $\alpha_{\epsilon}\mathbf{1}_{[\theta_{1},\infty)}+(1-\alpha_{\epsilon})\mathbf{1}_{[\theta_{2},\infty)}$. This latter fact also implies that the above inequality holds with equality whenever $\theta \leq \theta_1$ or $\theta \geq \theta_2$. It follows that for all such $\theta$, $I_{F_{\tilde{\gamma},\epsilon}}(\theta)=I_{F^*}(\theta)\geq 0$, with equality whenever $\theta \in \{\thmin,\thmax\}$. 

Next we show $I_{F_{\tilde\gamma,\epsilon}}(\theta) \geq 0$ for all  $\theta\in\left(\theta_{1},\theta_{2}\right)$ whenever $\tilde\gamma$ is sufficiently small. Towards this goal, we first argue one can find $\tilde{\theta}_1,\tilde\theta_2$ such that the following properties hold: (a) $\theta_1\leq \tilde\theta_1<\tilde\theta_2 \leq \theta_2$; (b) $I_{F^*}(\theta)>0$ for all $\theta \in [\tilde\theta_1,\tilde\theta_2]$; (c) $\zeta_1:=\inf_{\theta \in (\theta_1,\tilde\theta_1)} (F_0 - F^*)(\theta) >0$ and $\zeta_2:=\inf_{\theta \in (\tilde\theta_2,\theta_2)} (F^*_{-} - F_{0-})(\theta) >0$. We begin with defining $\tilde\theta_1$. If $I_{F^*}(\theta_1)>0$, set $\tilde\theta_1=\theta_1$. If $I_{F^*}(\theta_1)=0$, then $F^*(\theta_1)<F_0(\theta_1)$ but the Lemma's assumptions. By right continuity of $F^*$ and $F_0$, a small $\nu>0$ exists such that $F_0(\theta) - F^*(\theta) > \nu$ for all $\theta \in (\theta_1,\theta_1+\nu)$: otherwise, one could find a decreasing sequence $(\theta_n)_{n\in\mathbb{N}}$ that converges to $\theta_1$ and has $F_{0}(\theta_n) \leq F^{*}(\theta_n)+1/n$, delivering $F_{0}(\theta_1) \leq F^{*}(\theta_1)$, which is a contradiction. Defining $\tilde\theta_2$ analogously (and using left continuity of $F_{0-}$ and $F^{*}_{-}$) delivers a $\tilde{\theta_1}$ and $\tilde{\theta_2}$ as desired (where property (b) is guaranteed by $I{F^*}>0$ over $(\theta_1,\theta_2)$). 

Armed with $\tilde\theta_1$ and $\tilde\theta_2$ as above, we now find $\tilde\gamma>0$ small such that $I_{F_{\tilde\gamma,\epsilon}}(\theta)\geq 0$ holds for all $\theta \in (\theta_1,\theta_2)$. Define
\[
\zeta_3 : = \frac{
\min_{\theta \in [\tilde\theta_1,\tilde\theta_2]} I_{F^*}(\theta)
}
{
\max_{\theta \in [\tilde\theta_1,\tilde\theta_2]}\left[I_{G_{1,\epsilon}}(\theta) - \alpha_{\epsilon}I_{\mathbf{1}_{[\theta_{1},\infty)}}(\theta)+(1-\alpha_{\epsilon})I_{\mathbf{1}_{[\theta_{2},\infty)}}(\theta)\right].
}
\]
Note $\zeta_3$ is well-defined because $\alpha_{\epsilon}\mathbf{1}_{[\theta_{1},\infty)}+(1-\alpha_{\epsilon})\mathbf{1}_{[\theta_{2},\infty)}$ is a strict mean-preserving spread of $G_{1,\epsilon}$, and so the denominator is strictly positive. By choice of $\tilde\theta_1$ and $\tilde\theta_2$ and continuity of $I_{F^*}$, the numerator is also strictly positive, meaning $\zeta_3>0$. Take any $\tilde\gamma < \min\{\zeta_1,\zeta_2, \zeta_3\}/\beta$. We argue $I_{F_{\tilde\gamma,\epsilon}}(\theta)\geq 0$ for all $\theta \in (\theta_1,\theta_2)$ by dividing $(\theta_1,\theta_2)$ into three subintervals.

Consider first $\theta \in (\tilde\theta_1,\tilde\theta_2)$. For such $\theta$,
    \[
    \begin{split}
    I_{F_{\tilde\gamma,\beta}}(\theta) 
    & = I_{F^*}(\theta) +     \tilde{\gamma}\beta\left(\alpha_{\epsilon}I_{\mathbf{1}_{[\theta_{1},\infty)}}(\theta)+(1-\alpha_{\epsilon})I_{\mathbf{1}_{[\theta_{2},\infty)}}(\theta)-I_{G_{1,\epsilon}}(\theta)\right) 
    \\ & \geq I_{F^*}(\theta) - \min_{\theta \in [\tilde\theta_1,\tilde\theta_2]} I_{F^*}(\theta) \geq 0.
    \end{split}
    \]
Second, take any $\theta \in [\theta_1,\tilde\theta_1]$. For such $\theta$,
\[
(\alpha_\epsilon I_{\mathbf{1}_{[\theta_{1},\infty)}}+(1-\alpha_{\epsilon})I_{\mathbf{1}_{[\theta_{2},\infty)}}-I_{G_{1,\epsilon}})(\theta) \leq \alpha_\epsilon (\theta - \theta_1) \leq \theta - \theta_1.
\]
Moreover, by definition of $\zeta_1$, $I_{F^*}(\theta) \geq I_{F^*}(\theta_1) + \zeta_1(\theta - \theta_1)$. Therefore, by choice of $\tilde\gamma$,
\[
    \begin{split}
    I_{F_{\tilde\gamma,\beta}}(\theta) 
    & = I_{F^*}(\theta) +     \tilde{\gamma}\beta\left(\alpha_{\epsilon}I_{\mathbf{1}_{[\theta_{1},\infty)}}(\theta)+(1-\alpha_{\epsilon})I_{\mathbf{1}_{[\theta_{2},\infty)}}(\theta)-I_{G_{1,\epsilon}}(\theta)\right) 
    \\ & \geq I_{F^*}(\theta_1) + (\zeta_1 - \tilde\gamma\beta )(\theta - \theta_1) \geq 0.
    \end{split}
    \]
Third, take any $\theta \in [\tilde\theta_2,\theta_2]$. Some simple algebra reveals that for such $\theta$,
\[
(\alpha_\epsilon I_{\mathbf{1}_{[\theta_{1},\infty)}}+(1-\alpha_{\epsilon})I_{\mathbf{1}_{[\theta_{2},\infty)}}-I_{G_{1,\epsilon}})(\theta) \leq \alpha_\epsilon (\theta_2 - \theta) \leq \theta_2 - \theta.
\]
Moreover, 
\[
I_{F^*}(\theta) = I_{F^*}(\theta_2) - \int_{\theta}^{\theta_2}(F_{0-} - F^{*}_{-})(\hat\theta)\dd\hat\theta 
\geq I_{F^*}(\theta_2) + \zeta_2 (\theta_2 - \theta).
\]
Thus, 
\[
    \begin{split}
    I_{F_{\tilde\gamma,\beta}}(\theta) 
    & = I_{F^*}(\theta) +     \tilde{\gamma}\beta\left(\alpha_{\epsilon}I_{\mathbf{1}_{[\theta_{1},\infty)}}(\theta)+(1-\alpha_{\epsilon})I_{\mathbf{1}_{[\theta_{2},\infty)}}(\theta)-I_{G_{1,\epsilon}}(\theta)\right) 
    \\ & \geq I_{F^*}(\theta_2) + (\zeta_2 - \tilde\gamma\beta )(\theta_2 - \theta) \geq 0.
    \end{split}
    \]
It follows that $F_{\tilde{\gamma},\epsilon}\in\Signals$ for all sufficiently small $\tilde\gamma$. 

We now argue $Q^{*}$ is $F_{\tilde{\gamma},\epsilon}$-IC for all
above-mentioned $\epsilon$ and all sufficiently small $\tilde{\gamma}$.
To see this, observe that
\begin{multline*}
\int[\idu_{Q^*}(\theta) - \icost(\theta)]\left(F_{\tilde{\gamma},\epsilon}-F^{*}\right)(\dd\theta) \\
=\tilde{\gamma}\beta\int[\idu_{Q^*}(\theta) - \icost(\theta)]\left(\alpha_{\epsilon}\mathbf{1}_{[\theta_{1},\infty)}+(1-\alpha_{\epsilon})\mathbf{1}_{[\theta_{2},\infty)}-G_{1,\epsilon}\right)(\dd\theta)\\
=\tilde{\gamma}\beta\left(\alpha_{\epsilon}[\idu_{Q^*}(\theta_1) - \icost(\theta_1)]+\left(1-\alpha_{\epsilon}\right)[V_{Q^*}(\theta_2) - \icost(\theta_2)]-[\idu_{Q^*}(\theta_\epsilon) -\icost(\theta_\epsilon)]\right)
=0,
\end{multline*}

where the last equality follows from $\alpha_{\epsilon}\mathbf{1}_{[\theta_{1},\infty)}+(1-\alpha_{\epsilon})\mathbf{1}_{[\theta_{2},\infty)}\succeq G_{1,\epsilon}$,
the support of $\alpha_{\epsilon}\mathbf{1}_{[\theta_{1},\infty)}+(1-\alpha_{\epsilon})\mathbf{1}_{[\theta_{2},\infty)}$
and $G_{1,\epsilon}$ being contained in $\left[\theta_{*},\theta^{*}\right]$,
and $\idu_{Q^*}-\icost$ being affine on $\left[\theta_{*},\theta^{*}\right]$.

For the proof's last step, observe that because $Q^{*}$ is $F_{\tilde{\gamma},\epsilon}$-IC
for the buyer for all small $\epsilon$ and $\tilde{\gamma}$, seller
optimality of $\left(Q^{*},F^{*}\right)$ implies
\begin{align*}
0\geq\frac{1}{\tilde{\gamma}}\int\pi_{Q^{*}}(\theta)\left(F_{\tilde{\gamma},\epsilon}-F^{*}\right)(\dd\theta) & =\alpha_{\epsilon}\pi_{Q^{*}}\left(\theta_{1}\right)+\left(1-\alpha_{\epsilon}\right)\pi_{Q^{*}}\left(\theta_{2}\right)-\pi\left(\theta_{\epsilon}\right)\\
 & \overset{\epsilon\rightarrow0}{\longrightarrow}\alpha\pi_{Q^{*}}\left(\theta_{1}\right)+\left(1-\alpha\right)\pi_{Q^{*}}\left(\theta_{2}\right)-\pi\left(\theta_{\alpha}\right),
\end{align*}
where convergence follows from $\theta_{\epsilon}\rightarrow\theta_{\alpha}$,
$\alpha_{\epsilon}\rightarrow\alpha$, and $\pi_{Q^{*}}$ being continuous
on $\left[\theta_{*},\theta^{*}\right]$. The desired inequality follows. \hfill$\square$

\subsubsection{Proof of Theorem \ref{thm: Inefficiency}}

Without loss, we can assume $(Q,F)$ is a seller optimal outcome with the property that $Q$ is $F$-CC. 

We first argue that if \eqref{eq: avg MC below threshold type} holds at $\theta^* \in [\thmin_{F},\thmax_{F})$, then $\qcost'(Q(\theta^*)) < \theta^*$. To do so, note that because the allocation $Q$ is $F$-CC, the allocation is
strictly increasing over $[\thmin_{F},\thmax_{F})$. When $\qcost$ is strictly convex, the marginal cost for quality provision is strictly increasing as well, and so equation (\ref{eq: avg MC below threshold type}) implies that 
\[
\qcost'\left(Q\left(\theta^{*}\right)\right)<\int_{\theta\geq\theta^{*}}\qcost'\left(Q(\theta)\right)F^{*}\left(\dd\theta|\theta\geq\theta^{*}\right) \leq \theta^{*};
\]
that is, $Q\left(\theta^{*}\right)$ lies strictly below its efficient level.

Next, we argue that $\qcost'\circ Q(\theta) <\theta$ holds for every $\theta \in [\thmin_{F},\thmax_F)$ at which $\DMprice_Q$ is strictly increasing, and so $Q(\theta)$ lies below its efficient level. If $\theta > \thmin_{F}$, this claim directly follows from Lemma~\ref{lem: allocation perturbations summary}-\eqref{lem: allocation perturbations summary - low mc and strictly increasing}, which deliver the inequality \eqref{eq: avg MC below threshold type}. For $\theta = \thmin_{F}$, distinguish two cases: either $Q(\thmin_F) > 0$ and so we can apply Lemma~\ref{lem: allocation perturbations summary}-\eqref{lem: allocation perturbations summary - low mc and min-theta} to get that \eqref{eq: avg MC below threshold type} holds, or $Q(\thmin_F) = 0$, in which case $\qcost'\circ Q(\thmin_Q) = \qcost'(0)<\thmin \leq \thmin_F$. 
It follows quality is inefficiently low (strictly) at all such $\theta$. 

Next we argue that $\qcost'\circ Q (\theta^*)<\theta^*$ holds for any $\theta^* \in \supp(F)$ around which $\DMprice_Q$ is constant. Thus suppose $\theta^{*}\in\supp\,F$
is such that $\DMprice_{Q}$ is constant on $\left[\theta^{*}-\delta,\theta^{*}+\delta\right]\cap \Theta$
for some $\delta>0$. Our goal is to show $\qcost'\circ Q\left(\theta^{*}\right)<\theta^{*}$.

Let $\theta_{*}=\inf\left\{ \theta\geq\underline{\theta}_{F}:\DMprice_{Q+}(\theta)=\DMprice_{Q}\left(\theta^{*}\right)\right\} .$
We now argue $\qcost'\circ Q(\theta_{*}) <\theta_{*}$.
There are three cases to consider: (a) $\theta_{*}=\thmin_{F}$ and $Q(\thmin_{F})=0$, (b) $\theta_{*}=\thmin_{F}$ and $Q(\thmin_{F})>0$, and (c) $\theta_{*}>\thmin_{F}$. In the first case $\qcost'\circ Q(\theta_*) = \qcost'(0) < \thmin \leq \thmin_F$. In the second case, we can use inequality \eqref{eq: avg MC below threshold type} by appealing to Lemma \ref{lem: allocation perturbations summary}-\eqref{lem: allocation perturbations summary - low mc and min-theta}. In the third case, $\DMprice_Q$ must be strictly increasing at $\theta_*$, and so \eqref{eq: avg MC below threshold type} must hold by Lemma~\ref{lem: allocation perturbations summary}-\eqref{lem: allocation perturbations summary - low mc and strictly increasing}, giving $\qcost'\circ Q(\theta_*) < \theta_*$.

Define $\bar{\theta}^{*}=\min\{\theta^{*}+\delta,\bar{\theta}_{F}\}$,
and let $G:=F\left(\cdot|\theta\in\left[\theta_{*},\bar{\theta}^{*}\right]\right)$.
We claim \eqref{eq: avg MC below threshold type} holds for
\[
\theta':=\min\left(\supp\,G\right).
\]
Clearly, we are done if $\theta'=\theta_{*}$. If $\theta'>\theta_{*}$,
then $\theta_{*}\notin\supp\,G$, and so $F_{-}\left(\theta_{*}\right)=F\left(\theta_{*}\right)=F_{-}\left(\theta'\right)$.
We therefore have the following inequality chain: 
\begin{align*}
\left(1-F_{-}\left(\theta'\right)\right)\theta' & >\left(1-F_{-}\left(\theta'\right)\right)\theta_{*}=\left(1-F_{-}\left(\theta_{*}\right)\right)\theta_{*}\\
 & \geq\int_{\theta\geq\theta_{*}}\qcost'\circ Q(\theta)F\left(\dd\theta\right)=\int_{\theta\geq\theta'}\qcost'\circ Q(\theta)F\left(\dd\theta\right),
\end{align*}
where the weak inequality follows from \eqref{eq: avg MC below threshold type}
holding at $\theta_{*}$. Thus, we have shown  \eqref{eq: avg MC below threshold type strong}
holds at $\theta'$, and so \eqref{eq: avg MC below threshold type} holds
as well (see Lemma~\ref{lem: strong FOC implies weak FOC}).

If $\theta^{*}=\theta'$, then \eqref{eq: avg MC below threshold type} holds for $\theta^*$, and so 
$\qcost'\circ Q\left(\theta^{*}\right)<\theta^{*}$, as explained after Lemma~\ref{lem: allocation perturbations summary}. Hence, there is nothing left to prove in this case. Thus, hereafter, we suppose $\theta^{*}\neq\theta'$. 
Since $\theta^{*}\in\supp\,G$, we must have $\theta^{*}>\theta^{'}$.

We now argue $\DMprice_{Q}$ is constant on $\left[\theta',\bar{\theta}^{*}\right]$.
To do so, notice $\qcost'\circ Q\left(\theta'\right)<\theta'$ implies
$Q\left(\theta'\right)=Q_{+}\left(\theta'\right)$ since $Q$ jumps towards efficiency. Hence,
\[
\DMprice_{Q}\left(\theta'\right)=Q\left(\theta'\right)-\icost'\left(\theta'\right)=Q_{+}\left(\theta'\right)-\icost'\left(\theta'\right)=\DMprice_{Q+}\left(\theta'\right)=\DMprice_{Q}\left(\theta^{*}\right),
\]
where the last equality follows from $\theta'\geq\theta_{*}$. It
follows $\DMprice_{Q}$ is constant on $[\theta',\theta^{*}]\cup[\theta^{*}-\delta,\bar{\theta}^{*})=[\theta',\bar{\theta}^{*})$.
Recalling $\bar{\theta}^{*}=\min\left\{ \bar{\theta}_{F},\theta^{*}+\delta\right\} $,
it follows $\DMprice_{Q}\left(\bar{\theta}^{*}\right)=\DMprice_{Q}\left(\theta^{*}\right).$ 
Thus, we have shown $\DMprice_{Q}$ is constant on $\left[\theta',\bar{\theta}^{*}\right]$. 

Consider now the line segment connecting $\left(\theta',\pi_{Q}\left(\theta'\right)\right)$
with $\left(\theta^{*},\pi_{Q}\left(\theta^{*}\right)\right)$, 
\begin{align*}
\varphi:\left[\theta',\theta^{*}\right] & \rightarrow\mathbb{R},\\
\theta & \mapsto\pi_{Q}\left(\theta'\right)+\left(\frac{\pi_{Q}\left(\theta^{*}\right)-\pi_{Q}\left(\theta'\right)}{\theta^{*}-\theta'}\right)\left(\theta-\theta'\right).
\end{align*}
We claim $\varphi(\theta)\geq\pi_{Q}(\theta)$
for all $\theta\in\left[\theta',\theta^{*}\right]$. Obviously, $\varphi(\theta)=\pi_{Q}(\theta)$
whenever $\theta\in\left\{ \theta',\theta^{*}\right\} $. For $\theta\in\left(\theta',\theta^{*}\right)$,
we get the following inequality:
\begin{align*}
\varphi(\theta) & =\left(\frac{\theta^{*}-\theta}{\theta^{*}-\theta'}\right)\varphi\left(\theta'\right)+\left(\frac{\theta-\theta'}{\theta^{*}-\theta'}\right)\varphi\left(\theta^{*}\right)\\
 & =\left(\frac{\theta^{*}-\theta}{\theta^{*}-\theta'}\right)\pi_{Q}\left(\theta'\right)+\left(\frac{\theta-\theta'}{\theta^{*}-\theta'}\right)\pi_{Q}\left(\theta^{*}\right)\geq\pi_{Q}(\theta),
\end{align*}
where the inequality follows from Lemma~\ref{lem: ConvCav} part~\eqref{lem: ConvCav:Conv},
which applies because $\DMprice_{Q}$ is constant on $\left[\theta',\bar{\theta}^{*}\right]$.

Next, we show $\left(\frac{\pi_{Q}\left(\theta^{*}\right)-\pi_{Q}\left(\theta'\right)}{\theta^{*}-\theta'}\right)$
is strictly positive. For this purpose, fix any $\epsilon\in\left(0,\theta^{*}-\theta'\right)$.
Because $\DMprice_{Q}$ is constant on $\left[\theta',\bar{\theta}^{*}\right]$,
\[
Q\left(\theta'+\epsilon\right)-Q_{+}\left(\theta'\right)=Q\left(\theta'+\epsilon\right)-Q\left(\theta'\right)=c'\left(\theta'+\epsilon\right)-\icost'\left(\theta'\right).
\]
It follows $Q'_{+}\left(\theta'\right)=c''\left(\theta'\right)$,
delivering the following inequality chain, 
\begin{align*}
\left(\frac{\pi_{Q}\left(\theta^{*}\right)-\pi_{Q}\left(\theta'\right)}{\theta^{*}-\theta'}\right) & =\frac{1}{\epsilon}\left[\varphi\left(\theta'+\epsilon\right)-\varphi\left(\theta'\right)\right]\\
 & \geq\frac{1}{\epsilon}\left[\pi_{Q}\left(\theta'+\epsilon\right)-\pi_{Q}\left(\theta'\right)\right]\\
 & =\frac{1}{\epsilon}\theta'\left(Q\left(\theta'+\epsilon\right)-Q\left(\theta'\right)\right)-\frac{1}{\epsilon}\left[\qcost\circ Q\left(\theta'+\epsilon\right)-\qcost\circ Q\left(\theta'\right)\right]\\
 & \,\,\,+Q\left(\theta'+\epsilon\right)-\frac{1}{\epsilon}\left[V_{Q}\left(\theta'+\epsilon\right)-V_{Q}\left(\theta'\right)\right]\\
 & \rightarrow\left(\theta'-\kappa'\circ Q\left(\theta'\right)\right)c''\left(\theta'\right)>0,
\end{align*}
where convergence follows from the chain rule and $V_{Q+}^{'}\left(\theta'\right)=Q_{+}\left(\theta'\right)$,
and the strict inequality from $c$ being strictly convex.

We now turn to establishing $\qcost'\circ Q\left(\theta^{*}\right)<\theta^{*}$,
thereby concluding the proof. Toward this goal, notice again that
for any $\epsilon\in\left(0,\theta^{*}-\theta'\right)$, 
\[
Q\left(\theta^{*}\right)-Q\left(\theta^{*}-\epsilon\right)=c'\left(\theta^{*}\right)-\icost'\left(\theta^{*}-\epsilon\right),
\]
because $\DMprice_{Q}$ is constant on $\left[\theta',\bar{\theta}^{*}\right]$.
Therefore, $Q'_{-}\left(\theta^{*}\right)=c''\left(\theta'\right)$.
So, we obtain the following inequality chain:
\begin{align*}
0<\left(\frac{\pi_{Q}\left(\theta^{*}\right)-\pi_{Q}\left(\theta'\right)}{\theta^{*}-\theta'}\right) & =\frac{1}{\epsilon}\left[\varphi\left(\theta^{*}\right)-\varphi\left(\theta^{*}-\epsilon\right)\right]\\
 & \leq\frac{1}{\epsilon}\left[\pi_{Q}\left(\theta^{*}\right)-\pi_{Q}\left(\theta^{*}-\epsilon\right)\right]\\
 & =\frac{1}{\epsilon}\theta^{*}\left(Q\left(\theta^{*}\right)-Q\left(\theta^{*}-\epsilon\right)\right)-\frac{1}{\epsilon}\left[\qcost\circ Q\left(\theta^{*}\right)-\qcost\circ Q\left(\theta^{*}-\epsilon\right)\right]\\
 & \,\,\,+Q\left(\theta^{*}-\epsilon\right)-\frac{1}{\epsilon}\left[V_{Q}\left(\theta^{*}\right)-V_{Q}\left(\theta^{*}-\epsilon\right)\right]\\
 & \rightarrow\left(\theta^{*}-\kappa'\circ Q\left(\theta^{*}\right)\right)c''\left(\theta^{*}\right),
\end{align*}
where convergence follows from the chain rule and $V_{Q-}^{'}\left(\theta^{*}\right)=Q_{-}\left(\theta^{*}\right)$.
Since $c''\left(\theta^{*}\right)>0$, the above inequality implies
$\qcost'\circ Q\left(\theta^{*}\right)<\theta^{*}$, as required.

To conclude the proof, it remains to show that quality is efficient at $\thmax_F$ whenever $\thmax_F=\thmax$. Suppose $\thmax_{F} =\thmax$. Recall that $Q_{+}(\thmax)=\bar{q}$ by definition (see the beginning of the Online Appendix). Because $\qcost'(\bar{q})>\thmax$, and because $Q$ jumps towards efficiency, we need only to show that $\qcost'\circ Q_{-}(\thmax)\leq\thmax.$ If $\DMprice_Q$ is constant around $\thmax$, this inequality follows from the argument above (and $Q_{-}(\thmax)\leq Q(\thmax)$).

Suppose then that $\DMprice_Q$ is strictly increasing at $\thmax$, and consider two cases. For the first case, assume $F$ has a jump at $\thmax$. Since $\DMprice_Q$ is strictly increasing at $\thmax$, we can apply Lemma~\ref{lem: allocation perturbations summary}-\eqref{lem: allocation perturbations summary - low mc and strictly increasing} to get that \eqref{eq: avg MC below threshold type} must hold. Since $F_{-}(\thmax)< F(\thmax)$, evaluating this equation at $\theta^*=\thmax_F=\thmax$ gives $\qcost'\circ Q(\thmax) \leq \thmax$, as required. 

Assume now for the second case that $F$ is continuous at $\thmax$. Since $\thmax \in \supp \ F$, continuity of $F$ at $\thmax$ means one can approach $\thmax$ from below with a sequence of types that receive inefficiently low quality. Since the efficient quality allocation is strictly increasing, it follows that $Q_{-}(\thmax)$ lies (weakly) below the efficient level, as required.

\subsection{Proofs from Section~\ref{sec: Comparative Statics} and Proposition~\ref{prop: sufficient distortion}}

\subsubsection{Proof of Theorem~\ref{thm: large cost comparative statics}}
\begin{proof}[\textbf{Proof of Theorem~\ref{thm: large cost comparative statics}}-\eqref{thm: large cost comparative statics - uninformative F}]

It is straightforward to show that some $\hat\costcoef$ exists such that the steep slope condition holds for all $\costcoef\geq\hat\costcoef$. For such $\costcoef$, the MPS constraint only binds at $\thmin$ and $\thmax$, and there is a binary optimal signal---see Proposition~\ref{prop: binary signals}. 
We consider $\costcoef \geq \hat{\costcoef}$, and start by showing that for every $[\tilde{\theta}_L, \tilde{\theta}_H] \subseteq (\thmin,\thmax)$ with $\tilde{\theta}_L < \theta_0 <  \tilde{\theta}_H$, there is a $\costcoef'\geq \hat\costcoef$ such that $[\tilde{\theta}_L, \tilde{\theta}_H]$ contains the support of the optimal signal for every larger $\costcoef$. 

For any $\costcoef \geq \hat\costcoef$, let $(F_\costcoef,Q_\costcoef)$ be an optimal outcome with $F_\costcoef$ binary, and $Q_\costcoef$ $F_\costcoef$-CC. Define $\supp F_\costcoef = \{\theta^\costcoef_L, \theta^\costcoef_H \} $, with $\thmin \leq \theta^{\costcoef}_L\leq \theta_0 \leq \theta^{\costcoef}_H \leq \thmax$. $F$-CC implies: 

\[
\costcoef \left( \icost'(\theta^\costcoef_H) - \icost'(\theta^\costcoef_L )\right) \leq Q_\costcoef(\theta_H) - Q_\costcoef(\theta_L) \leq \bar{q},
\]
where the first inequality follows from $F$-CC and $p_\costcoef$ increasing, and the second inequality from $Q_\costcoef \in [0,\bar{q}]$.

Thus:

\[
\icost'(\theta^\costcoef_H) \leq \frac{\bar q}{ \costcoef} + \icost'(\theta^\costcoef_L ) \leq  \frac{\bar q}{ \costcoef} + \icost'(\theta_0 ),
\]

where the second inequality follows from convexity of $c$ and $\theta^\costcoef_L \leq \theta_0$. Thus, one can choose a sufficiently high $\costcoef$ so that $\theta^\costcoef_H \leq \tilde{\theta}_H < \bar{\theta}$. Similarly:

\[
\icost'(\theta^\costcoef_L) \geq  \icost'(\theta^\costcoef_H ) -\frac{\bar q}{ \costcoef} \geq  \icost'(\theta_0 ) -\frac{\bar q}{ \costcoef},
\]

implying that we can choose sufficiently large $\costcoef$ for $\theta^\costcoef_L \geq \tilde{\theta}_L > \underline{\theta}$. Therefore, we can find $\costcoef'$ such that, for $\costcoef \geq \costcoef'$, we can focus on $[\tilde{\theta}_L, \tilde{\theta}_H]$, where $\icost', \icost'', \icost'''$---which are assumed to be continuous---are bounded. We have then concluded that for any $\costcoef \geq \costcoef'$ there exist an optimal signal which is (i) binary, (ii) does not separate any points in $(\thmin,\thmax)$ and (iii) whose support lies within $[\tilde{\theta_L},\tilde{\theta_H}]$.

We now move on to the main argument. In particular we show a $\bar{\costcoef}$ exists such that for all $\costcoef>\bar\costcoef$, the principal's profit function $\pi_{Q}$ is concave whenever $Q$ is an $F$-CC allocation for an $F$ that satisfies (i), (ii) and (iii). Because we know there is an optimal signal satisfying (i), (ii) and (iii), showing that the profit is concave for any such $\DMprice$ is sufficient to establish optimality of no information.  
Start by recalling that any $\DMprice$ consistent with a signal $F$ which is $F$-pooling in $(\thmin,\thmax)$
must be constant in $(\thmin,\thmax)$.  Define $\Theta^\costcoef_\DMprice = \{\theta \in (\tilde{\theta}_L, \tilde{\theta}_H): \costcoef c'(\theta) + \DMprice(\theta) \in [0,\bar{q}] \}$.

Because $p$ is constant on the interval $\Theta^\costcoef_\DMprice$, the profit function $\pi_{Q_\DMprice}$ is twice differentiable in that interval. We consider the second derivative of $\pi_{Q_\DMprice}$ with respect to $\theta \in (\tilde{\theta}_L,\tilde{\theta}_H)$:

\[
\pi ''_{Q_\DMprice}(\theta)=(\theta - \qcost'\left(Q_p (\theta)\right)) \costcoef \icost'''(\theta) + \left(1-\qcost''\left(Q_p(\theta)\right) \costcoef \icost''(\theta)\right) \costcoef \icost''(\theta) ,
\]

Dividing this expression by $\costcoef c'' (\theta)$, which is strictly positive by strong convexity of $c$:

\begin{equation*}
    \begin{split}
     \frac{\pi_{Q_\DMprice}''(\theta)}{\costcoef c''(\theta)} =    (\theta - \qcost'\left(Q_\DMprice (\theta)\right)) \frac{\icost'''(\theta)}{\icost'' (\theta)} + 1 -\qcost''\left(Q_\DMprice(\theta)\right) \costcoef \icost''(\theta) \\  \leq \max_{q \in [0,\bar{q}], x \in [\tilde{\theta}_L, \tilde{\theta}_H]}| x - \qcost'\left(q\right) | \left| \frac{\icost'''(x)}{\icost''(x)} \right| + 1 - \costcoef \min_q \qcost''(q) \min_{x \in [\tilde{\theta}_L, \tilde{\theta}_H]} \icost'' (x) ,
    \end{split}
\end{equation*}

where the first term is bounded because $\icost$ is strongly convex, and the term multiplying $\costcoef$ is strictly positive, as $\icost$ and $\qcost$ are both strongly convex, ensuring that $\icost''$ and $\qcost''$ are positive and bounded away from zero. Thus, by choosing high enough $\costcoef$ we obtain that $\pi_{Q_\DMprice}''$ is negative on $\Theta^\costcoef_\DMprice$. Because the bound is uniform across $\DMprice$---satisfying (i), (ii) and (iii)---and $\theta \in [\tilde\theta_L, \tilde\theta_H]$, we conclude that there exists $\bar{\costcoef}$ such that, for $\costcoef \geq \bar{\costcoef}$, the principal's objective is concave for $\theta \in \Theta^\DMprice_\costcoef$, for any $\DMprice$. Finally, because $\supp F \subset \Theta^\costcoef_\DMprice$ for any $F$ consistent with $\DMprice$, the principal benefits from pooling together all types, regardless of the marginal price function. Therefore, $\supp F_\costcoef = \{\theta_0 \}$.

\end{proof}

\begin{proof}[\textbf{Proof of Theorem~\ref{thm: large cost comparative statics}}-\eqref{thm: large cost comparative statics - seller payoff}]

 We now consider $\costcoef \geq \bar{\costcoef}$, so $\supp F_\costcoef = \{ \theta_0 \}$. For any marginal price $\hat{p}$, re-normalize it as $\hat{p} = \costcoef p$. Notice that any feasible $\hat{p}$ for $F_\costcoef$ is constant in $[\thmin,\thmax)$ and therefore so is the associated normalization $p$. We use $p(\thmin)$ to denote that constant value.

Define $\thmin^\costcoef_p = \arg \min_{v \in [\thmin,\thmax]} \{ \costcoef \icost(v) - p(\thmin)(\theta_0-v) \} $. Note that, if $\costcoef\icost'(v) + p(\thmin) = 0$ for any $\theta \in [\thmin,\thmax]$, $\thmin^\costcoef_p = v$. Otherwise, $\thmin^\costcoef_p = \thmin$. In other words, $\thmin^\costcoef_p$ is the largest type such that the $Q_p(\theta) =0$---i.e. for a fixed $\costcoef$, $\thmin^\costcoef_p = \thmin_{Q_p}$.
 
We now rewrite the gross value for buyer $\theta_0$ using the observation above:

\begin{equation}
    \begin{split}
\idu^\costcoef_p (\theta_0) = \costcoef \ \icost(\theta_0) - \costcoef \ \icost(\thmin^\costcoef_p) + \ p(\thmin) \left(\theta_0 - \thmin^\costcoef_p \right)\\  = \costcoef \icost(\theta_0) -  \min_{v \in [\thmin,\thmax]}\{ \costcoef \icost(v) - p(\thmin) (\theta_0 - v) \}  \\ = 
\costcoef \icost(\theta_0) -  \min_{v \in [\thmin,\thmax]}\left\{ \costcoef \icost(v) + \costcoef\icost'(\theta_0)(\theta_0-v) - Q_p (\theta_0) (\theta_0-v)  \right\},      
    \end{split}
\end{equation}
where the last equality follows from noticing that, by $F_\costcoef$-CC, $Q_p(\theta_0) = \costcoef \icost'(\theta_0) + p(\thmin)$. Two observations about $V^\costcoef_p$ are in order. First, it depends of $p$ only through $Q_p(\theta_0)$. Second, it is decreasing in $\costcoef$. Indeed, by the envelope theorem, the derivative of $V^\costcoef_p$ with respect to $\costcoef$ is $\icost(\theta_0) - \icost(\thmin_p^\gamma)-\icost'(\theta_0)(\theta_0-\thmin_p^\gamma)$, which is negative by strong convexity of $\icost$.

By the first observation, we can write the problem of the seller, given $\supp F_\costcoef = \{\theta_0\}$, as an optimization by direct choice of  $Q_p (\theta_0)$:

\begin{equation*}
    \begin{split}
        U_S := \max_{Q_p (\theta_0) \in [0,\bar{q}]} \theta_0 Q_p (\theta_0) - \qcost\left(Q_p(\theta_0)\right) - V^\costcoef_p(\theta_0).
    \end{split}
\end{equation*}

By the second observation, this objective function is strictly increasing in $\costcoef$. Thus, the optimal profits must strictly increase in $\costcoef$.  
\end{proof}

\begin{proof}[\textbf{Proof of Theorem~\ref{thm: large cost comparative statics}}-\eqref{thm: large cost comparative statics - buyer payoff}]
 Consider $\costcoef \geq \bar{\costcoef}$ again such that, by Part~\eqref{thm: large cost comparative statics - uninformative F}, $\supp F = \{ \theta_0\}$, and define $\thmin_p$ as in the proof of Part~\eqref{thm: large cost comparative statics - seller payoff}.  Recall that $\icost'(\theta_0) = 0$, so we omit it moving forward to save notation, observing that $Q_p(\theta_0) = p(\thmin).$

We can write the seller's problem, given $\supp F_\costcoef = \{\theta_0\}$ as follows:
\[
\max_{p(\thmin)\geq 0} \theta_0 p(\thmin) - \qcost\left( p(\thmin)\right)  - \left(\costcoef \icost(\theta_0) -  \min_{v \in [\thmin,\thmax]}\{  \costcoef c(v) - p(\thmin) (\theta_0- v)\} \right) .
\]

We first argue that, if $\supp F_\gamma = \{\theta_0\}$, then $p(\thmin)>0$. Indeed, if $p(\thmin) = 0$, the seller makes zero profits. However, the seller can always guarantee positive profits by offering any contract that is accepted by all types---for example, offering only the efficient contract of type $\thmin$,  at a price that extracts all rents from that type. Therefore, $p(\thmin)>0$.

We can then apply the envelope theorem, to immediately obtain the following interior solution to the seller's problem:

\[
\qcost' \left(p(\thmin) \right) = \underline{\theta}^\costcoef_{p}.
\]

By total differentiation of the equation, we obtain the derivative of the optimal $p(\thmin)$ with respect to $\costcoef$:

\[
\frac{d p(\thmin)}{d\costcoef}= -\frac{c'(\underline{\theta}^\costcoef_{p})}{ \costcoef \qcost'' (\hat{p})\icost''(\underline{\theta}^\costcoef_{p})+1}>0.
\]

Because no information is acquired in the optimal, the consumer surplus is:

\[
U_B : = \idu(\theta_0) = \costcoef \icost(\theta_0) - \min_{v \in [\thmin,\thmax]}\{ \costcoef c(v)  - p(\thmin) (\theta_0 - v)\}. 
\]

Taking derivatives with respect to $\costcoef$ we obtain:

\begin{equation*}
    \begin{split}
        \frac{d U_B}{ d \costcoef} = \icost(\theta_0) - \icost(\underline{\theta}^\costcoef_{p}) + \frac{d p(\thmin)}{d\costcoef}(\theta_0-\underline{\theta}^\costcoef_{p}) \\ =  \icost(\theta_0) - \icost(\underline{\theta}^\costcoef_{p}) -\frac{c'(\underline{\theta}^\costcoef_{p})}{ \costcoef \qcost'' (\hat{p})\icost''(\underline{\theta}^\costcoef_{p})+1}(\theta_0-\underline{\theta}^\costcoef_{p}).
    \end{split}
\end{equation*}

We now apply two observations. First, by the mean value theorem, there exists some $x^\costcoef \in [\underline{\theta}^\costcoef_{p},\theta_0]$, such that $-\icost'(\underline{\theta}^\costcoef_{p}) = \icost''(x^\costcoef) \left(\theta_0 - \underline{\theta}^\costcoef_{p} \right) $. Second, by strong convexity of $\icost$, there exists some $\alpha >0$ such that $\icost(\underline{\theta}^\costcoef_{p}) \geq \icost(\theta_0) + \icost'(\theta_0)(\theta_0 -\underline{\theta}^\costcoef_{ p}) + \alpha (\theta_0 - \underline{\theta}^\costcoef_{ p})^2 $. These two observations imply, in the formula above:

\begin{equation*}
    \begin{split}
        \frac{d U_B}{ d \costcoef} \leq \left( -\alpha +\frac{ \icost''(x^\costcoef) }{ \costcoef \qcost'' (p(\thmin))\icost''(\underline{\theta}^\costcoef_{p})+1} \right)(\theta_0-\underline{\theta}^\costcoef_{ p})^2
    \end{split}
\end{equation*}

Because for $\costcoef>\bar{\costcoef}$, $\icost''$ is bounded by the proof Part~\eqref{thm: large cost comparative statics - uninformative F}, and because $\qcost, \icost$ are strongly convex, there exists $\bar{\bar{\costcoef}}\geq \hat{\costcoef}$ such that the term in the first parentheses is negative, which completes the proof.
\end{proof}

\subsubsection{Proof of Theorem~\ref{thm: low cost comparative statics}}

\begin{proof}[\textbf{Proof of Theorem~\ref{thm: low cost comparative statics}}-\eqref{thm: low cost comparative statics - fully informative F}]
 Below, we establish the existence of $\underline{\costcoef}>0$, such that for all $\costcoef \leq \underline{\costcoef}$, the optimal signal is $F_\gamma = F_0$. This is done in a sequence of Lemmas. Lemmas~\ref{lem: finite support binding MPS}-\ref{lem: finite support full learning is unique optimum} establish that, when $\costcoef=0$, the full learning outcome---$F=F_0$ and $Q$ being the optimal screening allocation for $F_0$---is uniquely optimal for the seller. Then, Lemmas~\ref{lem: finite support convergence to full learning}-\ref{lem: finite support full learning for low cost} establish, through an approximation argument, that there exists an information cost scale that is low enough that guarantees $F_0$ to be optimal.

\begin{lemma}\label{lem: finite support binding MPS}
    Let Condition~\ref{as: finitesupport} hold. Let $F \in \Signals$ be such that $I_F(\theta)=0$ for some $\theta \in (\theta_i,\theta_{i+1})$ for some $i$. Then $I_F([\theta_i,\theta_{i+1}]) = \{0\}$ and $F(\theta') = F_0(\theta')$ for all $\theta' \in [\theta_i,\theta_{i+1})$.
\end{lemma}
\begin{proof}
    Since $\theta \notin \supp F_0$, $F_0$ is continuous at $\theta$, and so the argument of Lemma 3 part (i) in \cite{ravid2022learning} delivers that $F(\theta)=F_0(\theta)$. Moreover, the same argument delivers that whenever $I_F([\theta_i,\theta_{i+1}]) = \{0\}$ holds, $F(\theta') = F_0(\theta')$ must also hold for all $\theta' \in [\theta_i,\theta_{i+1})$. Hence, proving that $I_F([\theta_i,\theta_{i+1}]) = \{0\}$ is sufficient. 
    
    We first argue that $I_F(\theta') =0$ for all $\theta' \in (\theta,\theta_{i+1}]$. This follows from noting that
    \[
    \begin{split}
    0 \leq I_F(\theta') = I_F(\theta) + \int_{\theta}^{\theta'}(F_0 - F)(\tilde{\theta})\dd\tilde{\theta} 
    & = \int_{\theta}^{\theta'}(F_0 - F)(\tilde{\theta})\dd\tilde{\theta} 
    \\ & \leq \int_{\theta}^{\theta'}(F_0(\tilde{\theta}) - F(\theta))\dd\tilde{\theta}
    \\&  = (\theta' - \theta)(F_0(\theta) - F(\theta))=0,
    \end{split}
    \]
    where the second equality follows from $I_F(\theta)=0$, the third equality from $F$ being increasing, and the fourth equality from $F_0$ being constant on $[\theta_i,\theta_{i+1})\supseteq [\theta,\theta']$. 

    To conclude the lemma, we show that $I_F(\theta') = 0$ for all $\theta' \in [\theta_i,\theta]$. We do so by way of contradiction---hence, suppose $I_F(\theta')>0$. Then,
    \[
    0 < I_{F}(\theta')=I_{F}(\theta) - \int_{\theta'}^{\theta}(F_0-F)(\tilde\theta)\dd \tilde{\theta} = \int_{\theta'}^{\theta}(F-F_0)(\tilde\theta)\dd \tilde{\theta}.
    \] 
    In particular, a $\theta'' \in (\theta',\theta)$ must exist such that $F(\theta'') > F_0(\theta'')$. To get a contradiction, notice that
    \[
    F(\theta) \geq F(\theta'') > F_0(\theta'') = F_0(\theta) = F(\theta),
    \]
    where the first inequality follows from $F$ being increasing, and the second equality from $F_0$ being constant on $[\theta_i,\theta_{i+1})\supseteq [\theta'',\theta)$. 
\end{proof}

\begin{lemma}\label{lem: finite support increasing p}
    Let Condition~\ref{as: finitesupport} hold. Assume $(Q,F)$ is seller optimal and $Q$ is $F$-CC. Then $\DMprice_Q$ is constant on $(\theta_i,\theta_{i+1})$ for all $i=1,\ldots,L-1$. 
\end{lemma}
\begin{proof}
    Suppose by way of a contradiction that $\DMprice_Q$ is strictly increasing at some $\hat\theta \in (\theta_i,\theta_{i+1})$. We argue that $\hat\theta > \thmin_Q$. To do so, note that $I_F(\hat\theta) = 0$, and so $F(\theta') = F_0(\theta') = F_0(\theta_i)>0$ for all $\theta' \in [\theta_i,\theta_{i+1})$. Hence, $\hat\theta > \theta_i \geq \thmin_F \geq \thmin_Q$, as required. 

    Thus, $\hat\theta > \thmin_Q$. Moreover, $I_F(\theta_{i+1}) = 0$ by Lemma~\ref{lem: finite support binding MPS}. Notice that Theorem~\ref{thm: Inefficiency} implies $Q(\thmax_F) < \bar{q}$. Hence, we can apply Lemma \ref{lem: allocation perturbations summary} to obtain the following contradiction:
    \[
    \begin{split}
    \int_{\theta \geq \hat\theta} \qcost'(Q(\theta))F(\dd \theta | \theta \geq \hat\theta)
    \leq \hat\theta 
    < \theta_{i+1} 
    & \leq \int_{\theta \geq \theta_{i+1}} \qcost'(Q(\theta))F(\dd \theta | \theta \geq \theta_{i+1}) 
    \\ & = \int_{\theta \geq \hat\theta} \qcost'(Q(\theta))F(\dd \theta | \theta \geq \hat\theta), 
    \end{split}
    \]
    where the first inequality follows from Lemma~\ref{lem: allocation perturbations summary} part~\eqref{lem: allocation perturbations summary - low mc and strictly increasing}, the third inequality from Lemma~\ref{lem: allocation perturbations summary} part~\eqref{lem: allocation perturbations summary - high mc}, and the last inequality from $F$ being constant on $[\theta_i,\theta_{i+1})$ (by Lemma~\ref{lem: finite support binding MPS}). 
\end{proof}

\begin{lemma}\label{lem: finite supp, free, full learning better}
Suppose Condition~\ref{as: finitesupport} holds, $\costcoef=0$, and let $Q$ be an $F$-CC allocation that is (weakly) below efficient for all $\theta \in \supp \ F$. Then,
\[
\int \pi_Q(\theta) F_0(\dd \theta) \geq \int \pi_Q(\theta) F(\dd \theta).
\]
Moreover, the inequality is strict whenever $F_{0-}(\theta) \neq F(\theta)$ holds either at some $\theta \in \{\thmin,\thmax\}$ or at some $\theta$ at which $\DMprice$ is strictly increasing.
\end{lemma}
\begin{proof}
    Let $\tilde{\Theta}_1$ be the set of $\theta$ at which $\DMprice$ is strictly increasing, and $\tilde{\Theta} = \tilde{\Theta}_1 \cup \{\thmin,\thmax\}$. By Lemma~\ref{lem: finite support increasing p},  $\tilde{\Theta} \subseteq \supp \ F_0$, and so we can order its elements, $\tilde{\Theta} = \{\tilde\theta_1,\ldots,\tilde\theta_K\}$, where $\theta_i < \theta_j$ for all $i<j$. Because $Q$ is $F$-CC and $\costcoef=0$, $\pi_Q$ is constant on $[\tilde\theta_i,\tilde{\theta}_{i+1})$ (that $\tilde{\theta}_i$ is in the interval follows from $Q$ jumping towards efficiency and quality being underprovided). Therefore, for every $G \in \Signal$,
    \[
    \int \pi_Q(\theta) G(\dd \theta) 
    = \sum_{i} \pi_Q(\tilde{\theta}_i)[G_{-}(\tilde{\theta}_{i+1}) - G_{-}(\tilde{\theta}_i)] = \sum_{i=2}^{K}[\pi_{Q}(\tilde{\theta}_{i-1}) - \pi_{Q}(\tilde{\theta}_{i})]G_{-}(\tilde{\theta}_i) + \pi_{Q}(\tilde{\theta}_{K}),
    \]
    where $G_{-}(\tilde{\theta}_{K+1}):=1$ (recall $\tilde{\theta}_1 = \thmin$ and $\tilde{\theta}_K = \thmax$). Therefore,
    \[
    \int \pi_Q(\theta) (F-F_0)(\dd \theta) 
    = \sum_{i=2}^{K}[\pi_{Q}(\tilde{\theta}_{i-1}) - \pi_{Q}(\tilde{\theta}_{i})][F_{-}(\tilde{\theta}_i) -  F_{0-}(\tilde{\theta}_i)].
    \]
    Now, since $I_F(\tilde{\theta}_i)=0$ for all $i$, the logic of \cite{ravid2022learning}, Lemma~3 implies $F_{-}(\tilde{\theta}_i) \geq F_{0-}(\tilde{\theta}_i)$.

    Therefore, showing $\pi_{Q}(\tilde{\theta}_i)< \pi_{Q}(\tilde{\theta}_j)$ for all $i<j$ is sufficient for proving the lemma. We prove this inequality by showing it holds for $j=i+1$. Since $Q$ is constant on $[\tilde{\theta}_i,\tilde{\theta}_{i+1})$, we have $Q_{-}(\tilde{\theta}_{i+1}) = Q(\tilde{\theta}_{i})$ and $T_{Q-}(\tilde\theta_{i+1})=T_{Q}(\tilde{\theta}_{i})$. We can therefore derive the following inequality chain,
    \[
    \pi_Q(\tilde{\theta}_{i+1}) = \Pi(Q(\tilde{\theta}_{i+1}),T_Q(\tilde{\theta}_{i+1}))> \Pi(Q(\tilde{\theta}_{i}),T_Q(\tilde{\theta}_{i})) = \pi_Q(\tilde{\theta}_i),
    \]
    where the inequality follows from $Q$ jumping towards efficiency. The lemma follows.
\end{proof}

In what follows, let $Q_0\in \mathcal{Q}$ be the unique optimal allocation when the buyer is restricted to full learning. This allocation is defined by two properties: it is constant around every $\theta \notin \supp F_0$, and has $Q_0(\theta_i) = (\qcost')^{-1}(v_i)$ for all $i$, where $v_i$ is the virtual value associated with type $\theta_i$ (see equation \eqref{eq: virtual values}).
 
\begin{lemma}\label{lem: finite support full learning is unique optimum}
   Suppose Condition~\ref{as: finitesupport} holds and $\costcoef=0$. Then $(Q_0,F_0)$ is the unique seller optimal outcome.
\end{lemma}
\begin{proof}
    Note that $Q_0$ is $F_0$-CC. Moreover, by standard arguments, $Q_0$ is the unique allocation in $\mathcal{Q}$ that maximizes $\int \pi_Q(\theta) F_0 (\dd \theta)$. Take the seller optimal CC $(Q,F)$. By Theorem~\ref{thm: Inefficiency}, $Q$ is below efficient for all $\theta \in \supp \ F$. Therefore, by Lemma~\ref{lem: finite supp, free, full learning better}, every CC $(Q,F)$ has
    \[
    \int \pi_{Q}(\theta) F(\dd \theta) \leq \int \pi_{Q}(\theta) F_0(\dd \theta) \leq \int \pi_{Q_0}(\theta) F_0 (\dd \theta),
    \]
    where the second inequality is strict whenever $Q\neq Q_0$. 
    
    Thus, it remains to show that $F_0$ is the unique maximizer of $\int \pi_{Q_0}(\theta) F(\dd \theta)$ across all $F \in \Signals$ for which $Q_0$ is $F$-CC. For this purpose, note that every $\theta \in \Theta$ has $Q_0(\theta) = Q_0(\theta_i)$ for some $\theta_i \leq \theta$. Since the efficient quality is increasing in type and $Q_0$ provides quality weakly below efficiency to all types in $\supp \ F_0$, it follows $Q_0$ is below efficient for all types. We can therefore apply Lemma~\ref{lem: finite supp, free, full learning better} to get that every $F$ for which $Q_0$ is $F$-CC has $\int \pi_{Q_0}(\theta) F(\dd \theta) < \int \pi_{Q_0}(\theta) F_0(\dd \theta)$ whenever $F_{-}(\theta) \neq F_{0-}(\theta)$ for some $\theta \in \supp \ F_0$. Moreover, since $\DMprice_{Q_0}$ is strictly increasing at every $\theta \in \{\theta_2,\ldots,\theta_L\}$, $Q_0$ is CC for some $F \in \Signals$ if and only if $I_F(\theta) = 0$ for all $\theta \in \supp \ F_0$. Next, we argue that the unique $F \in \Signals$ for which this equality holds and which has $F_{-}(\theta) = F_{0-}(\theta)$ for all $\theta \in \supp \ F_0$ is $F_0$. 

    Towards this goal, take any $i=1,\ldots,L-1$, and note that every $\theta \in [\theta_i,\theta_{i+1})$ has 
    $F(\theta) \leq F_{-}(\theta_{i+1}) = F_{0-}(\theta_{i+1}) = F_{0}(\theta)$. In other words, $F \leq F_{0}$---that is, $F$ first order stochastically dominates $F_0$. Since $F$ is also a mean-preserving contraction of $F_0$ (since $F \in \Signals$), we must have $F=F_0$, as required. 
\end{proof}

\begin{lemma}\label{lem: finite support convergence to full learning}
   Suppose Condition~\ref{as: finitesupport} holds. For every $\epsilon>0$, a $\costcoef_\epsilon>0$ exists such that, for every CC $(Q,F)$ that is seller optimal for $\costcoef < \costcoef_\epsilon$ one has $|Q(\theta) - Q_0(\theta)|< \epsilon$ for all $\theta$, and $|F(\theta) - F_0(\theta)|<\epsilon$ for every $\theta$ at which $F_0$ is continuous. 
\end{lemma}
\begin{proof}
    We prove the lemma by way of contradiction. If the lemma is false, one can find a sequence $(\costcoef_n,Q_n,F_n)_{n\in \mathbb{N}}$ such that $\costcoef_n\rightarrow 0$, $(Q_n,F_n)$ is a seller optimal CC outcome for $\costcoef_n$, and either $Q_n$ does not uniformly converge to $Q_0$ or $F_n$ does not converge in distribution to $F_0$. We obtain a contradiction by arguing that every subsequence $(\costcoef_n,Q_n,F_n)_{n\in N}$ for some countably infinite $N\subseteq \mathbb{N}$ admits a further subsequence $(\costcoef_n,Q_n,F_n)_{n\in K}, \ K\subseteq N$ where both convergences hold. We make this argument by passing to an appropriately convergent subsubsequence and showing the limit must be an optimal CC outcome at $\costcoef=0$. Lemma~\ref{lem: finite support full learning is unique optimum} then implies the limit is $(Q_0,F_0)$. 

    We begin by finding an appropriately convergent subsubsequence. Let $\DMprice_n:=\DMprice_{Q_n}$ be the marginal price function associated with $Q_n$ and $\costcoef_n$. Note $[-\costcoef_n\icost'(\thmin_{F_n}),\bar{q}-\icost'(\thmax_{F_n})] \subseteq [0,\bar{q}]$. Thus, by Helly's Selection Theorem, we can choose the subsubsequence such that $\DMprice_n$ converges pointwise to some $\DMprice_{\infty}$, and $F_n$ converges in distribution to some $F_\infty$. By Lemma~\ref{lem: finite support increasing p}, each $\DMprice_n$ is completely determined by its values on $\supp \ F_0$, which is finite. It follows $\DMprice_n$ converges to $\DMprice_\infty$ uniformly. Thus, as $\icost'$ is bounded, $\DMprice_n + \costcoef_n\icost'$ also converges uniformly, and so $Q_n=\min\{(\DMprice_n + \costcoef_n\icost')_{+},\bar{q}\}$ also converges uniformly to the CC allocation (for $\costcoef=0$) defined by $\DMprice_\infty$, $Q_\infty$. Note this uniform convergence also implies that $\pi_{Q_n}$ uniformly converges to $\pi_{Q_\infty}$.

    Next we show that the limit outcome $(Q_\infty,F_\infty)$ is an CC outcome for $\costcoef=0$. We achieve this goal by arguing that $\DMprice_\infty$ is an $F_\infty$-marginal price. For this purpose, note first that $\DMprice_n(\Theta) \subseteq [-\costcoef_n\icost'(\thmin_{F_n}),\bar{q} -\costcoef_n\icost'(\thmax_{F_n})] \subseteq [-0 \icost'(\thmin_{F_\infty}),\bar{q} -0\icost'(\thmax_{F_\infty})]$. Second, observe Lemma~\ref{lem: finite support increasing p} means that $\DMprice_\infty$ strictly increases at some $\theta$ only if $\theta \in \supp \ F_0$, and $\DMprice_{n}$ is strictly increasing at $\theta$ for all sufficiently large $n$. Consequently, $\DMprice_\infty$ strictly increases at $\theta$ only if $I_{F_n}(\theta) = 0$ holds for all large $n$, and so $I_{F_\infty}(\theta)=0$ must hold as well. 
    
    Now we prove a helpful intermediate step by showing that $Q_\infty$ jumps towards efficiency. This property follows from noting that $Q_\infty$ can only have jumps on $\supp \ F_0$, which is finite. Moreover, if $Q_\infty$ jumps at $\theta \in \supp \ F_0$, then $Q_n$ must jump at $\theta$ as well for all sufficiently large $n$. Therefore, $\DMprice_n$ is strictly increasing at each $\theta$ where $Q_\infty$ jumps. By Lemma~\ref{lem: allocation perturbations summary}, $Q_n(\theta)$ must be below efficient, and so $Q_\infty(\theta)$ must be weakly below efficient, too. Hence, $Q_\infty$ jumps towards efficiency if and only if it is right continuous at every $\theta \in \supp \ F_0$. All that remains is to observe that $Q_n$ is increasing, right continuous, and constant on $[\theta_i,\theta_{i+1})$ for every $i =1,\ldots,L-1$. We conclude $Q_\infty$ jumps towards efficiency. 

    We now claim $(Q_\infty,F_\infty)$ is seller optimal at $\costcoef=0$. Note this claim completes the proof, since then $(Q_\infty,F_\infty)=(Q_0,F_0)$ holds by Lemma~\ref{lem: finite support full learning is unique optimum}. Let $\tilde Q_{n}$ be the $F_0$-CC allocation for $\costcoef_n$ with the property that $\tilde Q_{n}(\theta) = Q_0(\theta)$ for all $\theta \in \supp \ F_0$ (such an allocation exists whenever $\costcoef$ is sufficiently small). To define this allocation, define $\tilde\DMprice_n:\Theta \rightarrow [0,\bar{q}]$ by setting it
    $\tilde\DMprice_n(\theta) = Q_0(\theta_i) - \costcoef_n\icost'(\theta_i)$ for every $\theta \in [\theta_i,\theta_{i+1})$, $i=1,\ldots ,L-1$. Observe $\tilde Q_n$ converges to $Q_0$, and so $\pi_{\tilde{Q}_n}$ converges pointwise to $\pi_{\tilde{Q}_0}$. By the Bounded Convergence Theorem, $\int \pi_{\tilde{Q}_n}(\theta) F_0 (\dd\theta)\rightarrow \int \pi_{Q_0}(\theta) F_0 (\dd\theta)$. Together with the above observations, we obtain that, for every $\epsilon>0$, the following inequality chain holds for all sufficiently large $n \in K$:
    \[
    \begin{split}
    \int \pi_{Q_0}(\theta) F_0 (\dd\theta) - \frac{\epsilon}{2} 
    \leq \int \pi_{\tilde{Q}_n}(\theta) F_0 (\dd\theta)
     & \leq \int \pi_{Q_n}(\theta) F_n(\dd \theta) 
    \\ & \leq \int \pi_{Q_\infty}(\theta) F_n (\dd \theta) + \frac{\epsilon}{3} 
    \\ & \leq \int \pi_{Q_\infty}(\theta) F_\infty (\dd \theta) + \frac{2\epsilon}{3},
    \end{split}
    \]
    where the first inequality is due to $\pi_{\tilde{Q}_n}$ converging pointwise to $\pi_{Q_0}$, the second inequality from $(Q_n,F_n)$ being seller optimal for $\costcoef_n$ and $\tilde{Q}_n$ being $F_n$-CC for $\costcoef_n$, the third from uniform convergence of $\pi_{Q_n}$ to $\pi_{Q_\infty}$, and the last inequality from $F_n$ converging to $F_\infty$ and $\pi_{Q_\infty}$ being upper semicontinuous (due to $Q_\infty$ jumping towards efficiency, see Lemma~\ref{lem: jumping towards effiency yields usc objective}). Since $\epsilon>0$ is arbitrary, it follows the seller's profit from $(Q_\infty,F_\infty)$ is the same as from $(Q_0,F_0)$. Since $(Q_0,F_0)$ is optimal at $\costcoef=0$ (Lemma~\ref{lem: finite support full learning is unique optimum}, and $(Q_\infty,F_\infty)$ is IC for $\costcoef=0$ (because it is CC), $(Q_\infty,F_\infty)$ must be seller optimal as well. Thus, $(Q_\infty,F_\infty)$ must equal the unique seller optimal outcome at $\costcoef=0$, which is $(Q_0,F_0)$. 
\end{proof}

\begin{lemma} \label{lem:F is F0 except at the bottom}
    Suppose Condition~\ref{as: finitesupport} holds. Then there exists $\tilde{\costcoef}>0$ such that for all $\costcoef \leq \tilde{\costcoef}$, $\supp F = \{\tilde{\theta}, \theta_2, ...,\theta_L \}$, for some $\tilde{\theta} \leq \theta_2$, and $I_F(\theta) = 0$ for all $\theta \geq \theta_2$.
\end{lemma}

\begin{proof}
    Fix some $0<\delta<\min_{i \geq 2}\{Q_0(\theta_i) - Q_0(\theta_{i-1})\}$, and any $\frac{\min_{i \geq 2}\{Q_0(\theta_i) - Q_0(\theta_{i-1})\}-\delta}{3}>\epsilon>0$, and consider $\costcoef \leq \costcoef_\epsilon$, where $\costcoef_\epsilon$ is defined as in Lemma \ref{lem: finite support convergence to full learning}. First, we prove that $\theta_i \in \supp F$ for all $i \geq 2$. By  Lemma \ref{lem: finite support convergence to full learning}, we know that $Q(\theta_{i-1}) \leq Q_0(\theta_{i-1}) + \epsilon$, and $Q(\theta_i) \geq Q_0(\theta_i) - \epsilon$, and thus:

    $$Q(\theta_i) - Q(\theta_{i-1}) \geq Q_0(\theta_i) - Q_0(\theta_{i-1}) - 2 \epsilon $$
    
    By F-CC:

    $$Q(\theta_i) - Q(\theta_{i-1}) = p(\theta_i) - p(\theta_{i-1})   + \costcoef \left( \icost'(\theta_i) - \icost'(\theta_{i-1}) \right).  $$

By choosing $\costcoef \leq \frac{\epsilon}{\icost'(\bar{\theta}) - \icost'(\underline{\theta})}$, and putting these two observations together we get:

$$ p(\theta_i) - p(\theta_{i-1}) \geq  Q_0(\theta_i) - Q_0(\theta_{i-1}) - 3\epsilon > \delta>0, $$

implying, through Lemma \ref{lem: finite support increasing p}, that $\DMprice$ jumps at $\theta_i$. We now use this observation combined with Lemma~\ref{lem: ConvCav}-\eqref{lem: ConvCav:Cav} to show $\theta \notin\supp F$ for all $\theta \in (\theta_{i-1},\theta_i)$. 

To that goal, observe that because $\DMprice$ jumps at $\theta_{i-1}$ and at $\theta_i$, those are $F$-separating: $I_F(\theta_{i-1})=I_F(\theta_i) = 0$. If $F(\theta_{i-1})=F_0(\theta_{i-1})$, then convexity and nonnegativity of $I_F$ imply that $I_F = I_{F_0}$ for all $\theta \in [\theta_{i-1},\theta_i)$, implying $F=F_0$ in that interval, and there is nothing to prove. Therefore we may assume $F(\theta_{i-1})<F_0(\theta_{i-1})$ by nonnegativity of $I_F$. If, at any point $\theta \in (\theta_{i-1},\theta_i)$, $I_F(\theta) = 0$, Lemma~\ref{lem: finite support binding MPS} implies again $I_F([\theta_{i-1},\theta_i])=0$, and there is nothing to prove. Therefore, we may assume $I_F(\theta)>0$ for all $\theta \in (\theta_{i-1},\theta_i)$, which implies, by convexity and $I_F(\theta_i)=0$ that $F_{-}(\theta_i) > F_{0-}(\theta_i)=F_0(\theta_{i-1}).$ We therefore have $\theta_{i-1}<\theta_i$ satisfying the assumptions of Lemma~\ref{lem: ConvCav}-\eqref{lem: ConvCav:Cav}. Invoking that lemma, take any $\theta \in (\theta_{i-1},\theta_i)$, and $\alpha \in (0,1)$ such that $\theta = \alpha \theta_{i-1}+(1-\alpha)\theta_i$.

Since $\DMprice$ is constant on $[\theta_{i-1},\theta_i)$, the profit at $\theta$ satisfies:

$$\pi(\theta) = \pi(\theta_{i-1}) + \costcoef\int^{\theta}_{\theta_{i-1}} (x - \qcost'(Q(x)) \icost''(x) dx \leq \pi(\tilde{\theta}) + \costcoef \max_{\theta \in [\theta_{i-1},\theta_i]} \{ \icost''(\theta) \} (\theta - \theta_{i-1}).   $$

Moreover, 

\begin{equation*}
\begin{split}
\pi(\theta_{i}) = \pi(\theta_{i} -) + \theta_i \left( Q(\theta_i) - Q(\theta_i -)\right) + \qcost\left(Q(\theta_i-) \right) - \qcost\left(Q(\theta_i)\right) \\  \geq \pi(\theta_i -) + \left(\theta_i-\qcost'(Q(\theta_i))\right) \left(p(\theta_i) - p(\theta_{i-1}) \right) \\ \geq \pi(\theta_i -) + \left(\theta_i-\qcost'(Q(\theta_i))\right) \delta \\ = \pi(\theta_{i-1}) + \costcoef\int^{\theta_i}_{\theta_{i-1}} (x - \qcost'(Q(x)) \icost''(x) dx+ \left(\theta_i-\qcost'(Q(\theta_i))\right) \delta ,     
\end{split}
\end{equation*}

where the first inequality used convexity of $\qcost$, and the second inequality used the fact that, at any optimum, $\theta_2 \geq \qcost'(Q(\theta_2))$. The last equality used the formula for $\pi(\theta)$ given $p$ constant in the relevant interval. We then have, for sufficiently low $\costcoef$: 

\begin{equation*}
\begin{split}
    \pi(\alpha \theta_{i-1}+(1-\alpha) \theta_i) \leq \pi(\theta_{i-1}) + \costcoef \max_{\theta \in [\theta_{i-1},\theta_i]} \{ \icost''(\theta) \} (\theta - \theta_{i-1}) \\ <\pi(\theta_{i-1}) + (1-\alpha) \left(\costcoef\int^{\theta_i}_{\theta_{i-1}} (x - \qcost'(Q(x)) \icost''(x) dx+ \left(\theta_i-\qcost'(Q(\theta_i))\right) \delta\right) \\ \leq \alpha \pi(\theta_{i-1}) + (1-\alpha) \pi(\theta_i),
\end{split}
\end{equation*}

where the first and third inequality use our previous observations about $\pi(\theta)$ and $\pi(\theta_{i})$. By the counterpositive of Lemma~\ref{lem: ConvCav}-\eqref{lem: ConvCav:Cav}, $\theta \notin \supp F$. Therefore, $I_F(\theta_{i-1})=I_F(\theta_i)=0$, and $F(\theta) = F(\theta_{i-1})$ for all $\theta \in (\theta_{i-1},\theta_i)$, which implies $F(\theta_{i-1})=F_0(\theta_{i-1})$. This concludes that, for all $\theta \geq \theta_2$, $F(\theta) = F_0(\theta).$

It remains to show that there is exactly one point $\tilde{\theta} \leq \theta_2$, with $\tilde{\theta} \in \supp F$. For that, consider any $F \in \mathcal{I}$ such that $F(\theta)=F_0(\theta)$ for all $\theta \geq \theta_2$, and an F-CC allocation $Q$. Let $\underline{\theta}_Q :=\tilde{\theta}$. We shall find a $\tilde{\costcoef}$ such that, for all $\costcoef \leq \tilde{\costcoef}$, the principal prefers to concentrate all the support in $\{\tilde{\theta}, \theta_2\}$, regardless of what $F$ and $\tilde{\theta}$ are. Recall that $p$ must be constant in $(\theta_1,\theta_2)$. Therefore, notice that, for any type $\theta \in [\tilde{\theta}, \theta_2)$, the profits obtained by the seller from that type are:

$$\pi(\theta) = \pi(\tilde{\theta}) + \costcoef\int^{\theta}_{\tilde{\theta}} (x - \qcost'(Q(x)) \icost''(x) dx \leq \pi(\tilde{\theta}) + \costcoef \max_{\theta \in [\underline{\theta},\theta_2]} \{ \icost''(\theta) \} (\theta - \tilde{\theta}),   $$

Moreover:

\begin{equation*}
\begin{split}
\pi(\theta_2) = \pi(\theta_2 -) + \theta_2 \left( Q(\theta_2) - Q(\theta_2 -)\right) + \qcost\left(Q(\theta_2-) \right) - \qcost\left(Q(\theta_2)\right) \\  \geq \pi(\theta_2 -) + \left(\theta_2-\qcost'(Q(\theta_2))\right) \left(p(\theta_2) - p(\theta_1) \right) \\ \geq \pi(\theta_2 -) + \left(\theta_2-\qcost'(Q(\theta_2))\right) \delta,     
\end{split}
\end{equation*}
where the first inequality used convexity of $\qcost$, and the second inequality used the fact that, at any optimum, $\theta_2 \geq \qcost'(Q(\theta_2))$. Additionally, we must have: 

\begin{equation*}
    \begin{split}
        \theta_2 - \qcost'\left(Q(\theta_2)\right) = \underbrace{\theta_2 - \qcost'\left(Q_0(\theta_2)\right)}_{v_2} + \qcost'\left(Q_0(\theta_2)\right) - \qcost'\left(Q(\theta_2)\right) \\ \geq  v_2 + \min_{x \leq Q_0 (\theta_2)+\epsilon}\{\qcost''(x)\} \epsilon,
    \end{split}
\end{equation*}

where the inequality follows from the mean-value theorem.

Now, define $\tilde{\costcoef} = \frac{\min_{x \leq Q_0 (\theta_2)+\epsilon}\{\qcost''(x)\}}{\qcost'\left(Q_0(\theta_2) + \epsilon \right) \left(\icost'(\theta_2) - \icost'(\underline{\theta}) \right)}$, and let $\costcoef \leq \tilde{\costcoef}$. Then:

\begin{equation*}
    \begin{split}
\pi(\theta_2) - \pi(\tilde{\theta}) \geq \costcoef \int_{\tilde{\theta}}^{\theta_2} \left(x-\qcost'(Q(x)) \right) \icost''(x) dx + v_2 \delta + \min_{x \leq Q_0(\theta_2)+\epsilon}\left\{ \qcost''(x) \right\} \epsilon \delta \\ > \costcoef\int_{\underline{\theta}}^{\theta_2} \left(-\qcost'(Q(x)) \right) \icost''(x) dx + v_2 \delta + \min_{x \leq Q_0(\theta_2)+\epsilon}\left\{ \qcost''(x) \right\} \epsilon \delta \\ \geq v_2 \delta > 0,        
    \end{split}
\end{equation*}

where the first inequality simply used the bounds derived for $\pi(\theta_2)$ above, the second inequality makes the integral term as negative as possible and the third uses the fact that $\costcoef \leq \tilde{\costcoef}$. We can now conclude that the line segment connecting $\pi(\theta_2)$ and $\pi(\tilde{\theta})$ is everywhere above $\pi$. Indeed, for all $\theta \in [\tilde{\theta},\theta_2)$:

\begin{equation*}
\begin{split}
    \pi(\tilde{\theta}) + \frac{\pi(\theta_2) - \pi(\tilde{\theta})}{\theta_2 - \tilde{\theta}} (\theta - \tilde{\theta}) \\ \geq 
 \pi(\tilde{\theta}) + \frac{v_2 \delta}{\theta_2 - \underline{\theta}} (\theta-\tilde{\theta}) \geq \pi(\tilde{\theta}) + \costcoef \max_{\theta \in [\underline{\theta},\theta_2]} \{ \icost''(\theta) \} (\theta - \tilde{\theta})\geq \pi(\theta),    
\end{split}  
\end{equation*}

for $\costcoef \leq \tilde{\costcoef}$. Therefore, the seller always prefer to concentrate any mass available for $[\theta_1,\theta_2]$ in  $\{\tilde{\theta}, \theta_2\}$. 
\end{proof}

\begin{lemma} \label{lem: finite support full learning for low cost}
   Suppose Condition~\ref{as: finitesupport} holds. There exists $\underline{\costcoef}>0$ such that for $\costcoef \leq \underline{\costcoef}$, $F = F_0$. 
\end{lemma}

\begin{proof}
By the previous result, $\supp F = \{\tilde\theta, \theta_2,...,\theta_L \}$, with $\tilde{\theta}<\theta_2$, $F(\theta)=F_0(\theta)$ for $\theta \geq \theta_2$, and $I_F(\theta_2) = I_{F_0}(\theta_2)$. We assume $\tilde{\theta}>\thmin$ and show a contradiction.  By Lemma~\ref{lem: finite support convergence to full learning}, for any $\varepsilon$ there is small enough $\costcoef_\varepsilon$ such that $\tilde{\theta}-\underline{\theta} < \varepsilon$ for $\costcoef<\costcoef_\varepsilon$. 

Let $\mu_1 : = \int\theta F\left(d\theta | \theta \in [\theta_1,\theta_2]\right)$. Because $F$ is a mean-preserving-contraction of $F_0$, $I_F(\theta_1)=I_F(\theta_2)=0$, and, by the proof of Lemma~\ref{lem:F is F0 except at the bottom}, $F_0 (\theta) = F(\theta)$ for $\theta \geq \theta_2$, we must have: 
\[
F(\tilde{\theta}) \tilde{\theta} + \left(F_0 (\theta_2) - F(\tilde{\theta})\right) \theta_2= \mu_1 F_0(\theta_2).
\]
Thus, $\frac{d F(\tilde{\theta})}{d\tilde{\theta}} = F_0(\theta_2)\frac{\theta_2 - \mu_1}{(\theta_2 - \tilde\theta)^2},$ which is positive and bounded away from zero since $\tilde{\theta} \in [\underline{\theta}, \underline{\theta}+\varepsilon]$.

Lemma~\ref{lem: allocation perturbations summary} Part~\eqref{lem: allocation perturbations summary - low mc and strictly increasing} implies that, for $i>2$, $Q_\costcoef(\theta_i) = Q_0 (\theta_i)$, where $Q_0 (\theta_i)$ is the optimal screening quality for type $\theta_i$ under $F_0$. Lemma~\ref{lem: allocation perturbations summary} Parts~\eqref{lem: allocation perturbations summary - high mc} and \eqref{lem: allocation perturbations summary - low mc and strictly increasing} imply:

\begin{equation} \label{eq: F foc}
\begin{split}
   \theta_2 = \int\qcost'(Q_\gamma(\theta)) F\left( d\theta|\theta \geq \theta_2\right) = \frac{F_0(\theta_2) - F(\tilde{\theta})}{1-F(\tilde{\theta})} \qcost'(Q_\gamma(\theta_2)) + \\  \frac{1-F_0(\theta_2)}{1-F(\tilde{\theta})} \int\qcost'(Q_\gamma(\theta)) F\left( d\theta|\theta > \theta_2\right) \\ = \frac{F_0(\theta_2) - F(\tilde{\theta})}{1-F(\tilde{\theta})} \qcost'(Q_\gamma(\theta_2)) + \frac{1-F_0(\theta_2)}{1-F(\tilde{\theta})} \theta_3,
    \end{split}
\end{equation}

where the second equality follows from applying  Lemma~\ref{lem: allocation perturbations summary} Parts~\eqref{lem: allocation perturbations summary - high mc} and \eqref{lem: allocation perturbations summary - low mc and strictly increasing} to $\theta_3$. Reorganizing the equation above obtains:

\begin{equation*}
     \qcost'(Q_\gamma(\theta_2)) = \theta_2 - \frac{1 - F_0 (\theta_2)}{F_0 (\theta_2) - F(\tilde{\theta})} \left(\theta_3 - \theta_2 \right),
\end{equation*}

whenever the right-hand-side of the equation above is positive. By choosing $\varepsilon$ small enough, we can make sure that this is the case since, by Condition~\ref{as: finitesupport} the right-hand-side is positive for $\tilde{\theta} = \underline{\theta}$, and it is continuous in $\tilde{\theta}$.

The above, in turn, implies (i) that $Q_\gamma (\theta_2)$ is the optimal screening quantity for $\theta_2$ under $F$, and (ii) that $Q_\gamma (\theta_2)$ is decreasing with $\tilde{\theta}$, with $Q_\gamma(\theta_2) = Q_0 (\theta_2)$, when $\tilde{\theta}=\thmin$, and it is bounded away from zero by choosing $\varepsilon$ small enough. Define $Q_2 (\tilde{\theta}) := Q_\gamma(\theta_2)$ to emphasize the dependence on $\tilde{\theta}.$

Let $S_i$ be the optimal screening surplus obtained from an agent with type $\theta_i$---that is, $S_i = \theta_i Q_0 (\theta_i) - \qcost(Q_0 (\theta_i))$. Noting that $Q_\gamma(\tilde{\theta})=0$, and therefore the seller makes no profits from $\tilde{\theta}$, the seller's profits are:

\begin{equation*}
\begin{split}
      (F(\theta_2) - F(\tilde{\theta})) \left( \theta_2 Q_2 (\tilde{\theta}) - \qcost(Q_2(\tilde{\theta}))  - \costcoef \left( \icost(\theta_2) - \icost(\tilde{\theta}) - \icost'(\tilde\theta) \left(\theta_{2}-\tilde{\theta} \right)\right) \right)  + \\  \sum_{i>2} \left(F(\theta_i) - F(\theta_{i-1})\right) \biggr\{ S_i - Q_2(\tilde{\theta}) (\theta_3 - \theta_2) + \sum_{2<j<L} q_j (\theta_{j+1}-\theta_j)  \\ -   \costcoef \left(\icost(\theta_i) - \icost(\tilde{\theta}) - \sum_{j<L} \icost'(\theta_j) \left(\theta_{j+1}-\theta_j \right)  \right)\biggr\},
\end{split}
\end{equation*}

where the first line describes the profits obtained from type $\theta_2$. Regrouping the terms that depend on $\tilde{\theta}$ and dropping the terms that do not, we obtain:

\begin{equation*}
\begin{split}
      (F(\theta_2) - F(\tilde{\theta})) \left( \theta_2 Q_2(\tilde{\theta}) - \qcost(Q_2(\tilde{\theta})) \right) - (1-F(\theta_2)) Q_2 (\tilde{\theta}) (\theta_3-\theta_2)  \\ -  \sum_{i>1} \left(F(\theta_i) - F(\theta_{i-1})\right)  \costcoef \left(\icost(\theta_i) - \icost(\tilde{\theta}) - \icost'(\tilde{\theta}) \left(\theta_2-\tilde{\theta} \right) \right).
\end{split}
\end{equation*}

We can then take derivatives on the expression above with respect to $\tilde{\theta}$ to find:

\begin{equation*}
\begin{split}
 \left(F(\theta_2) - F(\tilde{\theta})\right)  \underbrace{\left( \theta_2 - \qcost'(Q_2(\tilde{\theta}) - \sum_{i>2} \left(F(\theta_i) - F(\theta_{i-1}) (\theta_3 - \theta_2) \right) \right)}_{=0} \frac{d Q_2 (\tilde{\theta})}{d \tilde{\theta}}  \\  - \frac{d F (\tilde{\theta})}{d \tilde{\theta}} \left( \theta_2 Q_2 (\tilde{\theta)} - \qcost(Q_2(\tilde{\theta})) \right) + (1-F_0 (\tilde{\theta})) \costcoef \icost''(\tilde{\theta}), 
 \end{split}
\end{equation*}

where the first term equals zero because of \eqref{eq: F foc}. Now, the second term is negative and bounded away from zero due to $\frac{d F(\tilde\theta)}{d \tilde\theta}$ being positive and bounded away from zero, and due to $Q_2 (\tilde{\theta}) >0$. The third term can be made arbitrarily small for small $\costcoef$ since $\icost''$ is bounded. We then find $\underline{\costcoef}$ such that the expression is strictly decreasing in $\tilde{\theta}$, proving that it is optimal for the seller to choose $\tilde{\theta}=\underline{\theta}$ for sufficiently low $\costcoef$. 
\end{proof}

This completes the proof of Theorem~\ref{thm: low cost comparative statics}-\eqref{thm: low cost comparative statics - fully informative F}.
\end{proof}

\begin{proof}[\textbf{Proof of Theorem~\ref{thm: low cost comparative statics}}-\eqref{thm: low cost comparative statics - seller payoff}, \eqref{thm: low cost comparative statics - buyer payoff} and Corollary~\ref{cor: consumer surplus increasing for binary} ]

We now consider $\costcoef\leq \underline{\costcoef}$ which, by Part~\eqref{thm: low cost comparative statics - fully informative F}, implies $F_\costcoef=F_0$. By the argument in Step 1 of the proof of Part~\eqref{thm: low cost comparative statics - fully informative F}, $p(\theta)$ is constant around any $\theta \notin \supp F$. Moreover, by CC: $\costcoef p(\theta_i) = Q_0 (\theta_i) - \costcoef \icost'(\theta_i)$, for all $i=1,...,L$. We can then write the consumers' gross value as a function of his interim-type as:

\begin{equation}
\begin{split}
    \idu_\costcoef (\theta_i) = \costcoef \icost(\theta_i) - \costcoef \icost(\underline{\theta}) + \costcoef \int^\theta_{\underline{\theta}} p(x) dx  = \costcoef \left( \icost(\theta_i) - \icost(\underline{\theta}) + \sum_{j<i} p(\theta_j) (\theta_{j+1}-\theta_j) \right) \\ =  \sum_{j<i} q_j (\theta_{j+1}-\theta_j) + \costcoef \left( \icost(\theta_i) - \icost(\underline{\theta}) - \sum_{j<i} \icost'(\theta_j) (\theta_{j+1}-\theta_j) \right),
\end{split}
\end{equation}

where the first equality uses the fact that $p$ is constant outside the support of $F_0$. 

First, note that because the qualities in the contract are fixed for $\costcoef \leq \underline{\costcoef}$, the change in the seller's profit $U_S$ is the negative of the change in the expected gross consumer value, $\int V_\costcoef(\theta)F_0(d\theta)$. Furthermore, for $\costcoef<\costcoef'\leq \underline{\costcoef}$:

$$V_{\costcoef'} (\theta_i) - V_\costcoef (\theta_i) = (\costcoef' - \costcoef) \left( \icost(\theta_i) - \icost(\underline{\theta}) - \sum_{j<i} \icost'(\theta_j) (\theta_{j+1}-\theta_j) \right) = (\costcoef'-\costcoef)\left( \icost(\theta_i) - \tilde{\icost} (\theta_i) \right) > 0, $$

for all $i>1$. Therefore, gross surplus is increasing type by type for $\costcoef \leq \underline{\costcoef}$. Therefore, expected gross surplus is also increasing, which implies that profits are decreasing, proving part~\eqref{thm: low cost comparative statics - seller payoff} of the Theorem.

We now proceed to studying how consumer surplus,  $U_B:= \int \left[ V_\costcoef (\theta) - \costcoef \icost (\theta)\right] F_0(d\theta)$ , changes with $\costcoef$. Assuming again $\costcoef<\costcoef'\leq \underline{\costcoef}$ we have:

\begin{equation} \label{eq:csinequality}
\begin{split}
   \int \left[V_{\costcoef'} (\theta) - {\costcoef'} \icost(\theta) -  \left( V_\costcoef (\theta) - \costcoef \icost (\theta)\right) \right]F_0(d\theta) \\  = (\costcoef'-\costcoef) \sum^L_{i=1} \left(F_0(\theta_i)-F_0(\theta_{i-1}) \right) \left\{ \left( \icost(\theta_i) - \icost(\underline{\theta}) - \sum_{j<i} \icost'(\theta_j) (\theta_{j+1}-\theta_j) \right) - \icost(\theta_i) \right\} \\  = - (\costcoef'-\costcoef) \sum^L_{i=1} \left(F_0(\theta_i)-F_0(\theta_{i-1}) \right) \left(\icost(\underline{\theta}) + \sum_{j<i} \icost'(\theta_j) (\theta_{j+1}-\theta_j) \right) \\ = -(\costcoef' - \costcoef) \int \tilde{\icost}(\theta) F_0(d\theta), 
\end{split}
\end{equation}

proving the claim about consumer surplus. 

We finish by proving Corollary~\ref{cor: consumer surplus increasing for binary}. Assume $L=2$. In that case, $\tilde{\icost}$ is a linear function, and therefore:

\begin{equation*}
\begin{split}
    \int\left[V_{\costcoef'} (\theta) - {\costcoef'} \icost(\theta) -  \left( V_\costcoef (\theta) - \costcoef \icost (\theta)\right) \right]F_0(d\theta) \\ = - (\costcoef' - \costcoef) \int \tilde{\icost} (\theta) F_0(d\theta) = - (\costcoef' - \costcoef) \tilde{\icost}(\theta_0) \\ > - (\costcoef'-\costcoef) \icost(\theta_0) = 0,  
\end{split}
\end{equation*}
where the second equality follows from linearity of $\tilde{\icost}$, the inequality from $\tilde{\icost}(\theta) < \icost(\theta)$ for $\theta \neq \underline{\theta}$, and the last equality from $\icost(\theta_0)=0$.
\end{proof}

\subsubsection{Proof of Proposition~\ref{prop: sufficient distortion}}
\begin{proof}[Proof of Proposition~\ref{prop: sufficient distortion}-\eqref{prop: sufficient distortion - high costs}] Take $\costcoef \geq \bar{\costcoef}$ as defined in Theorem~\ref{thm: large cost comparative statics}. Part~\eqref{thm: large cost comparative statics - uninformative F} of the theorem says $\supp F = \{\theta_0\}$, so $\thmax \notin \supp F$, and $\theta_0=\thmin_F$. Moreover, $\thmin_F>\thmin_Q\geq \thmin$, since otherwise the seller's profits would be zero, which is suboptimal.

$$\theta_0 -\qcost'\left(Q^s(\theta_0)\right) = 0 < \theta_0 -  \thmin_Q = \theta_0- \int \qcost'\left(Q(\theta)\right) F(d\theta) , $$
where the first equality follows from efficiency of the screening contract when information is symmetric, the inequality follows from the observation that $\theta_0>\thmin_Q$, and the final equality from parts~\eqref{lem: allocation perturbations summary - high mc} and \eqref{lem: allocation perturbations summary - low mc and min-theta} of Lemma~\ref{lem: allocation perturbations summary}. By Theorem \ref{thm: Inefficiency}, there is strict distortion at $\thmax_{F^*}$.
\end{proof}

\begin{proof}[Proof of Proposition~\ref{prop: sufficient distortion}-\eqref{prop: sufficient distortion - low costs}] 
Take $\costcoef \leq \underline\costcoef$, as defined in Theorem~\ref{thm: low cost comparative statics}. Part~\eqref{thm: low cost comparative statics - fully informative F} of the Theorem says $F = F_0$, so $\thmax \in \supp F$. The proof of Part~\eqref{thm: low cost comparative statics - fully informative F} also shows that the outcome $Q = Q^s$, therefore efficiency is exactly the same as in the pure screening problem.
\end{proof}

\subsection{Proofs from Section \ref{section:Optimal Info}}

\subsubsection{Proof of Proposition~\ref{prop: binary signals}}
First, by Theorem~\ref{thm: FCC}, a signal $F$ can be made IC under some $Q$ if and only if the allocation $Q$ is $F$-almost surely equal to some $F$-CC allocation. Second, given an $F$-CC allocation $Q$, one must have $Q(\thmax_F) - Q(\thmin_F) \geq \icost'(\thmax_F) - \icost'(\thmin_F)$---see equation~\ref{eq: minimal differentiation}. Third, Theorem~\ref{thm: Inefficiency} implies the seller optimal outcome $(Q,F)$ never gives a quality above $\bar{q}_e$ to any realized type, meaning that $\bar{q}_e \geq Q(\thmax_F) - Q(\thmin_F)$. Therefore, no $F$ for which $\icost'(\thmax_F) - \icost'(\thmin_F) > \bar{q}_e$ can be part of a seller optimal outcome. 
Fourth, mean preserving spreads make the most extreme signal realizations more extreme; that is, $F \succeq G$ only if $\thmin_F \leq \thmin_G \leq \thmax_G \leq \thmax_F$---see Lemma 9 in \cite{ravid2022learning}. 
Because $\icost'$ and the efficient quality are both strictly increasing, we obtain that when learning costs satisfy the steep-slope condition, mean preserving spreads of cutoff signals must provide a quality that is above efficient, and hence cannot arise at the optimum. 
The fifth and final observation (which we prove below) is that any signal with more than two signal realizations is either a mean-preserving spread of a cutoff signal, or can be replaced with an optimal signal with binary support, thereby delivering a binary-menu outcome. 

More precisely, we now conclude the proof by showing that every signal with more than two realizations, 
either (a) separates some $\theta\in (\underline{\theta},\bar{\theta})$, and so is a 
mean-preserving spread of a cutoff signal; or (b) can be replaced with an IC signal that generates the same profit to the seller whose support is (at most) a binary signal. 

For case (a), let $\theta^*$ be $F$-separating. Let $\theta_{-}^*\coloneqq E_{F_0}[\theta\vert \theta<\theta^*]$ and $\theta_{+}^*\coloneqq E_{F_0}[\theta\vert \theta\geq \theta^*]$. Then the binary distribution which has support on $\{\theta^*_{-},\theta^*_{+}\}$ is a mean-preserving contraction of $F$ and is a cutoff signal.

For case (b), notice that by Lemma \ref{lem: ConvCav} and the upper-semicontinuity of $\pi_{Q^*}$, all points $\theta\in \mbox{supp}(F)$ lie along the same affine function. Consider now two cases. In the first case, $\theta_0 \in \supp \ F$. Then $\pi_{Q^*}(\theta_0)=E_F[\pi_{Q^*}(\theta)]$, and so $F$ and $\mathbf{1}_{[\theta_0,\infty)}$ (i.e, an atom at $\theta_0$) achieve the same value. In the second case, $\theta_0 \notin \supp \ F$. Let $\theta_{-}\coloneqq \sup\{\theta<\theta_0: \theta\in \supp (F)\}$ and $\theta_{+}\coloneqq \inf\{\theta>\theta_0: \theta\in \supp (F)\}$. Then there exists a mean-preserving contraction $G$ of $F$ which has support at $\{\theta_{-},\theta_{+}\}$, and preserves the expected profit of the seller. All that remains is to argue that $Q$ is cost-compensating for both $G$ and $\mathbf{1}_{[\theta_0,\infty)}$. For this purpose, note that $F$ has no separating types, and so every $F$-CC allocation is cost-compensating for any distribution whose support is contained in $[\thmin_F,\thmax_F]$---in particular, $Q$ is both $G$-CC and $\mathbf{1}_{[\theta_0,\infty)}$-CC. \hfill $\square$

\subsubsection{Proof of Proposition~\ref{prop: average efficiency}}

Recall that $F$ is at-most binary. Theorem~\ref{thm: Inefficiency} shows directly that the result is true if $\supp F=\{\theta_0\}$. So from now on assume $\supp F = \{\thmin_F,\thmax_F\}$, with $\thmin_F<\thmax_F$.

First, suppose $\thmin_F=\thmin_Q$. Indeed, recalling that the steep learning condition implies a strict distortion at the top, $Q(\thmax_F)<Q^s_F(\thmax_F)$. Moreover, because $\thmin_F=\thmin_Q$, $Q(\thmin_F) = 0 \leq Q^s_F(\thmin_F)$. Therefore:

$$\avgdist(Q,F) = \theta_0 - \int \qcost'\left(Q(\theta)\right) F(d\theta) > \theta_0 - \int \qcost'(Q^s_F(\theta)) F(d\theta), $$
where the strict inequality follows from one of the inequalities being strict, and for $F$ having positive mass of $\thmax_F$, since $\supp F = \{\thmin_F, \thmax_F \}$. This proves the result for $\thmin_F=\thmin_Q$.

Suppose now that $\thmin_Q<\thmin_F$.  In standard screening, we know that $\qcost'(Q^s_F(\thmax_F)) = \thmax_F$, and 

$$\qcost'\left(Q^s_F(\thmin_F)\right) = \max\left\{\thmin_F - \frac{1-F(\thmin_F)}{F(\thmin_F)} \left(\thmax_F - \thmin_F\right),0\right\}. $$

Therefore, it is easy to see that the wedge in standard screening satisfies:

$$\avgdist(Q^s_F,F) = \theta_0 - \max\{\thmin_F, \left(1-F(\thmin_F)\right) \thmax_F\}. $$

We can then obtain:
$$\int \qcost'\left(Q(\theta)\right) F(d\theta) \leq \thmin_Q < \thmin_F \leq \max\{\thmin_F,(1-F(\thmin_F))\thmax_F \} = \int \qcost'(Q^s_F(\theta)) F(d\theta),$$
where the first inequality follows from Lemma~\ref{lem: allocation perturbations summary}-\eqref{lem: allocation perturbations summary - low mc and min-theta}.

By multiplying the inequalities by minus one and adding $\theta_0$, we obtain the desired order of wedges. \hfill $\square$

\subsection{Extending Theorem~\ref{thm: Inefficiency} to Allow Efficient Exclusion}
In this section we explain the argument that extends Theorem~\ref{thm: Inefficiency} to the case where $\qcost'(0) \geq \thmin$.  Note that the argument establishing that $\thmax = \qcost'(Q(\thmax))$ whenever $\thmax \in \supp (F)$ easily extends to this case. Thus, all we need to do is argue that any $\theta\in \supp(F)$ such that $Q(\theta)>0$ and $\theta < \thmax$, $\theta > \qcost'(Q(\theta))$. For brevity, we provide the major steps only, and skip some of the details. So, suppose $(Q,F)$ is a seller optimal outcome in which $Q$ is $F$-ICC. Our goal is to argue that $\theta \in \supp(F)\setminus\{\thmax\}$ has $\qcost'(Q(\theta))\geq \theta$ only if $\theta = \thmin_Q$. Suppose then for a contradiction that another $\tilde\theta\in \supp(F)\setminus\{\thmin_Q,\thmax\}$ exists for which said inequality holds.

First, recall that (as shown in Section A.5.3) 
$\pi^\prime_Q$ is positive for all $\theta\in\mbox{supp}(F)$ 
on a given interval $(\theta_1,\theta_2)$ where $I_F(\theta)>0$ 
if $\kappa^\prime(Q(\theta))< \theta$, as by the envelope theorem, 
$$\pi^{\prime}_{Q}(\theta)=\theta Q^\prime(\theta)-\kappa^\prime(Q(\theta))Q^\prime(\theta)$$
(recall that an $F$-ICC $Q$ is differentiable whenever $I_F(\theta)>0$).
Conversely, if $\pi^\prime$ is negative, then $\kappa^\prime(Q(\theta))>\theta$.
In addition, if $p_Q$ is constant on some interval $(\theta_1,\theta_2)$, then one can apply an argument similar to that in Theorem~\ref{thm: Inefficiency}'s proof to show that Lemma~\ref{lem: ConvCav}-\eqref{lem: ConvCav:Conv} implies that $\pi^\prime_Q(\theta)\leq \pi^{\prime}_{Q}(\theta')$ for any $\theta<\theta'$ both in $\supp (F)$. Moreover, if $I_F$ is strictly positive over $(\theta_1,\theta_2)$, then a similar argument using Lemma~\ref{lem: ConvCav}-\eqref{lem: ConvCav:Cav} shows $\pi^\prime_Q(\theta)\geq \pi^{\prime}_{Q}(\theta')$, meaning $\pi^\prime_Q(\theta)= \pi^{\prime}_{Q}(\theta')$.

We now use the above to explain that 
$\DMprice_Q(\theta)=\DMprice_Q(\thmin_{Q})$ for any $\theta\in\supp(F)\setminus\{\thmax\}$ at which 
$\kappa^\prime(Q(\theta))\geq\theta$. Fix any $\tilde\theta\in \supp(F)$ such that $\kappa^\prime(Q(\tilde\theta))\geq\tilde\theta$. Suppose that $\DMprice_Q(\tilde\theta) > \DMprice_Q(\thmin_{Q})$. By Lemma~\ref{lem: allocation perturbations summary}-\eqref{lem: allocation perturbations summary - low mc and strictly increasing}, $\DMprice_Q$ must be constant around $\tilde\theta$ . Define 
$$\theta^\prime=\sup \{\theta \in \Theta:\,\DMprice_Q\mbox{ strictly increases at } \theta,\,\theta\leq\tilde\theta\}.$$
Observe $\DMprice_Q$ cannot be constant around $\theta^\prime$. Therefore, $\theta'<\tilde\theta$, and so $\kappa^\prime(Q(\theta^\prime))<\theta^\prime$ by Lemma~\ref{lem: allocation perturbations summary}-\eqref{lem: allocation perturbations summary - low mc and strictly increasing}. But then one can apply the argument in Theorem~\ref{thm: Inefficiency}'s proof to show that $\kappa^\prime(Q(\tilde\theta))<\tilde\theta$, a contradiction.

Next we establish that a $\theta^*$ exists with three properties. First, every $\theta \in \supp(F)$ has $\qcost'(Q(\theta)) \geq \theta$ only if $\theta \leq \theta^*$, and $\qcost'(Q(\theta)) \leq \theta$ only if $\theta \geq \theta^*$. Second, $I_{F}(\theta^*) = 0$. And third, $\DMprice_{Q-}(\theta^*)=\DMprice(\thmin_{Q})$. To do so, pick any $\tilde\theta \in \supp(F)\setminus\{\thmin_Q\}$ such that $\qcost(Q(\tilde\theta)) \geq \tilde\theta$. Let $\hat\theta = \sup\{\theta: \DMprice(\theta)=\DMprice(\thmin_Q)\}$. Note that $\hat\theta > \tilde\theta$. As explained earlier, $\pi^\prime_{Q}$ is increasing over $(\thmin_{Q},\hat\theta) \cap \supp(F)$. Since $\pi^\prime_{Q}$ has to be negative (positive) for quality to be above (below) efficient, we get that some $\theta^* \in [\thmin_Q,\hat\theta]$ exists such that for all $\theta \in \supp(F)\cap(\thmin_Q,\hat\theta)$, $\qcost'(Q(\theta)) \geq \theta$ only if $\theta \leq \theta^*$, and $\qcost'(Q(\theta)) \leq \theta$ only if $\theta \geq \theta^*$. 
We now explain that one can pick $\theta^*$ to satisfy $I_{F}(\theta^*)=0$. First, suppose that $\pi^{\prime}_{Q}(\theta)$ is constant at some value $x$ for all $\theta \in [\thmin_Q,\hat\theta]\cap \supp(F)$. Notice $x \leq 0$, because $\tilde\theta \in [\thmin_Q,\hat\theta]\cap \supp(F)$ and $\qcost'(Q(\tilde\theta))\geq \tilde\theta$. In fact, in this case $\qcost'(Q(\theta))\geq \theta$ for all $\theta \in [\thmin_Q,\hat\theta]\cap \supp(F)$, and so we can take $\theta^* =\hat\theta$. To get that $I_{F}(\theta^*)=0$ in this case, observe $\DMprice_Q$ must be strictly increasing at $\hat\theta$ whenever $\hat\theta\neq \thmax$. This concludes the constant $\pi^{'}_{Q}$ case. For the second case, suppose we have $\theta,\theta'\in [\thmin_Q,\hat\theta] \cap \supp(F)$ with $\theta<\theta'$ such that $\pi^{\prime}_{Q}(\theta)< \pi^{\prime}_{Q}(\theta')$, and $0 \in [\pi^{\prime}_{Q}(\theta),\pi^{\prime}_{Q}(\theta')]$. 
Note $\theta^* \in [\theta,\theta']$. Without loss, we can choose $\theta$ and $\theta'$ such that the sets $(\theta,\theta^*)\cap\supp(F)$ and $(\theta^*,\theta')\cap\supp(F)$ are both empty.
Notice $I_{F}$ cannot be strictly positive over $[\theta,\theta']$, because then one could apply the argument sketched in the second paragraph of this section to $(\theta-\epsilon,\theta'+\epsilon)$ for sufficiently small $\epsilon>0$ to get that $\pi^{\prime}_{Q}(\theta)=\pi^{\prime}_{Q}(\theta')$. Thus, $I_{F}$ must equal $0$ at some $\theta''\in [\theta,\theta']$, and we can choose $\theta^*=\theta''$.

We now claim that $\thmin_F=\thmin_Q$. If not, then since 
$Q(\bar{\theta}_F)<\bar{q}$, by Lemma~\ref{lem: allocation perturbations summary}-\eqref{lem: allocation perturbations summary - low mc and min-theta}, 
\[
\kappa^\prime(Q(\thmin_F))\leq \int \kappa^\prime(Q(\theta))dF(\theta)=\thmin_Q<\thmin_F
\]
contradicting the claim that $\thmin_F\leq\kappa^\prime(Q(\thmin_F))$.

We now show the seller can improve upon $(Q,F)$ by looking at 
\[
\Theta_{F-}\coloneqq\{\theta:\kappa^\prime(Q(\theta))>\theta\}
\]
and replacing $(Q,F)$ with $(\hat{Q},\hat{F})$, where, letting 
\[
\theta_{F-}=\frac{\int_{\Theta_{F-}} \theta dF(\theta)}{F_{-}(\theta^*)}
\]
we set 
\[
\hat{F}(\theta)=\begin{cases}
0, & \theta<\theta_{F-}\\
F_{-}(\theta^*), & \theta\in[\theta_{F-},\theta^*)\\
F(\theta), & \theta>\theta^*
\end{cases}
\]
and
\[
\hat{Q}(\theta)=\begin{cases}
0, & \theta\leq \theta_{F-}\\
c^\prime(\theta)-\icost^\prime(\theta_{F_{-}}), & \theta\in(\theta_{F_{-}},\theta^*)\\
Q(\theta), & \theta\geq \theta^* 
\end{cases}
\]
Notice that $\hat{Q}$ is a $\hat{F}$-ICC allocation for the 
marginal price function 
\[
\hat{p}(\theta)=\begin{cases}
-\icost^\prime(\theta_{F-}), & \theta< \theta^*\\
p_Q(\theta)+c^\prime(\theta_{F-})-\icost^\prime(\underline{\theta}_F), & \theta\geq \theta^*
\end{cases}
\]
The new profit $\Pi_{\hat{Q}}\coloneqq\int \pi_{\hat{Q}}(\theta)\hat{F}(\dd \theta)$ 
is now given by
\begin{align*}
\int \pi_{\hat{Q}}(\theta)\hat{F}(\dd\theta) &=\int_{\theta< \theta^*} \pi_{\hat{Q}}(\theta)\hat{F}(\dd\theta)+\int_{\theta\geq \theta^*}\pi_{\hat{Q}}(\theta)F(\dd\theta)\\
&= \int_{\theta< \theta^*} (0)\hat{F}(\dd\theta)+\int_{\theta\geq \theta^*}[\pi_Q(\theta)+V_Q(\theta_{F-})]F(\dd\theta)\\
&> \int_{\theta< \theta^*} (0)F(\dd\theta)+\int_{\theta\geq \theta^*}\pi_Q(\theta)F(\dd\theta)\\
&\geq \int_{\theta< \theta^*} \pi_Q(\theta) F(\dd\theta)+\int_{\theta\geq \theta^*}\pi_Q(\theta)F(\dd\theta)\\
&=\int \pi_Q(\theta)F(\dd\theta),
\end{align*}
where the weak inequality follows from noting that $\pi_{Q}(\thmin_Q) = 0$ and $\pi^{'}_{Q}(\theta)\leq 0$ for all $\theta \in [\thmin_Q,\theta^*)$. Thus, $(\hat{Q},\hat{F})$ is an improvement over $(Q,F)$. A contradiction. \hfill$\square$

\subsection{Cost Function Characterization}

In this section, we show a continuous cost function $C:\Signals\rightarrow\mathbb{R}$
is affine and strictly increasing in informativeness if and only if
a strictly convex continuous function $c:\Theta\rightarrow\mathbb{R}$
exists such that $C\left(F\right)=\int\icost(\theta)F\left(\dd \theta\right).$
To prove this result, note the Riesz representation theorem implies
$C$ is continuous and affine if and only if $C\left(F\right)=\int\tilde{\icost}(\theta)F\left(\dd \theta\right)$
for some continuous $\tilde{\icost}:\Theta\rightarrow\mathbb{R}$.
All that remains is to show $\tilde{\icost}$ must be strictly convex.
For this purpose, fix any $x,y,z\in\left(\thmin,\thmax\right)$ such
that $y=\beta x+\left(1-\beta\right)z$ for some $\beta\in\left(0,1\right)$.
By Lemma 6 in \citet{ravid2020learning}, one can find $F',F''\in\Signals$ and $\gamma>0$
such that $F'\succ F''$, and 
\[
F'-F''=\gamma\left(\beta\mathbf{1}_{\left[x,\thmax\right]}+\left(1-\beta\right)\mathbf{1}_{\left[z,\thmax\right]}-\mathbf{1}_{\left[y,\thmax\right]}\right),
\]
where for any $w\in \Theta$, $\mathbf{1}_{[w,\bar{\theta}]}$ is the CDF of the distribution that generates $w$ with probability $1$. Since $C$ is strictly increasing in $\succeq$, it follows that
\[
0<C\left(F'\right)-C\left(F''\right)
=\int \icost(\theta) (F' - F'')(\dd\theta)
= \gamma\left(\beta\tilde{c}\left(x\right)+\left(1-\beta\right)\tilde{c}\left(z\right)-\tilde{c}\left(y\right)\right).
\]
The claim follows. \hfill $\square$

\end{document}

%% file: quadraticqualities0.tex
%
%
\definecolor{mycolor1}{rgb}{0.00000,0.44700,0.74100}%
\definecolor{mycolor2}{rgb}{0.85000,0.32500,0.09800}%
\definecolor{mycolor3}{rgb}{0.92900,0.69400,0.12500}%
\begin{tikzpicture}

\begin{axis}[
axis x line=bottom,
axis y line=left,
xlabel near ticks,
ylabel near ticks,
trim axis left,
trim axis right,
width=\linewidth,
height=0.75\linewidth,
clip=true,
clip mode=individual,
at={(0\linewidth,0\linewidth)},
scale only axis,
xmin=0.45,
xmax=1.25,
xlabel style={ at={(axis description cs:1,0)}, anchor=north west, yshift=-2pt, font=\color{white!15!black}},
xlabel={$\gamma$},
ymin=0,
ymax=0.6,
ylabel style={  at={(axis description cs:0,1)}, anchor=south east, xshift=-4pt, rotate=-90, font=\color{white!15!black}},
ylabel={$Q$},
xtick={0.512525252525252,0.994949494949495,1.0150505050505},
xticklabels={$\gamma_l$,\hspace{-1em}$\gamma_m$, \hspace{1em}$\gamma_h$},
ytick={.1,.5},
yticklabels={$\thmin$,$\theta_0$},
axis background/.style={fill=white},
legend style={at={(0.98,0.02)}, anchor=south east, draw=none, row sep=1.75pt}
]

\addplot [color=mycolor3, line width=2.0pt]
  table[row sep=crcr]{%
0.01	0.1\\
0.0301010101010101	0.1\\
0.0502020202020202	0.1\\
0.0703030303030303	0.1\\
0.0904040404040404	0.1\\
0.11050505050505	0.1\\
0.130606060606061	0.1\\
0.150707070707071	0.1\\
0.170808080808081	0.1\\
0.190909090909091	0.1\\
0.211010101010101	0.1\\
0.231111111111111	0.1\\
0.251212121212121	0.1\\
0.271313131313131	0.1\\
0.291414141414141	0.1\\
0.311515151515152	0.1\\
0.331616161616162	0.1\\
0.351717171717172	0.1\\
0.371818181818182	0.1\\
0.391919191919192	0.1\\
0.412020202020202	0.1\\
0.432121212121212	0.1\\
0.452222222222222	0.1\\
0.472323232323232	0.1\\
0.492424242424242	0.1\\
0.512525252525252	0.1\\
0.532626262626263	0.112772687080097\\
0.552727272727273	0.120374706905443\\
0.572828282828283	0.127782418869379\\
0.592929292929293	0.135003171854303\\
0.613030303030303	0.142043961709998\\
0.633131313131313	0.148911431031897\\
0.653232323232323	0.155611902635299\\
0.673333333333333	0.162151393646764\\
0.693434343434343	0.16853564250639\\
0.713535353535354	0.174770102785675\\
0.733636363636364	0.180859991371627\\
0.753737373737374	0.18681027585721\\
0.773838383838384	0.192625704695804\\
0.793939393939394	0.198310809221408\\
0.814040404040404	0.203869925111153\\
0.834141414141414	0.20930719265408\\
0.854242424242424	0.21462657196314\\
0.874343434343434	0.219831860283607\\
0.894444444444444	0.224926687370185\\
0.914545454545455	0.229914530472518\\
0.934646464646465	0.234798728409641\\
0.954747474747475	0.239582470895141\\
0.974848484848485	0.244268836750467\\
0.994949494949495	0.248860760597984\\
1.0150505050505	0.499999999639219\\
1.03515151515152	0.499999999996607\\
1.05525252525253	0.499999999903773\\
1.07535353535354	0.49999999999105\\
1.09545454545455	0.5\\
1.11555555555556	0.499999999977471\\
1.13565656565657	0.499999999988522\\
1.15575757575758	0.499999999996347\\
1.17585858585859	0.499999999976159\\
1.1959595959596	0.499999999991181\\
1.21606060606061	0.499999999997412\\
1.23616161616162	0.499999999885077\\
1.25626262626263	0.499999999989548\\
1.27636363636364	0.499999999996253\\
1.29646464646465	0.499999999998058\\
1.31656565656566	0.5\\
1.33666666666667	0.499999999998644\\
1.35676767676768	0.5\\
1.37686868686869	0.5\\
1.3969696969697	0.499999999941375\\
1.41707070707071	0.499999999968385\\
1.43717171717172	0.499999999994891\\
1.45727272727273	0.499999999997611\\
1.47737373737374	0.499999999990782\\
1.49747474747475	0.5\\
1.51757575757576	0.499999999960713\\
1.53767676767677	0.499999999976923\\
1.55777777777778	0.5\\
1.57787878787879	0.499999999988733\\
1.5979797979798	0.499999999993697\\
1.61808080808081	0.49999999999662\\
1.63818181818182	0.499999998413861\\
1.65828282828283	0.499999998617608\\
1.67838383838384	0.499999999221995\\
1.69848484848485	0.499999999698256\\
1.71858585858586	0.499999999889889\\
1.73868686868687	0.49999999999895\\
1.75878787878788	0.499999999993269\\
1.77888888888889	0.499999999997288\\
1.7989898989899	0.49999999999884\\
1.81909090909091	0.5\\
1.83919191919192	0.5\\
1.85929292929293	0.499999999841616\\
1.87939393939394	0.499999999998927\\
1.89949494949495	0.5\\
1.91959595959596	0.5\\
1.93969696969697	0.5\\
1.95979797979798	0.5\\
1.97989898989899	0.5\\
2	0.49999999999816\\
};
\addlegendentry{$Q^e_L$}

\addplot [color=mycolor1, line width=2.0pt]
  table[row sep=crcr]{%
0.01	0\\
0.0301010101010101	0\\
0.0502020202020202	0\\
0.0703030303030303	0\\
0.0904040404040404	0\\
0.11050505050505	0\\
0.130606060606061	0\\
0.150707070707071	0\\
0.170808080808081	0\\
0.190909090909091	0\\
0.211010101010101	0\\
0.231111111111111	0\\
0.251212121212121	0\\
0.271313131313131	0\\
0.291414141414141	0\\
0.311515151515152	0\\
0.331616161616162	0\\
0.351717171717172	0\\
0.371818181818182	0\\
0.391919191919192	0\\
0.412020202020202	0\\
0.432121212121212	0\\
0.452222222222222	0\\
0.472323232323232	0\\
0.492424242424242	0\\
0.512525252525252	0\\
0.532626262626263	3.10450318651054e-16\\
0.552727272727273	3.06825272260043e-17\\
0.572828282828283	1.58991787188113e-17\\
0.592929292929293	0\\
0.613030303030303	0\\
0.633131313131313	0\\
0.653232323232323	9.06541957102385e-17\\
0.673333333333333	0\\
0.693434343434343	0\\
0.713535353535354	0\\
0.733636363636364	8.14499982611365e-17\\
0.753737373737374	2.09204146710932e-17\\
0.773838383838384	6.4434989330707e-17\\
0.793939393939394	4.40724897654229e-17\\
0.814040404040404	0\\
0.834141414141414	1.15760375471648e-16\\
0.854242424242424	1.18549951000694e-16\\
0.874343434343434	0\\
0.894444444444444	1.73780742882299e-16\\
0.914545454545455	5.07674710351322e-17\\
0.934646464646465	0\\
0.954747474747475	0\\
0.974848484848485	0\\
0.994949494949495	0\\
1.0150505050505	0.25186726348733\\
1.03515151515152	0.254318048397065\\
1.05525252525253	0.256720891576234\\
1.07535353535354	0.259077195981415\\
1.09545454545455	0.261388284821023\\
1.11555555555556	0.263655463302713\\
1.13565656565657	0.265879960430021\\
1.15575757575758	0.268062973404245\\
1.17585858585859	0.270205652618597\\
1.1959595959596	0.272309106857841\\
1.21606060606061	0.274374401171136\\
1.23616161616162	0.27640256667206\\
1.25626262626263	0.278394592000368\\
1.27636363636364	0.280351434689387\\
1.29646464646465	0.282274027611023\\
1.31656565656566	0.284163250331507\\
1.33666666666667	0.286019971381582\\
1.35676767676768	0.287845021378335\\
1.37686868686869	0.289639198427564\\
1.3969696969697	0.29140328790033\\
1.41707070707071	0.293138028785231\\
1.43717171717172	0.294844165332227\\
1.45727272727273	0.296522381164808\\
1.47737373737374	0.298173368049284\\
1.49747474747475	0.299797777465712\\
1.51757575757576	0.301396240781145\\
1.53767676767677	0.302969390603653\\
1.55777777777778	0.30451781148815\\
1.57787878787879	0.306042084159906\\
1.5979797979798	0.307542768066849\\
1.61808080808081	0.309020413030516\\
1.63818181818182	0.310475530294807\\
1.65828282828283	0.31190865090224\\
1.67838383838384	0.313320256590208\\
1.69848484848485	0.314710836488346\\
1.71858585858586	0.31608084916365\\
1.73868686868687	0.317430752386331\\
1.75878787878788	0.318760983250089\\
1.77888888888889	0.320071970795476\\
1.7989898989899	0.321364128096755\\
1.81909090909091	0.322637858474035\\
1.83919191919192	0.323893551323415\\
1.85929292929293	0.325131593340565\\
1.87939393939394	0.326352348364135\\
1.89949494949495	0.327556174161037\\
1.91959595959596	0.328743429033241\\
1.93969696969697	0.329914443486105\\
1.95979797979798	0.331069553060144\\
1.97989898989899	0.332209076322311\\
2	0.333333333207223\\
};
\addlegendentry{$Q_L$}

\addplot [color=mycolor2, style=dashed, line width=2.0pt]
  table[row sep=crcr]{%
0.01	0\\
0.0301010101010101	0\\
0.0502020202020202	0\\
0.0703030303030303	0\\
0.0904040404040404	0\\
0.11050505050505	0\\
0.130606060606061	0\\
0.150707070707071	0\\
0.170808080808081	0\\
0.190909090909091	0\\
0.211010101010101	0\\
0.231111111111111	0\\
0.251212121212121	0\\
0.271313131313131	0\\
0.291414141414141	0\\
0.311515151515152	0\\
0.331616161616162	0\\
0.351717171717172	0\\
0.371818181818182	0\\
0.391919191919192	0\\
0.412020202020202	0\\
0.432121212121212	0\\
0.452222222222222	0\\
0.472323232323232	0\\
0.492424242424242	0\\
0.512525252525252	0\\
0.532626262626263	0\\
0.552727272727273	0\\
0.572828282828283	0\\
0.592929292929293	0\\
0.613030303030303	0\\
0.633131313131313	0\\
0.653232323232323	0\\
0.673333333333333	0\\
0.693434343434343	0\\
0.713535353535354	0\\
0.733636363636364	0\\
0.753737373737374	0\\
0.773838383838384	0\\
0.793939393939394	0\\
0.814040404040404	0\\
0.834141414141414	0\\
0.854242424242424	0\\
0.874343434343434	0\\
0.894444444444444	0\\
0.914545454545455	0\\
0.934646464646465	0\\
0.954747474747475	0\\
0.974848484848485	0\\
0.994949494949495	0\\
1.0150505050505	0.499999999278413\\
1.03515151515152	0.499999999993215\\
1.05525252525253	0.499999999807542\\
1.07535353535354	0.499999999982101\\
1.09545454545455	0.5\\
1.11555555555556	0.499999999954941\\
1.13565656565657	0.499999999977045\\
1.15575757575758	0.499999999992695\\
1.17585858585859	0.499999999952319\\
1.1959595959596	0.499999999982363\\
1.21606060606061	0.499999999994825\\
1.23616161616162	0.499999999770128\\
1.25626262626263	0.499999999979095\\
1.27636363636364	0.499999999992507\\
1.29646464646465	0.499999999996115\\
1.31656565656566	0.5\\
1.33666666666667	0.499999999997288\\
1.35676767676768	0.5\\
1.37686868686869	0.5\\
1.3969696969697	0.499999999882744\\
1.41707070707071	0.499999999936768\\
1.43717171717172	0.499999999989782\\
1.45727272727273	0.499999999995222\\
1.47737373737374	0.499999999981564\\
1.49747474747475	0.5\\
1.51757575757576	0.499999999921423\\
1.53767676767677	0.499999999953845\\
1.55777777777778	0.5\\
1.57787878787879	0.499999999977466\\
1.5979797979798	0.499999999987394\\
1.61808080808081	0.499999999993239\\
1.63818181818182	0.499999996562011\\
1.65828282828283	0.49999999698672\\
1.67838383838384	0.499999998375958\\
1.69848484848485	0.49999999938926\\
1.71858585858586	0.499999999778912\\
1.73868686868687	0.499999999997899\\
1.75878787878788	0.499999999986537\\
1.77888888888889	0.499999999994575\\
1.7989898989899	0.49999999999768\\
1.81909090909091	0.5\\
1.83919191919192	0.5\\
1.85929292929293	0.499999999682639\\
1.87939393939394	0.499999999997854\\
1.89949494949495	0.5\\
1.91959595959596	0.5\\
1.93969696969697	0.5\\
1.95979797979798	0.5\\
1.97989898989899	0.5\\
2	0.499999999996319\\
};
\addlegendentry{$Q^s_L$}

\addplot[gray, densely dashed, thick] coordinates {(1.0150505050505,.5) (1.0150505050505,0)};
\addplot[gray, densely dashed, thick] coordinates {(0,0.5) (1.0150505050505,0.5)}; 

\addplot[gray, densely dashed, thick] coordinates {(0.994949494949495,0.248860760597984) (0.994949494949495,0)}; 

\addplot[gray, densely dashed, thick] coordinates {(0.512525252525252,.1) (0.512525252525252,0)}; 

\end{axis}
\end{tikzpicture}%

%% file: quadraticsignal0.tex
%
%
\definecolor{mycolor1}{rgb}{0.00000,0.44700,0.74100}%
\definecolor{mycolor2}{rgb}{0.85000,0.32500,0.09800}%
\definecolor{mycolor3}{rgb}{0.92900,0.69400,0.12500}%
\begin{tikzpicture}

\begin{axis}[
axis x line=bottom,
axis y line=left,
xlabel near ticks,
ylabel near ticks,
trim axis left,
trim axis right,
width=\linewidth,
height=0.75\linewidth,
clip=true,
clip mode=individual,
at={(0\linewidth,0\linewidth)},
scale only axis,
xmin=0.45,
xmax=1.25,
xlabel style={ at={(axis description cs:1,0)}, anchor=north west, yshift=-2pt, font=\color{white!15!black}},
xlabel={$\gamma$},
ymin=0,
ymax=1.1,
ylabel style={  at={(axis description cs:0,1)}, anchor=south east, xshift=-4pt, rotate=-90, font=\color{white!15!black}},
ylabel={$\theta$},
xtick={0.512525252525252,0.994949494949495,1.0150505050505},
xticklabels={$\gamma_l$,\hspace{-1em}$\gamma_m$, \hspace{1em}$\gamma_h$},
ytick={.1,.5,.9},
yticklabels={$\thmin$,$\theta_0$, $\thmax$},
axis background/.style={fill=white},
legend style={at={(0.98,0.55)}, anchor=south east, draw=none, row sep=1.75pt}
]

\addplot [color=mycolor2, line width=2.0pt]
  table[row sep=crcr]{%
0.01	0.9\\
0.0301010101010101	0.9\\
0.0502020202020202	0.9\\
0.0703030303030303	0.9\\
0.0904040404040404	0.9\\
0.11050505050505	0.9\\
0.130606060606061	0.9\\
0.150707070707071	0.9\\
0.170808080808081	0.9\\
0.190909090909091	0.9\\
0.211010101010101	0.9\\
0.231111111111111	0.9\\
0.251212121212121	0.9\\
0.271313131313131	0.9\\
0.291414141414141	0.9\\
0.311515151515152	0.9\\
0.331616161616162	0.9\\
0.351717171717172	0.9\\
0.371818181818182	0.9\\
0.391919191919192	0.9\\
0.412020202020202	0.9\\
0.432121212121212	0.9\\
0.452222222222222	0.9\\
0.472323232323232	0.9\\
0.492424242424242	0.9\\
0.512525252525252	0.9\\
0.532626262626263	0.9\\
0.552727272727273	0.9\\
0.572828282828283	0.9\\
0.592929292929293	0.9\\
0.613030303030303	0.9\\
0.633131313131313	0.9\\
0.653232323232323	0.9\\
0.673333333333333	0.9\\
0.693434343434343	0.9\\
0.713535353535354	0.9\\
0.733636363636364	0.9\\
0.753737373737374	0.9\\
0.773838383838384	0.9\\
0.793939393939394	0.9\\
0.814040404040404	0.9\\
0.834141414141414	0.9\\
0.854242424242424	0.9\\
0.874343434343434	0.9\\
0.894444444444444	0.9\\
0.914545454545455	0.9\\
0.934646464646465	0.9\\
0.954747474747475	0.9\\
0.974848484848485	0.9\\
0.994949494949495	0.9\\
1.0150505050505	0.500005085868379\\
1.03515151515152	0.500002033704599\\
1.05525252525253	0.500002796282631\\
1.07535353535354	0.500022954123783\\
1.09545454545455	0.5\\
1.11555555555556	0.500012635998733\\
1.13565656565657	0.50002824034875\\
1.15575757575758	0.500062360478939\\
1.17585858585859	0.500003459978871\\
1.1959595959596	0.500005998686485\\
1.21606060606061	0.500009754707998\\
1.23616161616162	0.500000507943998\\
1.25626262626263	0.50000819078743\\
1.27636363636364	0.500012592381081\\
1.29646464646465	0.500015305001232\\
1.31656565656566	0.5\\
1.33666666666667	0.500026544913496\\
1.35676767676768	0.5\\
1.37686868686869	0.5\\
1.3969696969697	0.500000484901817\\
1.41707070707071	0.500000664732318\\
1.43717171717172	0.500004365004783\\
1.45727272727273	0.500006250782472\\
1.47737373737374	0.500002880153774\\
1.49747474747475	0.5\\
1.51757575757576	0.500000532741243\\
1.53767676767677	0.500000885529635\\
1.55777777777778	0.5\\
1.57787878787879	0.50000122200311\\
1.5979797979798	0.500000922184616\\
1.61808080808081	0.50000125765079\\
1.63818181818182	0.500000009468296\\
1.65828282828283	0.500000007690307\\
1.67838383838384	0.500000008897196\\
1.69848484848485	0.500000012556101\\
1.71858585858586	0.500000014006379\\
1.73868686868687	0.500002940771906\\
1.75878787878788	0.500001259469526\\
1.77888888888889	0.500001057913655\\
1.7989898989899	0.500000820193661\\
1.81909090909091	0.5\\
1.83919191919192	0.5\\
1.85929292929293	0.50000004221379\\
1.87939393939394	0.500005133156294\\
1.89949494949495	0.5\\
1.91959595959596	0.5\\
1.93969696969697	0.5\\
1.95979797979798	0.5\\
1.97989898989899	0.5\\
2	0.500000738870032\\
};
\addlegendentry{$\theta_H$}

\addplot [color=mycolor3, line width=2.0pt, forget plot]
  table[row sep=crcr]{%
0.01	0.1\\
0.0301010101010101	0.1\\
0.0502020202020202	0.1\\
0.0703030303030303	0.1\\
0.0904040404040404	0.1\\
0.11050505050505	0.1\\
0.130606060606061	0.1\\
0.150707070707071	0.1\\
0.170808080808081	0.1\\
0.190909090909091	0.1\\
0.211010101010101	0.1\\
0.231111111111111	0.1\\
0.251212121212121	0.1\\
0.271313131313131	0.1\\
0.291414141414141	0.1\\
0.311515151515152	0.1\\
0.331616161616162	0.1\\
0.351717171717172	0.1\\
0.371818181818182	0.1\\
0.391919191919192	0.1\\
0.412020202020202	0.1\\
0.432121212121212	0.1\\
0.452222222222222	0.1\\
0.472323232323232	0.1\\
0.492424242424242	0.1\\
0.512525252525252	0.1\\
0.532626262626263	0.112772687080096\\
0.552727272727273	0.120374706905443\\
0.572828282828283	0.127782418869379\\
0.592929292929293	0.135003171854303\\
0.613030303030303	0.142043961709998\\
0.633131313131313	0.148911431031897\\
0.653232323232323	0.155611902635299\\
0.673333333333333	0.162151393646764\\
0.693434343434343	0.16853564250639\\
0.713535353535354	0.174770102785675\\
0.733636363636364	0.180859991371627\\
0.753737373737374	0.18681027585721\\
0.773838383838384	0.192625704695804\\
0.793939393939394	0.198310809221408\\
0.814040404040404	0.203869925111153\\
0.834141414141414	0.20930719265408\\
0.854242424242424	0.21462657196314\\
0.874343434343434	0.219831860283607\\
0.894444444444444	0.224926687370184\\
0.914545454545455	0.229914530472518\\
0.934646464646465	0.234798728409641\\
0.954747474747475	0.239582470895141\\
0.974848484848485	0.244268836750467\\
0.994949494949495	0.248860760597984\\
1.0150505050505	0.251867259214842\\
1.03515151515152	0.254318044577897\\
1.05525252525253	0.256720893308126\\
1.07535353535354	0.259077189525523\\
1.09545454545455	0.261388287715249\\
1.11555555555556	0.263655461160298\\
1.13565656565657	0.265879960118629\\
1.15575757575758	0.268062975288951\\
1.17585858585859	0.270205655768264\\
1.1959595959596	0.272309108277277\\
1.21606060606061	0.274374402223988\\
1.23616161616162	0.276402564842309\\
1.25626262626263	0.278394591868325\\
1.27636363636364	0.280351440054426\\
1.29646464646465	0.282274026227186\\
1.31656565656566	0.284163251627902\\
1.33666666666667	0.28601997153362\\
1.35676767676768	0.287845018489762\\
1.37686868686869	0.289639200026932\\
1.3969696969697	0.291403286260011\\
1.41707070707071	0.293138036537364\\
1.43717171717172	0.294844163841598\\
1.45727272727273	0.29652238347778\\
1.47737373737374	0.298173366346046\\
1.49747474747475	0.299797774238749\\
1.51757575757576	0.301396247049815\\
1.53767676767677	0.302969390571669\\
1.55777777777778	0.304517810029005\\
1.57787878787879	0.306042082238073\\
1.5979797979798	0.30754276839687\\
1.61808080808081	0.30902040769984\\
1.63818181818182	0.310475534859873\\
1.65828282828283	0.311908651603405\\
1.67838383838384	0.313320260401397\\
1.69848484848485	0.314710836377731\\
1.71858585858586	0.316080850558742\\
1.73868686868687	0.317430752998155\\
1.75878787878788	0.318760984706347\\
1.77888888888889	0.320071971442679\\
1.7989898989899	0.321364128681721\\
1.81909090909091	0.322637858909826\\
1.83919191919192	0.323893554585797\\
1.85929292929293	0.325131592492695\\
1.87939393939394	0.326352346080571\\
1.89949494949495	0.327556175262203\\
1.91959595959596	0.32874342520369\\
1.93969696969697	0.329914440946079\\
1.95979797979798	0.331069550804276\\
1.97989898989899	0.332209078333203\\
2	0.333333333394548\\
};

\addplot [color=mycolor1, style=dashed, line width=2.0pt, forget plot]
  table[row sep=crcr]{%
0.01	0.1\\
0.0301010101010101	0.1\\
0.0502020202020202	0.1\\
0.0703030303030303	0.1\\
0.0904040404040404	0.1\\
0.11050505050505	0.1\\
0.130606060606061	0.1\\
0.150707070707071	0.1\\
0.170808080808081	0.1\\
0.190909090909091	0.1\\
0.211010101010101	0.1\\
0.231111111111111	0.1\\
0.251212121212121	0.1\\
0.271313131313131	0.1\\
0.291414141414141	0.1\\
0.311515151515152	0.1\\
0.331616161616162	0.1\\
0.351717171717172	0.1\\
0.371818181818182	0.1\\
0.391919191919192	0.1\\
0.412020202020202	0.1\\
0.432121212121212	0.1\\
0.452222222222222	0.1\\
0.472323232323232	0.1\\
0.492424242424242	0.1\\
0.512525252525252	0.1\\
0.532626262626263	0.112772687080097\\
0.552727272727273	0.120374706905443\\
0.572828282828283	0.127782418869379\\
0.592929292929293	0.135003171854303\\
0.613030303030303	0.142043961709998\\
0.633131313131313	0.148911431031897\\
0.653232323232323	0.155611902635299\\
0.673333333333333	0.162151393646764\\
0.693434343434343	0.16853564250639\\
0.713535353535354	0.174770102785675\\
0.733636363636364	0.180859991371627\\
0.753737373737374	0.18681027585721\\
0.773838383838384	0.192625704695804\\
0.793939393939394	0.198310809221408\\
0.814040404040404	0.203869925111153\\
0.834141414141414	0.20930719265408\\
0.854242424242424	0.21462657196314\\
0.874343434343434	0.219831860283607\\
0.894444444444444	0.224926687370185\\
0.914545454545455	0.229914530472518\\
0.934646464646465	0.234798728409641\\
0.954747474747475	0.239582470895141\\
0.974848484848485	0.244268836750467\\
0.994949494949495	0.248860760597984\\
1.0150505050505	0.499999999639219\\
1.03515151515152	0.499999999996607\\
1.05525252525253	0.499999999903773\\
1.07535353535354	0.49999999999105\\
1.09545454545455	0.5\\
1.11555555555556	0.499999999977471\\
1.13565656565657	0.499999999988522\\
1.15575757575758	0.499999999996347\\
1.17585858585859	0.499999999976159\\
1.1959595959596	0.499999999991181\\
1.21606060606061	0.499999999997412\\
1.23616161616162	0.499999999885077\\
1.25626262626263	0.499999999989548\\
1.27636363636364	0.499999999996253\\
1.29646464646465	0.499999999998058\\
1.31656565656566	0.5\\
1.33666666666667	0.499999999998644\\
1.35676767676768	0.5\\
1.37686868686869	0.5\\
1.3969696969697	0.499999999941375\\
1.41707070707071	0.499999999968385\\
1.43717171717172	0.499999999994891\\
1.45727272727273	0.499999999997611\\
1.47737373737374	0.499999999990782\\
1.49747474747475	0.5\\
1.51757575757576	0.499999999960713\\
1.53767676767677	0.499999999976923\\
1.55777777777778	0.5\\
1.57787878787879	0.499999999988733\\
1.5979797979798	0.499999999993697\\
1.61808080808081	0.49999999999662\\
1.63818181818182	0.499999998413861\\
1.65828282828283	0.499999998617608\\
1.67838383838384	0.499999999221995\\
1.69848484848485	0.499999999698256\\
1.71858585858586	0.499999999889889\\
1.73868686868687	0.49999999999895\\
1.75878787878788	0.499999999993269\\
1.77888888888889	0.499999999997288\\
1.7989898989899	0.49999999999884\\
1.81909090909091	0.5\\
1.83919191919192	0.5\\
1.85929292929293	0.499999999841616\\
1.87939393939394	0.499999999998927\\
1.89949494949495	0.5\\
1.91959595959596	0.5\\
1.93969696969697	0.5\\
1.95979797979798	0.5\\
1.97989898989899	0.5\\
2	0.49999999999816\\
};

\addlegendimage{dashed, mycolor1, line width=2pt}
\addlegendentry{$\theta_L$}

\addlegendimage{mycolor3, line width=2pt}    
\addlegendentry{$\underline{\theta}_Q$}

\addplot[gray, densely dashed, thick] coordinates {(0.512525252525252,0) (0.512525252525252,.1)};

\addplot[gray, densely dashed, thick] coordinates {(1.0150505050505,.5) (1.0150505050505,0)};
\addplot[gray, densely dashed, thick] coordinates {(0,0.5) (1.0150505050505,0.5)};

\addplot[gray, densely dashed, thick] coordinates {(0.994949494949495,0.248860760597984) (0.994949494949495,0)}; 
\end{axis}
\end{tikzpicture}%

%% file: cubicqualities0.tex
%
%
\definecolor{mycolor1}{rgb}{0.00000,0.44700,0.74100}%
\definecolor{mycolor2}{rgb}{0.85000,0.32500,0.09800}%
\definecolor{mycolor3}{rgb}{0.92900,0.69400,0.12500}%
\definecolor{mycolor4}{rgb}{0.49400,0.18400,0.55600}%
\begin{tikzpicture}

\begin{axis}[%
axis x line=bottom,
axis y line=left,
xlabel near ticks,
ylabel near ticks,
trim axis left,
trim axis right,
width=\linewidth,
height=0.75\linewidth,
clip=true,
clip mode=individual,
at={(0\linewidth,0\linewidth)},
scale only axis,
xmin=.5,
xmax=2,
xlabel style={ at={(axis description cs:1,0)}, anchor=north west, yshift=-2pt, font=\color{white!15!black}},
xlabel={$\gamma$},
ylabel style={  at={(axis description cs:0,1)}, anchor=south east, xshift=-4pt, rotate=-90, font=\color{white!15!black}},
ymin=0,
ymax=1.1,
ylabel={$Q$},
axis background/.style={fill=white},
title style={font=\bfseries},
xtick={0.572828282828283,1.23616161616162,1.64},
xticklabels={${\gamma}'_l$,${\gamma}'_m$, ${\gamma}'_h$},
ytick={0.316230875128364,.7071,0.948683298050514},
yticklabels={$\sqrt{\thmin}$,$\sqrt{\theta_0}$, $\sqrt{\thmax}$}, 
legend style={at={(0.98,0.0)}, anchor=south east, fill=none, draw=none, row sep=1.6pt}
]

\addplot [color=mycolor4, line width=2.0pt]
  table[row sep=crcr]{%
0.01	0.948683298050514\\
0.0301010101010101	0.948683298050514\\
0.0502020202020202	0.948683298050514\\
0.0703030303030303	0.948683298050514\\
0.0904040404040404	0.948683298050514\\
0.11050505050505	0.948683298050514\\
0.130606060606061	0.948683298050514\\
0.150707070707071	0.948683298050514\\
0.170808080808081	0.948683298050514\\
0.190909090909091	0.948683298050514\\
0.211010101010101	0.948683298050514\\
0.231111111111111	0.948683298050514\\
0.251212121212121	0.948683298050514\\
0.271313131313131	0.948683298050514\\
0.291414141414141	0.948683298050514\\
0.311515151515152	0.948683298050514\\
0.331616161616162	0.948683298050514\\
0.351717171717172	0.948683298050514\\
0.371818181818182	0.948683298050514\\
0.391919191919192	0.948683298050514\\
0.412020202020202	0.948683298050514\\
0.432121212121212	0.948683298050514\\
0.452222222222222	0.948683298050514\\
0.472323232323232	0.948683298050514\\
0.492424242424242	0.948683298050514\\
0.512525252525252	0.948683298050514\\
0.532626262626263	0.948683298050514\\
0.552727272727273	0.948683298050514\\
0.572828282828283	0.948683298050514\\
0.592929292929293	0.948683298050514\\
0.613030303030303	0.948683298050514\\
0.633131313131313	0.948683298050514\\
0.653232323232323	0.948683298050514\\
0.673333333333333	0.948683298050514\\
0.693434343434343	0.948683298050514\\
0.713535353535354	0.948683298050514\\
0.733636363636364	0.948683298050514\\
0.753737373737374	0.948683298050514\\
0.773838383838384	0.948683298050514\\
0.793939393939394	0.948683298050514\\
0.814040404040404	0.948683298050514\\
0.834141414141414	0.948683298050514\\
0.854242424242424	0.948683298050514\\
0.874343434343434	0.948683298050514\\
0.894444444444444	0.948683298050514\\
0.914545454545455	0.948683298050514\\
0.934646464646465	0.948683298050514\\
0.954747474747475	0.948683298050514\\
0.974848484848485	0.948683298050514\\
0.994949494949495	0.948683298050514\\
1.0150505050505	0.948683298050514\\
1.03515151515152	0.948683298050514\\
1.05525252525253	0.948683298050514\\
1.07535353535354	0.948683298050514\\
1.09545454545455	0.948683298050514\\
1.11555555555556	0.948683298050514\\
1.13565656565657	0.948683298050514\\
1.15575757575758	0.948683298050514\\
1.17585858585859	0.948683298050514\\
1.1959595959596	0.948683298050514\\
1.21606060606061	0.948683298050514\\
1.23616161616162	0.948683298050514\\
1.25626262626263	0.815979734323104\\
1.27636363636364	0.807978191907457\\
1.29646464646465	0.800271382159939\\
1.31656565656566	0.792844999190364\\
1.33666666666667	0.785685575860548\\
1.35676767676768	0.77878046666127\\
1.37686868686869	0.772117794955018\\
1.3969696969697	0.765686346103801\\
1.41707070707071	0.759475591941582\\
1.43717171717172	0.75347557665276\\
1.45727272727273	0.747676903977815\\
1.47737373737374	0.742070709134718\\
1.49747474747475	0.736648623405117\\
1.51757575757576	0.731402710507706\\
1.53767676767677	0.726325481626622\\
1.55777777777778	0.721409832547298\\
1.57787878787879	0.71664902144084\\
1.5979797979798	0.712036673777911\\
1.61808080808081	0.707566738360573\\
1.63818181818182	0.707106781186548\\
1.65828282828283	0.707106781186548\\
1.67838383838384	0.707106781186548\\
1.69848484848485	0.707106781186548\\
1.71858585858586	0.707106781186548\\
1.73868686868687	0.707106781186548\\
1.75878787878788	0.707107010808574\\
1.77888888888889	0.707106945049782\\
1.7989898989899	0.707106781186548\\
1.81909090909091	0.707106781186548\\
1.83919191919192	0.707106781186548\\
1.85929292929293	0.707106781186548\\
1.87939393939394	0.707106792208705\\
1.89949494949495	0.707106790088909\\
1.91959595959596	0.707106784521919\\
1.93969696969697	0.707106781186548\\
1.95979797979798	0.707106781186548\\
1.97989898989899	0.707106843732324\\
2	0.707106781186548\\
};
\addlegendentry{$Q^e_H$}

\addplot [color=mycolor2, style=dashed, line width=2.0pt, forget plot]
  table[row sep=crcr]{%
0.01	0.948683298006654\\
0.0301010101010101	0.948683298006654\\
0.0502020202020202	0.948683298006654\\
0.0703030303030303	0.948683298006654\\
0.0904040404040404	0.948683298006654\\
0.11050505050505	0.948683298006654\\
0.130606060606061	0.948683298006654\\
0.150707070707071	0.948683298006654\\
0.170808080808081	0.948683298006654\\
0.190909090909091	0.948683298006654\\
0.211010101010101	0.948683298006654\\
0.231111111111111	0.948683298006654\\
0.251212121212121	0.948683298006654\\
0.271313131313131	0.948683298006654\\
0.291414141414141	0.948683298006654\\
0.311515151515152	0.948683298006654\\
0.331616161616162	0.948683298006654\\
0.351717171717172	0.948683298006654\\
0.371818181818182	0.948683298006654\\
0.391919191919192	0.948683298006654\\
0.412020202020202	0.948683298006654\\
0.432121212121212	0.948683298006654\\
0.452222222222222	0.948683298006654\\
0.472323232323232	0.948683298006654\\
0.492424242424242	0.948683298006654\\
0.512525252525252	0.948683298006654\\
0.532626262626263	0.948683298006654\\
0.552727272727273	0.948683298006654\\
0.572828282828283	0.948683298006654\\
0.592929292929293	0.948683298006654\\
0.613030303030303	0.948683298006654\\
0.633131313131313	0.948683298006654\\
0.653232323232323	0.948683298006654\\
0.673333333333333	0.948683298006654\\
0.693434343434343	0.948683298006654\\
0.713535353535354	0.948683298006654\\
0.733636363636364	0.948683298006654\\
0.753737373737374	0.948683298006654\\
0.773838383838384	0.948683298006654\\
0.793939393939394	0.948683298006654\\
0.814040404040404	0.948683298006654\\
0.834141414141414	0.948683298006654\\
0.854242424242424	0.948683298006654\\
0.874343434343434	0.948683298006654\\
0.894444444444444	0.948683298006654\\
0.914545454545455	0.948683298006654\\
0.934646464646465	0.948683298006654\\
0.954747474747475	0.948683298006654\\
0.974848484848485	0.948683298006654\\
0.994949494949495	0.948683298006654\\
1.0150505050505	0.948683298006654\\
1.03515151515152	0.948683298006654\\
1.05525252525253	0.948683298006654\\
1.07535353535354	0.948683298006654\\
1.09545454545455	0.948683298006654\\
1.11555555555556	0.948683298006654\\
1.13565656565657	0.948683298006654\\
1.15575757575758	0.948683298006654\\
1.17585858585859	0.948683298006654\\
1.1959595959596	0.948683298006654\\
};

\addplot [color=mycolor2, line width=2.0pt]
  table[row sep=crcr]{%
1.21606060606061	0.948683298006654\\
1.23616161616162	0.948683298006654\\
1.25626262626263	0.597008919683998\\
1.27636363636364	0.587606834797256\\
1.29646464646465	0.578496284986302\\
1.31656565656566	0.569663955684566\\
1.33666666666667	0.561097260500571\\
1.35676767676768	0.552784390047799\\
1.37686868686869	0.54471425642763\\
1.3969696969697	0.536876339542762\\
1.41707070707071	0.529260809992938\\
1.43717171717172	0.521858309069276\\
1.45727272727273	0.514660017584876\\
1.47737373737374	0.507657589819345\\
1.49747474747475	0.500843173102377\\
1.51757575757576	0.494209257618824\\
1.53767676767677	0.487748797037724\\
1.55777777777778	0.481455070239811\\
1.57787878787879	0.475321684019229\\
1.5979797979798	0.469342599211238\\
1.61808080808081	0.463512084331565\\
1.63818181818182	0.464950482096571\\
1.65828282828283	0.46719150539431\\
1.67838383838384	0.469393373986991\\
1.69848484848485	0.471557052363953\\
1.71858585858586	0.473683467173658\\
1.73868686868687	0.475773511077002\\
1.75878787878788	0.477828649529511\\
1.77888888888889	0.479848402537231\\
1.7989898989899	0.481834091867768\\
1.81909090909091	0.483787176157398\\
1.83919191919192	0.485708017087822\\
1.85929292929293	0.487597362219089\\
1.87939393939394	0.489455980959374\\
1.89949494949495	0.491284522043908\\
1.91959595959596	0.493083703330744\\
1.93969696969697	0.494854196409685\\
1.95979797979798	0.496596659853927\\
1.97989898989899	0.49831189672248\\
2	0.500000001912869\\
};
\addlegendentry{$Q_H$}

\addplot [color=mycolor3, line width=2.0pt]
  table[row sep=crcr]{%
0.01	0.316879380067812\\
0.0301010101010101	0.316230875128364\\
0.0502020202020202	0.316227770378247\\
0.0703030303030303	0.316227766020603\\
0.0904040404040404	0.316227766016842\\
0.11050505050505	0.316227766016838\\
0.130606060606061	0.316227766016838\\
0.150707070707071	0.316227766016838\\
0.170808080808081	0.316227766016838\\
0.190909090909091	0.316227766016838\\
0.211010101010101	0.316227766016838\\
0.231111111111111	0.316227766016838\\
0.251212121212121	0.316227766016838\\
0.271313131313131	0.316227766016838\\
0.291414141414141	0.316227766016838\\
0.311515151515152	0.316227766016838\\
0.331616161616162	0.316227766016838\\
0.351717171717172	0.316227766016838\\
0.371818181818182	0.316227766016838\\
0.391919191919192	0.316227766016838\\
0.412020202020202	0.316227766016838\\
0.432121212121212	0.316227766016838\\
0.452222222222222	0.316227766016838\\
0.472323232323232	0.316227766016838\\
0.492424242424242	0.316227766016838\\
0.512525252525252	0.316227766016838\\
0.532626262626263	0.316227766016838\\
0.552727272727273	0.316227766016838\\
0.572828282828283	0.316227766016838\\
0.592929292929293	0.318023612966261\\
0.613030303030303	0.331064579010487\\
0.633131313131313	0.34308415187328\\
0.653232323232323	0.354223370644981\\
0.673333333333333	0.364595336520277\\
0.693434343434343	0.374292281900384\\
0.713535353535354	0.383390602929683\\
0.733636363636364	0.391954320064486\\
0.753737373737374	0.40003768198195\\
0.773838383838384	0.407687112620103\\
0.793939393939394	0.414942632142799\\
0.814040404040404	0.421839047969072\\
0.834141414141414	0.428406769441548\\
0.854242424242424	0.434672595562801\\
0.874343434343434	0.440660209680894\\
0.894444444444444	0.446390670916855\\
0.914545454545455	0.451882801265659\\
0.934646464646465	0.45715346562552\\
0.954747474747475	0.462217843009892\\
0.974848484848485	0.46708966876393\\
0.994949494949495	0.471781369780663\\
1.0150505050505	0.476304249371525\\
1.03515151515152	0.480668614622865\\
1.05525252525253	0.484883893218383\\
1.07535353535354	0.488958717897345\\
1.09545454545455	0.492901022107155\\
1.11555555555556	0.49671811279094\\
1.13565656565657	0.50041671935915\\
1.15575757575758	0.504003085080192\\
1.17585858585859	0.507482976148706\\
1.1959595959596	0.510861740578171\\
1.21606060606061	0.514144362225425\\
1.23616161616162	0.51733547354546\\
1.25626262626263	0.436573845260795\\
1.27636363636364	0.438694692616105\\
1.29646464646465	0.440708148132929\\
1.31656565656566	0.442621541996377\\
1.33666666666667	0.444441573796696\\
1.35676767676768	0.446174370338326\\
1.37686868686869	0.447825556785396\\
1.3969696969697	0.449400300839683\\
1.41707070707071	0.450903348737752\\
1.43717171717172	0.452339090360037\\
1.45727272727273	0.453711558129091\\
1.47737373737374	0.455024499481889\\
1.49747474747475	0.456281365129848\\
1.51757575757576	0.457485372666243\\
1.53767676767677	0.45863949230248\\
1.55777777777778	0.459746492642923\\
1.57787878787879	0.460808944420289\\
1.5979797979798	0.461829248072763\\
1.61808080808081	0.462809632408382\\
1.63818181818182	0.707106781186548\\
};
\addlegendentry{$Q^e_L$}

\addplot [color=mycolor3, style=dashed, line width=2.0pt, forget plot]
  table[row sep=crcr]{%
1.65828282828283	0.707106781186548\\
1.67838383838384	0.707106781186548\\
1.69848484848485	0.707106781186548\\
1.71858585858586	0.707106781186548\\
1.73868686868687	0.707106781186548\\
1.75878787878788	0.707106781181445\\
1.77888888888889	0.707106781185706\\
1.7989898989899	0.707106781186548\\
1.81909090909091	0.707106781186548\\
1.83919191919192	0.707106781186548\\
1.85929292929293	0.707106781186548\\
1.87939393939394	0.70710678035996\\
1.89949494949495	0.707106781155715\\
1.91959595959596	0.707106781185406\\
1.93969696969697	0.707106781186548\\
1.95979797979798	0.707106781186548\\
1.97989898989899	0.707106780970266\\
2	0.707106781186548\\
};

\addplot [color=mycolor1, line width=2.0pt]
  table[row sep=crcr]{%
0.01	4.12541512160497e-06\\
0.0301010101010101	5.91901582221657e-08\\
0.0502020202020202	1.38477121969014e-10\\
0.0703030303030303	1.67410899957569e-13\\
0.0904040404040404	2.53430834118664e-16\\
0.11050505050505	1.686922206861e-17\\
0.130606060606061	1.99377551505601e-17\\
0.150707070707071	2.30062882325102e-17\\
0.170808080808081	2.60748213144603e-17\\
0.190909090909091	5.29879170843825e-18\\
0.211010101010101	5.85670681424735e-18\\
0.231111111111111	6.41462192005646e-18\\
0.251212121212121	6.97253702586557e-18\\
0.271313131313131	7.53045213167467e-18\\
0.291414141414141	8.08836723748378e-18\\
0.311515151515152	8.64628234329289e-18\\
0.331616161616162	9.20419744910199e-18\\
0.351717171717172	9.7621125549111e-18\\
0.371818181818182	1.03200276607202e-17\\
0.391919191919192	1.08779427665293e-17\\
0.412020202020202	1.14358578723384e-17\\
0.432121212121212	1.19937729781475e-17\\
0.452222222222222	1.25516880839566e-17\\
0.472323232323232	1.31096031897657e-17\\
0.492424242424242	1.36675182955748e-17\\
0.512525252525252	1.42254334013839e-17\\
0.532626262626263	0\\
0.552727272727273	0\\
0.572828282828283	0\\
0.592929292929293	0\\
0.613030303030303	0\\
0.633131313131313	0\\
0.653232323232323	0\\
0.673333333333333	0\\
0.693434343434343	0\\
0.713535353535354	0\\
0.733636363636364	0\\
0.753737373737374	0\\
0.773838383838384	0\\
0.793939393939394	0\\
0.814040404040404	0\\
0.834141414141414	0\\
0.854242424242424	0\\
0.874343434343434	0\\
0.894444444444444	0\\
0.914545454545455	0\\
0.934646464646465	0\\
0.954747474747475	0\\
0.974848484848485	0\\
0.994949494949495	1.1046158376321e-16\\
1.0150505050505	1.12693244186446e-16\\
1.03515151515152	0\\
1.05525252525253	1.17156565032919e-16\\
1.07535353535354	1.19388225456156e-16\\
1.09545454545455	1.21619885879392e-16\\
1.11555555555556	0\\
1.13565656565657	0\\
1.15575757575758	1.28314867149101e-16\\
1.17585858585859	0\\
1.1959595959596	1.68628298754379e-14\\
1.21606060606061	4.05029545256432e-16\\
1.23616161616162	2.26448489589378e-15\\
1.25626262626263	6.97365846326418e-17\\
1.27636363636364	4.9596690390982e-16\\
1.29646464646465	9.35587185726417e-16\\
1.31656565656566	1.09626112901245e-16\\
1.33666666666667	0\\
1.35676767676768	2.25947207072199e-16\\
1.37686868686869	5.35020961316457e-16\\
1.3969696969697	7.75473961139693e-17\\
1.41707070707071	3.1465290530235e-16\\
1.43717171717172	7.97790565372057e-17\\
1.45727272727273	4.0447443374412e-16\\
1.47737373737374	5.33069660242874e-16\\
1.49747474747475	7.27357287755528e-15\\
1.51757575757576	1.89545349113277e-15\\
1.53767676767677	1.70716415190594e-15\\
1.55777777777778	1.29711056710372e-16\\
1.57787878787879	8.75898680185332e-17\\
1.5979797979798	4.03167398456038e-14\\
1.61808080808081	3.9386740221716e-14\\
1.63818181818182	0.464950482096571\\
};
\addlegendentry{$Q_L$}

\addplot [color=mycolor1, style=dashed, line width=2.0pt, forget plot]
  table[row sep=crcr]{%
1.63818181818182	0.464950482096571\\
1.65828282828283	0.46719150539431\\
1.67838383838384	0.469393373986991\\
1.69848484848485	0.471557052363953\\
1.71858585858586	0.473683467173658\\
1.73868686868687	0.475773511077002\\
1.75878787878788	0.477828078377475\\
1.77888888888889	0.479847990299609\\
1.7989898989899	0.481834091867768\\
1.81909090909091	0.483787176157398\\
1.83919191919192	0.485708017087822\\
1.85929292929293	0.487597362219089\\
1.87939393939394	0.489455949467017\\
1.89949494949495	0.491284498046746\\
1.91959595959596	0.493083694273051\\
1.93969696969697	0.494854196409685\\
1.95979797979798	0.496596659853927\\
1.97989898989899	0.49831172098871\\
2	0.500000001912869\\
};

\addplot[gray, densely dashed, thick] coordinates {(1.23616161616162,0) (1.23616161616162,0.948683298006654)};

\addplot[gray, densely dashed, thick] coordinates {(0.572828282828283,0) (0.572828282828283,0.316227766016838)};

\addplot[gray, densely dashed, thick] coordinates {(1.64,0) (1.64,0.7071)};
\addplot[gray, densely dashed, thick] coordinates {(0,0.7071) (1.61,0.7071)};

\end{axis}
\end{tikzpicture}%

%% file: cubicsignal0.tex
%
%
\definecolor{mycolor1}{rgb}{0.00000,0.44700,0.74100}%
\definecolor{mycolor2}{rgb}{0.85000,0.32500,0.09800}%
\definecolor{mycolor3}{rgb}{0.92900,0.69400,0.12500}%
\begin{tikzpicture}

\begin{axis}[%
axis x line=bottom,
axis y line=left,
xlabel near ticks,
ylabel near ticks,
trim axis left,
trim axis right,
width=\linewidth,
height=0.75\linewidth,
clip=true,
clip mode=individual,
at={(0\linewidth,0\linewidth)},
scale only axis,
xmin=0.5,
xmax=2,
xlabel style={ at={(axis description cs:1,0)}, anchor=north west, yshift=-2pt, font=\color{white!15!black}},
xlabel={$\gamma$},
ylabel style={  at={(axis description cs:0,1)}, anchor=south east, xshift=-4pt, rotate=-90, font=\color{white!15!black}},
ymin=0,
ymax=1.1,
ylabel={$\theta$},
axis background/.style={fill=white},
title style={font=\bfseries},
xtick={0.572828282828283,1.23616161616162,1.64},
xticklabels={${\gamma}'_l$,${\gamma}'_m$, ${\gamma}'_h$},
ytick={.1,.5, .9},
yticklabels={$\thmin$,$\theta_0$,$\thmax$},
extra y ticks={.7071},
extra y tick labels={$\sqrt{\theta_0}$},
extra y tick style={
    yticklabel style={opacity=0},
    tick style={draw=none}
},
legend style={at={(0.98,0.5)}, anchor=south east, fill=none, draw=none, row sep=1.6pt}
]

\addplot [color=mycolor3, line width=2.0pt]
  table[row sep=crcr]{%
0.01	0.1\\
0.0301010101010101	0.1\\
0.0502020202020202	0.1\\
0.0703030303030303	0.1\\
0.0904040404040404	0.1\\
0.11050505050505	0.1\\
0.130606060606061	0.1\\
0.150707070707071	0.1\\
0.170808080808081	0.1\\
0.190909090909091	0.1\\
0.211010101010101	0.1\\
0.231111111111111	0.1\\
0.251212121212121	0.1\\
0.271313131313131	0.1\\
0.291414141414141	0.1\\
0.311515151515152	0.1\\
0.331616161616162	0.1\\
0.351717171717172	0.1\\
0.371818181818182	0.1\\
0.391919191919192	0.1\\
0.412020202020202	0.1\\
0.432121212121212	0.1\\
0.452222222222222	0.1\\
0.472323232323232	0.1\\
0.492424242424242	0.1\\
0.512525252525252	0.1\\
0.532626262626263	0.1\\
0.552727272727273	0.1\\
0.572828282828283	0.1\\
0.592929292929293	0.101139018404114\\
0.613030303030303	0.109603755475391\\
0.633131313131313	0.117706735266608\\
0.653232323232323	0.125474196311092\\
0.673333333333333	0.132929759412334\\
0.693434343434343	0.140094712290197\\
0.713535353535354	0.146988354414786\\
0.733636363636364	0.153628189017213\\
0.753737373737374	0.160030147005491\\
0.773838383838384	0.166208781796516\\
0.793939393939394	0.172177387969594\\
0.814040404040404	0.177948182391453\\
0.834141414141414	0.183532360103343\\
0.854242424242424	0.188940265333302\\
0.874343434343434	0.194181420396009\\
0.894444444444444	0.1992646310816\\
0.914545454545455	0.204198066079699\\
0.934646464646465	0.208989291133424\\
0.954747474747475	0.213645334396717\\
0.974848484848485	0.218172758665998\\
0.994949494949495	0.222577660872118\\
1.0150505050505	0.226865737969372\\
1.03515151515152	0.231042317083465\\
1.05525252525253	0.235112389902616\\
1.07535353535354	0.239080627807816\\
1.09545454545455	0.242951417594278\\
1.11555555555556	0.246728883574593\\
1.13565656565657	0.250416893014174\\
1.15575757575758	0.254019109770351\\
1.17585858585859	0.257538971080748\\
1.1959595959596	0.260979717986544\\
1.21606060606061	0.264344425208189\\
1.23616161616162	0.267635992188503\\
1.25626262626263	0.190596722365797\\
1.27636363636364	0.192453033329539\\
1.29646464646465	0.194223671830755\\
1.31656565656566	0.19591382943925\\
1.33666666666667	0.197528312518884\\
1.35676767676768	0.199071568746802\\
1.37686868686869	0.20054772931015\\
1.3969696969697	0.201960630394798\\
1.41707070707071	0.203313829902919\\
1.43717171717172	0.204610652667746\\
1.45727272727273	0.205854177979928\\
1.47737373737374	0.207047295128743\\
1.49747474747475	0.208192684164753\\
1.51757575757576	0.20929286620357\\
1.53767676767677	0.210350183899475\\
1.55777777777778	0.211366837497469\\
1.57787878787879	0.212344883257741\\
1.5979797979798	0.213286254375428\\
1.61808080808081	0.214192755849957\\
1.63818181818182	0.216178950995434\\
1.65828282828283	0.218267898921626\\
1.67838383838384	0.220330139475734\\
1.69848484848485	0.222366053023899\\
1.71858585858586	0.224376024155448\\
1.73868686868687	0.226360439222546\\
1.75878787878788	0.228319665973888\\
1.77888888888889	0.230254096644875\\
1.7989898989899	0.232164092673166\\
1.81909090909091	0.234050028099381\\
1.83919191919192	0.235912271025404\\
1.85929292929293	0.237751187810671\\
1.87939393939394	0.239567133103658\\
1.89949494949495	0.241360460968745\\
1.91959595959596	0.243131521083252\\
1.93969696969697	0.244880667371979\\
1.95979797979798	0.246608239740548\\
1.97989898989899	0.248314573553209\\
2	0.249999999043566\\
};
\addlegendentry{$\underline{\theta}_Q$}

\addplot [color=mycolor2, line width=2.0pt]
  table[row sep=crcr]{%
0.01	0.9\\
0.0301010101010101	0.9\\
0.0502020202020202	0.9\\
0.0703030303030303	0.9\\
0.0904040404040404	0.9\\
0.11050505050505	0.9\\
0.130606060606061	0.9\\
0.150707070707071	0.9\\
0.170808080808081	0.9\\
0.190909090909091	0.9\\
0.211010101010101	0.9\\
0.231111111111111	0.9\\
0.251212121212121	0.9\\
0.271313131313131	0.9\\
0.291414141414141	0.9\\
0.311515151515152	0.9\\
0.331616161616162	0.9\\
0.351717171717172	0.9\\
0.371818181818182	0.9\\
0.391919191919192	0.9\\
0.412020202020202	0.9\\
0.432121212121212	0.9\\
0.452222222222222	0.9\\
0.472323232323232	0.9\\
0.492424242424242	0.9\\
0.512525252525252	0.9\\
0.532626262626263	0.9\\
0.552727272727273	0.9\\
0.572828282828283	0.9\\
0.592929292929293	0.9\\
0.613030303030303	0.9\\
0.633131313131313	0.9\\
0.653232323232323	0.9\\
0.673333333333333	0.9\\
0.693434343434343	0.9\\
0.713535353535354	0.9\\
0.733636363636364	0.9\\
0.753737373737374	0.9\\
0.773838383838384	0.9\\
0.793939393939394	0.9\\
0.814040404040404	0.9\\
0.834141414141414	0.9\\
0.854242424242424	0.9\\
0.874343434343434	0.9\\
0.894444444444444	0.9\\
0.914545454545455	0.9\\
0.934646464646465	0.9\\
0.954747474747475	0.9\\
0.974848484848485	0.9\\
0.994949494949495	0.9\\
1.0150505050505	0.9\\
1.03515151515152	0.9\\
1.05525252525253	0.9\\
1.07535353535354	0.9\\
1.09545454545455	0.9\\
1.11555555555556	0.9\\
1.13565656565657	0.9\\
1.15575757575758	0.9\\
1.17585858585859	0.9\\
1.1959595959596	0.9\\
1.21606060606061	0.9\\
1.23616161616162	0.9\\
1.25626262626263	0.665822926826003\\
1.27636363636364	0.652828758598044\\
1.29646464646465	0.640434285104179\\
1.31656565656566	0.628603192741169\\
1.33666666666667	0.617301824115321\\
1.35676767676768	0.606499015253145\\
1.37686868686869	0.5961658892862\\
1.3969696969697	0.58627558060979\\
1.41707070707071	0.576803174755017\\
1.43717171717172	0.56772544461221\\
1.45727272727273	0.559020752741851\\
1.47737373737374	0.550668937355703\\
1.49747474747475	0.542651194364654\\
1.51757575757576	0.534949924938019\\
1.53767676767677	0.527548705260144\\
1.55777777777778	0.520432146495921\\
1.57787878787879	0.513585819932113\\
1.5979797979798	0.506996224804712\\
1.61808080808081	0.500650689234219\\
1.63818181818182	0.5\\
1.65828282828283	0.5\\
1.67838383838384	0.5\\
1.69848484848485	0.5\\
1.71858585858586	0.5\\
1.73868686868687	0.5\\
1.75878787878788	0.500000324734637\\
1.77888888888889	0.500000231737635\\
1.7989898989899	0.5\\
1.81909090909091	0.5\\
1.83919191919192	0.5\\
1.85929292929293	0.5\\
1.87939393939394	0.500000015587685\\
1.89949494949495	0.500000012589841\\
1.91959595959596	0.500000004716927\\
1.93969696969697	0.5\\
1.95979797979798	0.5\\
1.97989898989899	0.500000088453089\\
2	0.5\\
};
\addlegendentry{$\theta_H$}

\addplot [color=mycolor1, style=dashed, line width=2.0pt]
  table[row sep=crcr]{%
0.01	0.100412541512161\\
0.0301010101010101	0.100001966384451\\
0.0502020202020202	0.100000002758397\\
0.0703030303030303	0.100000000002381\\
0.0904040404040404	0.100000000000003\\
0.11050505050505	0.1\\
0.130606060606061	0.1\\
0.150707070707071	0.1\\
0.170808080808081	0.1\\
0.190909090909091	0.1\\
0.211010101010101	0.1\\
0.231111111111111	0.1\\
0.251212121212121	0.1\\
0.271313131313131	0.1\\
0.291414141414141	0.1\\
0.311515151515152	0.1\\
0.331616161616162	0.1\\
0.351717171717172	0.1\\
0.371818181818182	0.1\\
0.391919191919192	0.1\\
0.412020202020202	0.1\\
0.432121212121212	0.1\\
0.452222222222222	0.1\\
0.472323232323232	0.1\\
0.492424242424242	0.1\\
0.512525252525252	0.1\\
0.532626262626263	0.1\\
0.552727272727273	0.1\\
0.572828282828283	0.1\\
0.592929292929293	0.101139018404114\\
0.613030303030303	0.109603755475391\\
0.633131313131313	0.117706735266608\\
0.653232323232323	0.125474196311092\\
0.673333333333333	0.132929759412334\\
0.693434343434343	0.140094712290197\\
0.713535353535354	0.146988354414786\\
0.733636363636364	0.153628189017213\\
0.753737373737374	0.160030147005491\\
0.773838383838384	0.166208781796516\\
0.793939393939394	0.172177387969594\\
0.814040404040404	0.177948182391453\\
0.834141414141414	0.183532360103343\\
0.854242424242424	0.188940265333302\\
0.874343434343434	0.194181420396009\\
0.894444444444444	0.1992646310816\\
0.914545454545455	0.204198066079699\\
0.934646464646465	0.208989291133424\\
0.954747474747475	0.213645334396717\\
0.974848484848485	0.218172758665998\\
0.994949494949495	0.222577660872118\\
1.0150505050505	0.226865737969372\\
1.03515151515152	0.231042317083465\\
1.05525252525253	0.235112389902616\\
1.07535353535354	0.239080627807816\\
1.09545454545455	0.242951417594278\\
1.11555555555556	0.246728883574593\\
1.13565656565657	0.250416893014174\\
1.15575757575758	0.254019109770351\\
1.17585858585859	0.257538971080748\\
1.1959595959596	0.260979717986558\\
1.21606060606061	0.26434442520819\\
1.23616161616162	0.267635992188505\\
1.25626262626263	0.190596722365797\\
1.27636363636364	0.192453033329539\\
1.29646464646465	0.194223671830756\\
1.31656565656566	0.19591382943925\\
1.33666666666667	0.197528312518884\\
1.35676767676768	0.199071568746802\\
1.37686868686869	0.20054772931015\\
1.3969696969697	0.201960630394798\\
1.41707070707071	0.203313829902919\\
1.43717171717172	0.204610652667746\\
1.45727272727273	0.205854177979928\\
1.47737373737374	0.207047295128744\\
1.49747474747475	0.208192684164758\\
1.51757575757576	0.209292866203571\\
1.53767676767677	0.210350183899476\\
1.55777777777778	0.211366837497469\\
1.57787878787879	0.212344883257741\\
1.5979797979798	0.213286254375454\\
1.61808080808081	0.214192755849982\\
1.63818181818182	0.5\\
1.65828282828283	0.5\\
1.67838383838384	0.5\\
1.69848484848485	0.5\\
1.71858585858586	0.5\\
1.73868686868687	0.5\\
1.75878787878788	0.499999999992784\\
1.77888888888889	0.499999999998809\\
1.7989898989899	0.5\\
1.81909090909091	0.5\\
1.83919191919192	0.5\\
1.85929292929293	0.5\\
1.87939393939394	0.499999998831029\\
1.89949494949495	0.499999999956397\\
1.91959595959596	0.499999999998386\\
1.93969696969697	0.5\\
1.95979797979798	0.5\\
1.97989898989899	0.499999999694132\\
2	0.5\\
};
\addlegendentry{$\theta_L$}

\addplot[gray, densely dashed, thick] coordinates {(1.23616161616162,0) (1.23616161616162,0.9)};

\addplot[gray, densely dashed, thick] coordinates {(0.572828282828283,0) (0.572828282828283,0.1)};

\addplot[gray, densely dashed, thick] coordinates {(1.64,0) (1.64,0.5)};
\addplot[gray, densely dashed, thick] coordinates {(0,0.5) (1.61,0.5)};

\end{axis}
\end{tikzpicture}%